\documentclass[aps, pra, reprint, superscriptaddress, floatfix]{revtex4-1}

\usepackage[free-standing-units=true]{siunitx}
\usepackage{xspace} 
\usepackage{graphicx}
\usepackage{dsfont}
\usepackage{epstopdf}
\usepackage{bm, amsmath, amssymb, bbold}
\usepackage{float}
\usepackage{placeins}
\usepackage{xcolor}
\usepackage[T1]{fontenc}
\usepackage[utf8]{inputenc}
\usepackage[separate-uncertainty=true]{siunitx}
\usepackage{boldline,multirow}
\usepackage{ifthen}
\usepackage[hidelinks]{hyperref}

\newcommand{\ket}[1]{\left|#1\right\rangle}
\newcommand{\bra}[1]{\left\langle#1\right|}
\newcommand{\braket}[2]{\left\langle#1\middle|#2\right\rangle}
\newcommand{\mr}[1]{\mathrm{#1}}

\DeclareSIUnit{\torr}{Torr}

\newcommand{\nocontentsline}[3]{}
\newcommand{\stoptoc}{\let\addcontentsline\nocontentsline}
\newcommand{\resumetoc}{\let\addcontentsline\origcontentsline}
\let\origcontentsline\addcontentsline

\newcommand{\Twchange}{\tau_{\delta \omega}}

\newcommand{\Ttrace}{\tau_\mathrm{trace}}
\newcommand{\Ttraceval}{\SI{1}{\milli\second}}
\newcommand{\dt}{\Delta t}
\newcommand{\troval}{\SI{20}{\micro\second}}      
\newcommand{\sigmagainval}{\SI{20}{\nano\second}} 
\newcommand{\twaitgain}{\SI{1}{\micro\second}}    

\newcommand{\epsgain}{\epsilon_\mr{gain}}
\newcommand{\epstwo}{\epsilon_2}
\newcommand{\etwo}{\varepsilon_2}
\newcommand{\omegatwo}{\omega_2}
\newcommand{\gfour}{g_4}

\newcommand{\diss}[1]{\mathcal{D}\,[#1]}

\newcommand{\omegaqubit}{\omega_\mathrm{a}}
\newcommand{\omegacav}{\omega_\mathrm{b}}
\newcommand{\omegar}{\omega_\mathrm{r}}
\newcommand{\omegagain}{\omega_\mathrm{gain}}
\newcommand{\omegap}{\omega_\mathrm{p}}        
\newcommand{\Deltaab}{\Delta_\mr{ab}}
\newcommand{\wRa}{\omega_\mr{R,a}}             
\newcommand{\wRb}{\omega_\mr{R,b}}             
\newcommand{\Obig}[1]{\mathcal{O}\!\left(#1\right)}
\newcommand{\Ec}{E_\mr{C}}                     
\newcommand{\Ej}{E_\mr{J}}                     

\newcommand{\kappaqubit}{\kappa_\mathrm{a}}
\newcommand{\kappacav}{\kappa_\mathrm{b}}
\newcommand{\kappacavcoup}{\kappa_\mathrm{b,c}}
\newcommand{\kappacavint}{\kappa_\mathrm{b,int}}
\newcommand{\kappacavpin}{\kappa_\mathrm{b,pin}}
\newcommand{\kappacavlosses}{\kappa_\mathrm{b,\ell}}

\newcommand{\kappaone}[1]{\kappa_\mathrm{a}^{#1}}
\newcommand{\kappadeph}[1]{\kappa_\mr{\phi}^{#1}}
\newcommand{\kappadephgen}[1]{\kappa_\mr{\phi,(E)}^{#1}}
\newcommand{\Gammadec}[1]{\Gamma_{#1}}
\newcommand{\Gammaecho}[1]{\Gamma_{#1,\mathrm{E}}}
\newcommand{\Gammagen}[1]{\Gamma_{#1,(\mathrm{E})}}
\newcommand{\kappag}[1]{\kappa_\mathrm{gain}^{#1}}
\newcommand{\kappagain}{\kappa_\mathrm{gain}}

\newcommand{\Tone}[1]{T_1^{#1}}

\newcommand{\Ttwo}[1]{T_2^{#1}}
\newcommand{\Ttwoe}[1]{T_{2\mathrm{E}}^{#1}}

\newcommand{\Tphi}[1]{T_{\mr{\phi}}^{#1}}
\newcommand{\Tphie}[1]{T_{\mr{\phi,E}}^{#1}}

\newcommand{\gstate}{\ket{\mr{g}}}
\newcommand{\estate}{\ket{\mr{e}}}
\newcommand{\fstate}{\ket{\mr{f}}}

\newcommand{\transge}{\ket{\mr{g}}\text{-}\ket{\mr{e}}}
\newcommand{\transef}{\ket{\mr{e}}\text{-}\ket{\mr{f}}}
\newcommand{\transgf}{\ket{\mr{g}}\text{-}\ket{\mr{f}}}
\newcommand{\transfh}{\ket{\mr{f}}\text{-}\ket{\mr{h}}}
\newcommand{\transfhh}{\ket{\mr{f}}\text{-}\ket{\mr{h}}}

\newcommand{\chidisp}{\chi}
\newcommand{\nth}[1]{n_\mr{th,{#1}}}    
\newcommand{\nbar}{\bar{n}}
\newcommand{\xigain}{\xi_\mathrm{gain}}   
\newcommand{\domegageac}{\delta\omega_\mathrm{ge}^\mathrm{ac}}  
\newcommand{\xitwo}{\xi_2}
\newcommand{\Ap}{A_\mathrm{p}}            

\newcommand{\ggain}{g_\mathrm{gain}}

\newcommand{\aop}{\hat{a}}
\newcommand{\aod}{\hat{a}^\dagger}
\newcommand{\bop}{\hat{b}}
\newcommand{\bod}{\hat{b}^\dagger}
\newcommand{\dm}{\hat{\rho}}
\newcommand{\rhoa}{\hat{\rho}_\mathrm{a}}
\newcommand{\Ha}{\hat{H}_\mathrm{a}}
\newcommand{\Hb}{\hat{H}_\mathrm{b}}
\newcommand{\Hgain}{\hat{H}_\mathrm{gain}}

\newcommand{\Hstark}{\hat{H}_{\mathrm{Stark}}}
\newcommand{\Htwo}{\hat{H}_2}
\newcommand{\Hchi}{\hat{H}_\mr{disp}}
\newcommand{\Omegagf}{\Omega_\mr{gf}}           
\newcommand{\sigmaxgf}{\hat{\sigma}_x^\mr{gf}}  

\newcommand{\msig}[1]{\mathcal{M}_{#1}}
\newcommand{\rcon}[1]{A_{#1}}
\newcommand{\omegaro}{\omega_\mathrm{RO}}

\newcommand{\dtgaincalib}{\SI{2}{\micro\second}}
\newcommand{\roamp}{RO Amp.}

\newcommand{\rot}[2]{R^{#2}\!\left(#1\right)}

\newcommand{\ggainvalone}{\SI{31}{\kilo\hertz}}
\newcommand{\ggainvaltwo}{\SI{261}{\kilo\hertz}}
\newcommand{\ggainvalthree}{\SI{1.5}{\mega \hertz}}
\newcommand{\ggaintracea}{\SI{0.19}{\mega\hertz}}
\newcommand{\ggaintraceb}{\SI{0.56}{\mega\hertz}}
\newcommand{\ggaintracec}{\SI{0.97}{\mega\hertz}}
\newcommand{\ggaintraced}{\SI{2.13}{\mega\hertz}}
\newcommand{\xigainconv}{\SI[separate-uncertainty=true]{3.219(6)}{\per\volt}}
\newcommand{\Nfloquet}{50}
\newcommand{\wpi}{\omega_\pi}   

\newcommand{\fqubith}{\SI{4.872}{\giga\hertz}}

\newcommand{\anharmh}{\SI{172}{\mega\hertz}}  
\newcommand{\anharm}{\alpha}                   
\newcommand{\Ljvalh}{\SI{10.81}{\nano\henry}}
\newcommand{\nthhval}{(2.5 \pm 0.08)\cdot 10^{-2}}
\newcommand{\nthhvalsim}{0.025} 
\newcommand{\omegajumph}{\SI{6}{\kilo\hertz}}

\newcommand{\fresh}{\SI{7.442}{\giga\hertz}}
\newcommand{\kappacavtotal}{\SI{858}{\kilo\hertz}}
\newcommand{\kappacawavval}{\SI{791\pm1}{\kilo\hertz}}
\newcommand{\kappacavintval}{\SI{67\pm1}{\kilo\hertz}}
\newcommand{\kappalorfit}{(858 \pm 36)\, \text{kHz}}
\newcommand{\ggaindwval}{\SI{155}{\kilo\hertz}}
\newcommand{\kappagaindwval}{\SI{222}{\kilo\hertz}}
\newcommand{\nthhcavval}{$(9.9 \pm 1)\cdot 10^{-4}$}

\newcommand{\ggainsimonetwolevel}{\SI{60}{\kilo\hertz}}
\newcommand{\ggainsimonemax}{\SI{3}{\mega\hertz}}
\newcommand{\ggainsimonepeak}{\SI{290}{\kilo\hertz}}
\newcommand{\Tonefsimpeak}{\SI{15.4}{\milli\second}}
\newcommand{\Tonefsimceil}{\SI{42}{\milli\second}}
\newcommand{\Tonefsimceiltwokappa}{\SI{82}{\milli\second}}

\newcommand{\dispshifth}{(2.013 \pm 0.006) \, \text{MHz}}

\newcommand{\Tonege}{\SI{133.6 \pm 2.2}{\micro\second}}
\newcommand{\Toneef}{\SI{59.1 \pm 3}{\micro\second}}
\newcommand{\Ttwoge}{\SI{107.2 \pm 2.8}{\micro\second}}
\newcommand{\Ttwogeecho}{\SI{138 \pm 6}{\micro\second}}
\newcommand{\Ttwogf}{\SI{71.2 \pm 3.5}{\micro\second}}
\newcommand{\Ttwogfecho}{\SI{114 \pm 9}{\micro\second}}

\newcommand{\Tonefvalsel}{\SI{2.22 \pm 0.06}{\milli\second}}
\newcommand{\Tonegvalsel}{\SI{1.67 \pm 0.34}{\milli\second}}
\newcommand{\Ttwogfvalsel}{\SI{7.49 \pm 0.47}{\micro\second}}
\newcommand{\Ttwoegfvalsel}{\SI{65 \pm 4}{\micro\second}}
\newcommand{\Ttwogfmedvalsel}{\SI{22.9}{\micro\second}}

\newcommand{\ggainvalsel}{\SI{133}{\kilo\hertz}}
\newcommand{\kappagainsel}{\SI{165}{\kilo\hertz}}
\newcommand{\gatefiedelitysel}{99.58 \pm 0.01 \,\%}

\newcommand{\TtwoeRatioText}{2.1}

\newcommand{\TonefRatio}{$\times16.7$}
\newcommand{\ToneavgRatio}{$\times14.7$}
\newcommand{\TonegRatio}{$\times12.6$}
\newcommand{\TtwoRatio}{$\div 14$}
\newcommand{\TtwoeRatio}{$\div 2.1$}

\newcommand{\sigmageval}{\SI{8}{\nano\second}}
\newcommand{\sigmagfval}{\SI{20}{\nano\second}}

\newcommand{\taugeval}{\SI{48}{\nano\second}}
\newcommand{\taugfval}{\SI{120}{\nano\second}}

\newcommand{\taugfsigmaval}{\SI{20}{\nano\second}}

\newcommand{\taulimgf}{\SI{1.85}{\nano\second}}
\newcommand{\fid}{\mathcal{F}}

\newcommand{\ggainsimtwomax}{\SI{200}{\kilo\hertz}}

\newcommand{\figwidth}{90mm}
\newcommand{\figwidthWide}{180mm}

\newcommand{\Lj}{L_\mathrm{J}}
\newcommand{\bareToneh}{\SI{133.57 \pm 2.2}{\micro\second}}
\newcommand{\bareTonehef}{\SI{59.1 \pm 3}{\micro\second}}
\newcommand{\kappadephecho}[1]{\kappa_{\phi,\mathrm{E}}^{#1}}
\newcommand{\gstateh}{\ket{\mathrm{g}}}
\newcommand{\estateh}{\ket{\mathrm{e}}}
\newcommand{\fstateh}{\ket{\mathrm{f}}}
\newcommand{\hstateh}{\ket{\mathrm{h}}}
\newcommand{\gstatehb}{\bra{\mathrm{g}}}
\newcommand{\estatehb}{\bra{\mathrm{e}}}
\newcommand{\fstatehb}{\bra{\mathrm{f}}}
\newcommand{\hstatehb}{\bra{\mathrm{h}}}
\newcommand{\transgeh}{\ket{\mathrm{g}}\text{-}\ket{\mathrm{e}}}
\newcommand{\transefh}{\ket{\mathrm{e}}\text{-}\ket{\mathrm{f}}}
\newcommand{\transgfh}{\ket{\mathrm{g}}\text{-}\ket{\mathrm{f}}}

\newcommand{\Vint}{\hat{V}_\mr{int}}        
\newcommand{\Rrot}{\hat{R}(\theta_\mr{err})}
\newcommand{\etaint}{\eta}                  
\newcommand{\taudwell}{\tau_\mr{dwell}}     
\newcommand{\taudwellavg}{\langle \taudwell \rangle}  
\newcommand{\thetaerravg}{\langle \thetaerr \rangle} 
\newcommand{\thetaerr}{\theta_\mr{err}}     
\newcommand{\rhob}{\hat{\rho}_\mathrm{b}}   
\newcommand{\Sstab}{\hat{S}_\mr{b}}         
\newcommand{\Pstab}{\hat{P}_\mr{b}}         
\newcommand{\backact}{B}                    
\newcommand{\detinf}{E}                     
\newcommand{\ploss}{p}                      
\newcommand{\psiloss}{\psi_\mr{loss}}       

\newcommand{\zerol}{\ket{0_\mr{L}}}
\newcommand{\onel}{\ket{1_\mr{L}}}
\newcommand{\zeroe}{\ket{0_\mr{E}}}
\newcommand{\onee}{\ket{1_\mr{E}}}
\newcommand{\logstate}{\ket{\Psi_\mr{L}}}
\newcommand{\alphacat}{\alpha_\mr{cat}}
\newcommand{\catp}{\ket{\mathcal{C}^+_{\alphacat}}}
\newcommand{\catm}{\ket{\mathcal{C}^-_{\alphacat}}}
\newcommand{\catpj}{\ket{\mathcal{C}^+_{i\alphacat}}}
\newcommand{\catmj}{\ket{\mathcal{C}^-_{i\alphacat}}}
\newcommand{\Nasim}{3}
\newcommand{\Nbsimbin}{5}
\newcommand{\Nbsimcat}{30}
\newcommand{\Nbsimgkp}{150}
\newcommand{\anharmsimval}{\SI{180}{\mega\hertz}}
\newcommand{\chisimval}{\SI{1}{\mega\hertz}}
\newcommand{\betagkpval}{\SI{200}{\kilo\hertz}}
\newcommand{\alphacatval}{3.5}
\newcommand{\rgkpval}{1.4}                  
\newcommand{\kappaonesimval}{1/\SI{140}{\micro\second}}

\newenvironment{linenomath}{}{}  

\newenvironment{ac}
{\begin{center}
		\textsc{author contributions}\\[1ex]
	\end{center}
}

\newenvironment{ack}
{\begin{center}
		\textsc{acknowledgements}\\[1ex]
	\end{center}
}

\newenvironment{datav}
{\begin{center}
		\textsc{data and code availability statement}\\[1ex]
	\end{center}
}

\newenvironment{sicon}
{\begin{center}
		\textsc{supplementary information content}\\[1ex]
	\end{center}
}

\newcommand*\patchAmsMathEnvironmentForLineno[1]{%
  \expandafter\let\csname old#1\expandafter\endcsname\csname #1\endcsname
  \expandafter\let\csname oldend#1\expandafter\endcsname\csname end#1\endcsname
  \renewenvironment{#1}%
     {\linenomath\csname old#1\endcsname}%
     {\csname oldend#1\endcsname\endlinenomath}}%
\newcommand*\patchBothAmsMathEnvironmentsForLineno[1]{%
  \patchAmsMathEnvironmentForLineno{#1}%
  \patchAmsMathEnvironmentForLineno{#1*}}%
\AtBeginDocument{%
\patchBothAmsMathEnvironmentsForLineno{equation}%
\patchBothAmsMathEnvironmentsForLineno{align}%
\patchBothAmsMathEnvironmentsForLineno{flalign}%
\patchBothAmsMathEnvironmentsForLineno{alignat}%
\patchBothAmsMathEnvironmentsForLineno{gather}%
\patchBothAmsMathEnvironmentsForLineno{multline}%
}

\begin{document}

\title{The Gain-Engineered Transmon}

\author{I.~Yang}
\thanks{These authors contributed equally.}
\affiliation{PSI Center for Photon Science, 5232 Villigen PSI, Switzerland}
\author{F.~Adinolfi}
\thanks{These authors contributed equally.}
\affiliation{PSI Center for Photon Science, 5232 Villigen PSI, Switzerland}
\author{A.~Bruno}
\affiliation{PSI Center for Photon Science, 5232 Villigen PSI, Switzerland}
\affiliation{Swiss Nanoscience Institute, University of Basel, Klingelbergstrasse 82, 4056 Basel, Switzerland}
\author{V.~Hasanuzzaman Kamrul}
\affiliation{PSI Center for Photon Science, 5232 Villigen PSI, Switzerland}
\author{D.~Z.~Haxell}
\affiliation{PSI Center for Photon Science, 5232 Villigen PSI, Switzerland}
\author{A.~Grimm}
\email{alexander.grimm@psi.ch}
\affiliation{PSI Center for Photon Science, 5232 Villigen PSI, Switzerland}

\date{\today}
\begin{abstract}
The interaction between a qubit and its environment can be engineered such that one error channel dominates over all others, resulting in noise bias.
This property enables error correction codes to focus on the dominant error type, thereby significantly reducing the number of physical systems required for fault-tolerant quantum computation.
However, engineering noise bias typically introduces complexity at the physical system level, which decreases its usefulness by limiting scalability. 
Here, we introduce and experimentally realize a noise-biased qubit in a standard transmon-readout resonator circuit, one of the most common superconducting architectures, by only adding a single microwave tone.
We encode the qubit in the transmon $\gstate$- and $\fstate$-states, and engineer a frequency-selective gain channel that counteracts single-photon loss errors between the computational states.
We demonstrate an order-of-magnitude enhancement in relaxation time compared to the $\transge$ encoding, conceding only a factor-of-two decrease in the echo-coherence time. 
Furthermore, we show that this qubit is compatible with fast, high-fidelity operations.
Our results open a path towards using this system as a simple building-block for hardware-efficient quantum error detection and correction schemes.
\end{abstract}
\maketitle

Quantum computers hold the potential to solve problems that are fundamentally intractable for classical machines~\cite{Dalzell2025}, with promising applications ranging from the simulation of complex quantum systems to cryptography and optimization~\cite{Reiher2017,Lee21,Goings22,Shor,Gidney25}. Quantum states are however highly susceptible to interactions with a noisy environment, which induce errors that corrupt the quantum information they encode. Practical quantum computation therefore requires quantum error correction (QEC) protocols that detect and correct these errors. These protocols typically encode each logical qubit into many physical ones, an approach that imposes a large hardware overhead~\cite{Shor95,Steane96,nielsen_quantum_2011,Terhal15}. Despite rapid progress in the control of large-scale quantum devices~\cite{Acharya24,Bluvstein25,Team26,Daguerre25,RyanAnderson24,Besedin26}, it is an outstanding challenge to reach the number of physical qubits needed for practical applications.
A promising route to reduce this overhead exploits noise-biased qubits, in which certain error channels are dominant relative to others, enabling tailored QEC codes that achieve the same logical error rate with substantially fewer physical components~\cite{Tuckett18,Tuckett19,BonillaAtaides2021,Darmawan2021}.

Although noise bias arises naturally in some systems~\cite{Noiri22,Pi26,Harty14,Balasubramanian09}, actively engineering the bias can yield noise-bias ratios that are orders of magnitude beyond those of conventional qubits~\cite{cochrane_macroscopically_1999,Mirrahimi2014, Goto16, puri_engineering_2017}. This requires careful design of the system Hamiltonian and its dissipative interactions with the environment~\cite{Harrington22}, a capability well-suited to the superconducting circuits platform~\cite{Blais2021, Kapit17}. However, to date, most experimental realizations of noise-biased qubits have relied on specialized circuit elements~\cite{Grimm2020,Lescanne20,Marquet2024}, whose intricate design and complex control can make it challenging to scale to large system sizes. Consequently, an open question remains: can noise bias be engineered in a simple qubit platform, while remaining compatible with fast and high-fidelity operations?

Here, we answer the question in the affirmative by introducing a noise-biased qubit, called the gain-engineered transmon, realized within the multilevel structure of a standard superconducting transmon. We encode the qubit in the $\gstate$ and $\fstate$ transmon states and implement a frequency-selective single-photon gain channel on the $\transef$ transition. When a single-photon relaxation event maps $\fstate \to \estate$, the engineered gain rapidly drives $\estate \to \fstate$, autonomously correcting loss errors. As a consequence, loss errors are strongly suppressed compared to dephasing errors, generating a qubit with an engineerable noise bias. The large anharmonicity of the transmon ensures that the gain acts selectively on the $\transef$ transition without coupling the computational states, and simultaneously enables fast single-qubit control via a two-photon drive on the $\transgf$ transition.

\begin{figure}
    \includegraphics[angle = 0, width = \figwidth]{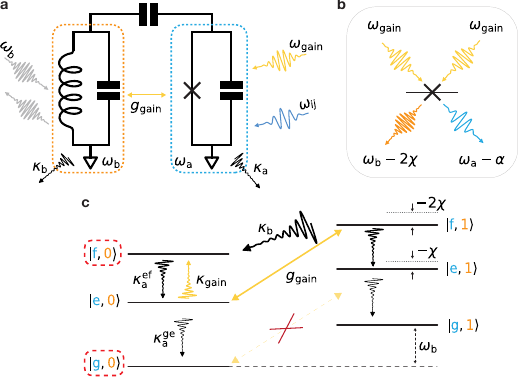}
    \caption{\label{fig:one}\textbf{Concept of the gain-engineered transmon.}
    \textbf{a,} Circuit schematic of the transmon-cavity system. The transmon (cavity) is shown inside the blue (orange) dashed box, with associated frequency $\omegaqubit$ ($\omegacav$) and single-photon loss rate $\kappaqubit$ ($\kappacav$). The two-mode squeezing interaction of rate $\ggain$ is shown as a yellow double arrow. Colored pulse arrows indicate the readout tone at $\omegacav$ (gray), the gain pump at $\omegagain$ (yellow), and the qubit control drive at $\omega_{ij}$ (blue).
    \textbf{b,} Schematic representation of the four-wave mixing process generating the engineered single-photon gain. The Josephson junction nonlinearity converts two pump photons at $\omegagain$ into one transmon photon at $\omegaqubit - \anharm$ and one cavity photon at $\omegacav-2\chidisp$.
    \textbf{c,} Energy-level diagram of the joint transmon-cavity system. Transmon (cavity) states are labeled in blue (orange), and the computational states $\ket{\mr{g},0}$ and $\ket{\mr{f},0}$ have dashed frames. A state-dependent dispersive shift is indicated by the difference between solid and dotted lines, giving $-\chidisp$ and $-2\chidisp$ for the $\ket{\mr{e},1}$ and $\ket{\mr{f},1}$ states, respectively. Black decaying arrows denote single-photon loss on the transmon-cavity state. The two-mode squeezing interaction $\ggain$ (solid yellow arrow) coherently couples $\ket{\mr{e},0}$ and $\ket{\mr{f},1}$, and photon loss from the cavity at rate $\kappacav$ irreversibly maps $\ket{\mr{f},1} \to \ket{\mr{f},0}$, resulting in an effective gain with rate $\kappagain$ (yellow decaying arrow). The yellow dashed arrow marks suppressed, off-resonant gain on the $\transge$ transition.}
\end{figure}

Our experimental platform consists of a standard superconducting transmon coupled to a microwave cavity, as shown in Fig.~\ref{fig:one}a. The transmon has a fundamental frequency $\omegaqubit/2\pi = \fqubith$, an anharmonicity $\anharm/2\pi = \anharmh$, and a residual thermal population $\nth{a}=\nthhval$. For the $\transge$ transition, the single-photon lifetime is $\Tone{\mr{e}} = 1/\kappaone{\mr{ge}} =  \Tonege$, Ramsey coherence time is $\Ttwo{\mr{ge}} =  \Ttwoge$ and echo coherence time is $\Ttwoe{\mr{ge}}=\Ttwogeecho$. The single photon lifetime for the $\transef$ transition is $\Tone{\mr{ef}} = 1/\kappaone{\mr{ef}} =  \Toneef$, which approximately follows $\kappaone{\mr{ef}}\approx 2\kappaone{\mr{ge}}$. The cavity has a frequency $\omegacav/2\pi = \fresh$ and a single-photon loss rate $\kappacav/2\pi = \kappacavtotal$. In our experiment, this mode serves a dual purpose as the readout resonator for qubit state measurement and as the lossy buffer mode. For details on the experimental setup and system parameters, see Supplementary Information Secs.~I and~II. 

The resulting system Hamiltonian is
\begin{equation}
    \hat{H}/\hbar = \Delta_\mr{a}\aod \aop - \frac{\anharm}{2}\aop^{2\dagger} \aop^2 -\chidisp \aod\aop \bod\bop.
    \label{eq:hamiltonian_main}
\end{equation}
Here, $\aop$ ($\bop$) is the annihilation operator of the transmon (cavity) and $\chidisp=2\anharm\lambda^2$ is the dispersive-shift interaction strength, with $\lambda$ the hybridization factor between the transmon and the cavity~\cite{Blais2021}. The Hamiltonian is defined in a frame rotating at $\omegaqubit-\Delta_\mr{a}$ for the transmon and $\omegacav$ for the cavity.

To realize the engineered gain, we use an incoherent single-photon excitation process which is frequency-selective on the $\transef$ transition. Specifically, we apply a single-photon parametric pump at frequency ${\omegagain = (\omegaqubit - \anharm + \omegacav - 2 \chidisp)/2}$. This activates a four-wave mixing process that converts two pump photons into one transmon photon at the $\transef$ frequency $\omegaqubit - \anharm$, and one cavity photon at $\omegacav - 2\chidisp$, generating a two-mode squeezing interaction of strength $\ggain$ between the transmon and cavity modes (Fig.~\ref{fig:one}b). In the limit $\ggain \ll \anharm$, this interaction coherently couples only the joint states $\ket{\mr{e},0}$ and $\ket{\mr{f},1}$, where the first (second) index labels the transmon (cavity) state. This realizes the Hamiltonian term
\begin{equation}
    \Hgain/\hbar = \sqrt{2}\ggain \left(\ket{\mr{f},1}\bra{\mr{e},0} + \ket{\mr{e},0}\bra{\mr{f},1}\right).
\end{equation}
For $\Delta_\mr{a} =\alpha +2\chidisp$ in Eq.~\eqref{eq:hamiltonian_main}, $\Hgain$ is a time-independent interaction (see Supplementary Information Sec.~IV.A).
Single-photon loss from the cavity at rate $\kappacav$ renders the engineered exchange unidirectional such that the intrinsic loss of the readout cavity is converted into an incoherent gain in the transmon. In the adiabatic elimination limit $\ggain \ll \kappacav$, the transmon-cavity dynamics reduce to an engineered gain with rate $\kappagain = 8\ggain^2/\kappacav$~\cite{Li24,Gertler21} and linewidth $\kappacav$ (Fig.~\ref{fig:one}c). This results in an engineered dissipator $\kappagain\diss{\fstateh\estatehb}$ (see Supplementary Information Secs.~IV.B and V.C).

\begin{figure*}
    \includegraphics[angle = 0, width = \figwidthWide]{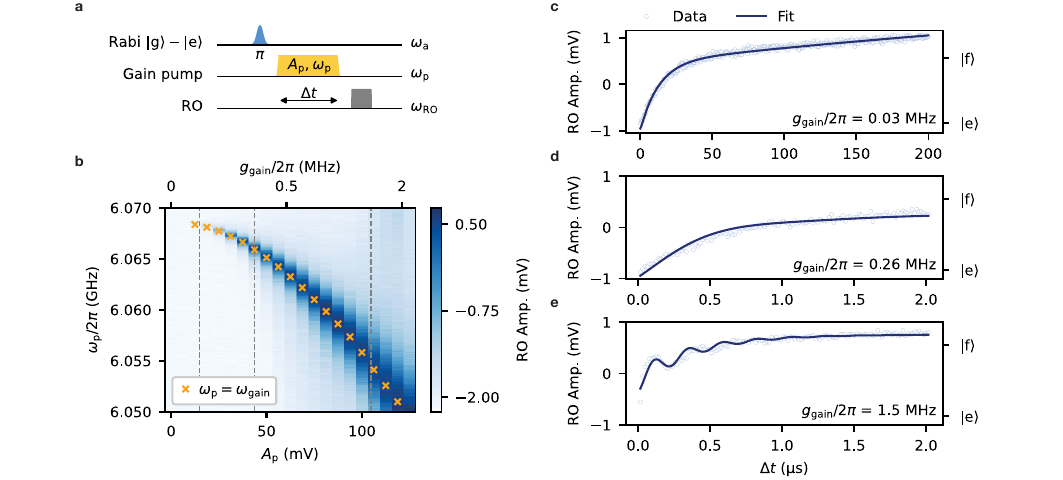}
    \caption{\label{fig:two}\textbf{Calibration of the engineered gain.}
    \textbf{a,} Pulse sequence for calibrating the gain-pump frequency $\omegagain$ and engineered interaction strength $\ggain$.
    \textbf{b,} Readout amplitude (\roamp) as a function of pump frequency $\omegap$ and pump amplitude $\Ap$. Orange crosses mark the resonance condition $\omegap = \omegagain$, at which the engineered gain drives the $\estate \to \fstate$ transition. 
    The calibrated top axis shows the corresponding $\ggain$ for $\omegap=\omegagain$. Here, the readout is performed at $\omegaro = \omegacav - 2 \chidisp$. The vertical dashed lines indicate the $\ggain$ values for the data plotted in panels \textbf{c}, \textbf{d}, and \textbf{e}.
    \textbf{c, d, e} Readout signal as a function of $\dt$ at $\ggain/2\pi = \ggainvalone$ (\textbf{c}), $\ggainvaltwo$ (\textbf{d}) and $\ggainvalthree$ (\textbf{e}). The left axis shows the \roamp\ and the right axis marks the RO value for the transmon population in $\estate$ and $\fstate$ calibrated from Rabi measurements. Open circles are experimental data and solid lines are numerical simulations using the predicted $\ggain$, with only the readout contrast as a free parameter. Here, the readout is performed at $\omegaro = \omegacav - \chidisp$.}
\end{figure*}

To begin, we calibrate the single-photon pump that activates the engineered gain. We find the resonant pump frequency $\omegagain$ as a function of the applied pump amplitude following the pulse sequence in Fig.~\ref{fig:two}a. After preparing the transmon in $\estate$, we apply the gain pump at frequency $\omegap$ and amplitude $\Ap$ for a fixed duration $\dt = \dtgaincalib$, before reading out the transmon state. The gain-pump pulse has a flat-top Gaussian envelope with peak output voltage $\Ap$ and rise time of $3\sigma = \SI{30}{\nano\second}$. On resonance, the engineered gain excites the transmon from $\estateh$ to $\fstateh$, changing the readout amplitude. Sweeping $\omegap$ and $\Ap$, we map the resonance condition $\omegap = \omegagain$ (orange crosses in Fig.~\ref{fig:two}b). The resonance shifts with $\Ap$ due to the Stark shift of the pump on the transmon frequency. To obtain the interaction strength $\ggain=\lambda\,|\domegageac|/2$, we measure the pump-induced shift of the $\transge$ transition $\domegageac$ as a function of $\Ap$ and $\omegagain$ (see Supplementary Information Sec.~V.A).

We validate the calibrated $\ggain$ and our description of the process by measuring the time dynamics of the system and comparing it against numerical simulations. Fixing $\omegap = \omegagain$, we record the readout signal versus pump duration $\dt$ at several values of $\Ap$, allowing us to infer the transmon population dynamics (Fig.~\ref{fig:two}a).  Figures~\ref{fig:two}c, d and e show results for $\ggain/2\pi = \ggainvalone$, $\ggainvaltwo$ and $\ggainvalthree$ respectively. For reference, we include the readout signals corresponding to $\estate$- or $\fstate$-state populations obtained from independent Rabi measurements. The readout signal increases from the $\estate$- to the $\fstate$-value while the gain pump is applied, at a rate which increases with pump amplitude. For $\ggain/2\pi = \ggainvalthree$, the interaction is sufficiently strong to induce coherent oscillations in population between $\estate$ and $\fstate$. The measurement data (open circles) at each $\ggain$ agree well with a master equation simulation (solid line) which takes the predicted $\ggain$ and other independently measured system parameters as input, leaving the readout contrast as the only free parameter per trace. Note that saturation in the readout contrast above the $\fstate$-value can be attributed to leakage to higher transmon levels, induced when the gain pump is applied for a long time (Fig.~\ref{fig:two}c) or with a large amplitude (Fig.~\ref{fig:two}e). For details of the numerical simulation and the full range of measured dynamics, see Supplementary Information Sec.~V.D.1-2.

\begin{figure*}
    \includegraphics[angle = 0, width = \figwidthWide]{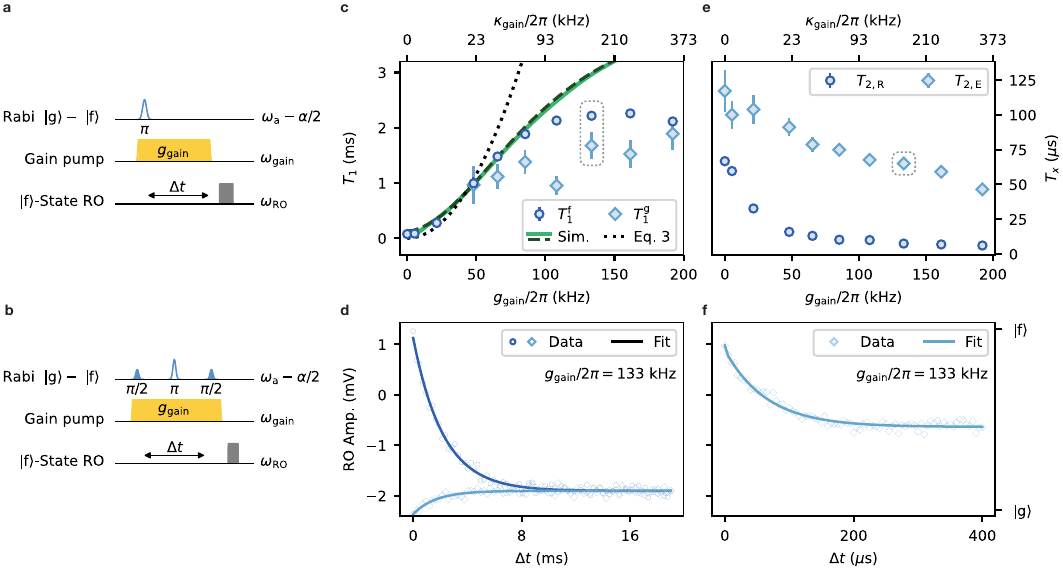}
    \caption{\label{fig:three}\textbf{Relaxation and coherence times of the gain-engineered transmon as a function of engineered gain rate $\ggain$.}
    \textbf{a,} Pulse sequence for $\Tone{\mr{f}}$ measurements in \textbf{c},\textbf{d}. The white fill in the $\pi_{\mr{gf}}$ pulse indicates its use for the $\Tone{\mr{f}}$ measurement only.
    \textbf{b,} Pulse sequence for $\Ttwo{\mr{gf}}$ and $\Ttwoe{\mr{gf}}$ measurements in \textbf{e},\textbf{f}. The white fill in the $\pi_{\mr{gf}}$ pulse indicates that it is used only for the $\Ttwoe{\mr{gf}}$ measurement.
    \textbf{c,} Decay times $\Tone{\mr{f}}$ ($\Tone{\mr{g}}$) out of the $\fstate$ ($\gstate$) state as a function of $\ggain$. The $\Tone{\mr{f}}$ begins to saturate at $\ggain/2\pi = \ggainvalsel$ ($\kappagain/2\pi = \kappagainsel$), which we indicate by the dotted gray box. The solid light green line (dashed dark green line) is a numerical simulation of $\Tone{\mr{f}}$ ($\Tone{\mr{g}}$) with experimentally measured parameters, while the dotted black line is an analytical theory prediction derived from the multilevel transmon decoherence model (Eq.~\eqref{eq:t1_analytic}). Note that $\Tone{\mr{g}}$ for the lowest $\ggain$ values is omitted because the measurement contrast is too low to extract a decay time.
    \textbf{d,} Relaxation measurement traces for $\ggain/2\pi = \ggainvalsel$. Circles (diamonds) are data for the $\Tone{\mr{f}}$ ($\Tone{\mr{g}}$) measurement, and the solid line is a simple exponential fit which yields $\Tone{\mr{f}} = \Tonefvalsel$ ($\Tone{\mr{g}} = \Tonegvalsel$).
    \textbf{e,} Ramsey coherence time $\Ttwo{\mr{gf}}$ (dark blue circles) and echo coherence time $\Ttwoe{\mr{gf}}$ (light blue diamonds) as a function of $\ggain$.
    \textbf{f,} Echo measurement trace for $\ggain/2\pi = \ggainvalsel$. Circles are data and solid line is a fit (see Supplementary Information Sec..~III.C), which yields $\Ttwoe{\mr{gf}} = \Ttwoegfvalsel$.}
\end{figure*}

Next, we apply the calibrated engineered gain and evaluate the enhancement of the relaxation times. To do so, we measure the decay time $\Tone{\mr{f}}$ ($\Tone{\mr{g}}$) out of the $\fstate$ ($\gstate$) state for different values of $\ggain$ (Figs.~\ref{fig:three}a,c,d). The data are fitted to simple exponential functions to obtain $\Tone{\mr{g,f}}$ (see Supplementary Information Sec.~III.C). To gain insight into the observed trend of relaxation times, we also perform numerical simulations of the engineered gain process in the transmon-cavity system using only independently-measured parameters (see Supplementary Information Sec.~IV.D.3). 
We find two main features in the data and simulations (Fig.~\ref{fig:three}c), which we discuss in the following. 

Firstly, $\Tone{\mr{g,f}}$ rise as $\ggain$ increases in both the measurement (circles and diamonds) and simulation (dashed and solid lines).
Both measurement and simulations also initially follow a simple analytical expression (dotted line) derived from the decoherence model in the $\{\gstate,\estate,\fstate\}$ subspace for the regime $\kappaone{\mr{ef}} \ll \kappagain$ (see Supplementary Information Sec.~III.D):
\begin{equation}
    \Tone{\mr{f}} \approx \frac{\kappag{}}{\kappaone{\mr{ge}}\kappaone{\mr{ef}}}.
    \label{eq:t1_analytic}
\end{equation}
From Eq.~\eqref{eq:t1_analytic}, we see that an increase in the engineered gain rate $\kappagain$ directly translates to an increase in $\Tone{\mr{f}}$, since the engineered gain counteracts relaxation on the $\fstate\to\estate$ decay channel $\kappaone{\mr{ef}}$. 

Secondly, the measured $\Tone{\mr{g,f}}$ reach a maximum and start to decrease for large $\ggain$. At $\ggain/2\pi=\ggainvalsel$, we measure $\Tone{\mr{f}} = \Tonefvalsel$ and $\Tone{\mr{g}} = \Tonegvalsel$ (Fig.~\ref{fig:three}d). Here, our simulation yields $\Tone{\mr{g}}=\Tone{\mr{f}}=\SI{3}{\milli\second}$. This value is mainly limited by thermal excitation from $\gstate$ to $\estate$, which the gain process converts into bit flips. This indicates that thermal excitations also constrain our experimental relaxation times.
We attribute the difference between the experimental and simulated values to noise processes not included in our model, such as additional pump-induced heating or multi-photon transitions~\cite{Dai2026,Dumas24}. 
Note that the latter is expected to affect the computational states differently, which could explain the measured difference between $\Tone{\mr{f}}$ and $\Tone{\mr{g}}$ at large $\ggain$.

\begin{figure}
    \includegraphics[angle = 0, width = \figwidth]{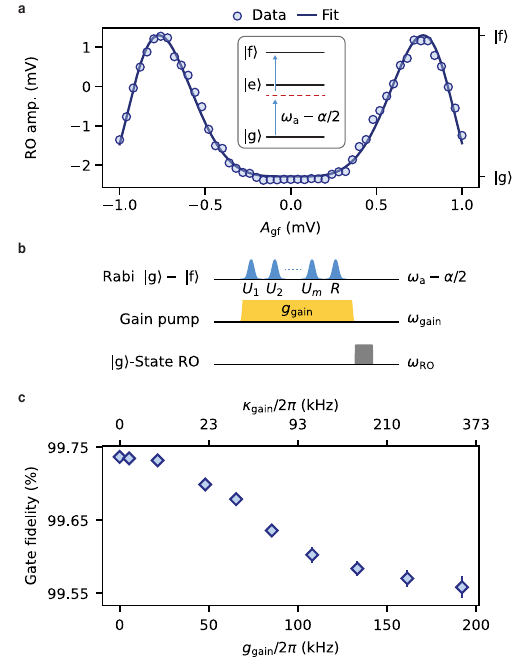}
    \caption{\label{fig:four}\textbf{Two-photon gate calibration and randomized benchmarking.}
    \textbf{a,} Readout amplitude as a function of the drive amplitude $A_{\mr{gf}}$, showing Rabi oscillations of the $\transgf$ transition. 
    Circles are data and the solid line is a fit. The right axis marks the readout values corresponding to $\gstate$ and $\fstate$. The inset shows the process underlying the two-photon $\transgf$ drive (see text).
    \textbf{b,} Pulse sequence for randomized benchmarking measurement used in \textbf{c}.
    \textbf{c,} Average gate fidelity as a function of $\ggain$ (bottom axis) and $\kappagain$ (top axis).}
\end{figure}

We now present Ramsey and echo coherence measurements (Figs.~\ref{fig:three}b,e,f).
We extract $\Ttwo{\mr{gf}}$ and $\Ttwoe{\mr{gf}}$ by fitting to an analytical decoherence model for the $\{\gstate,\estate,\fstate\}$ subspace (see Supplementary Information Sec.~III.C).
The measured $\Ttwo{\mr{gf}}$ drops sharply with $\ggain$, while $\Ttwoe{\mr{gf}}$ decreases more slowly (Fig.~\ref{fig:three}e). 
In contrast to the experimental result, simulated coherence times do not change with $\ggain$ (see Supplementary Information Sec.~V.D.4). In principle, the engineered gain should not degrade coherence times since it acts on the $\transgf$ superposition only after single-photon loss has already collapsed the superposition into $\estate$.
We therefore attribute the difference to processes that are not captured in the simulation. 
Notably, the data presented in Figs.~\ref{fig:two}c--e,~\ref{fig:three} and \ref{fig:four}, and the qubit characterization, were acquired in an interleaved manner over a continuous four-day period. This sets a very low frequency bound on the noise spectrum sampled by the $\Ttwo{\mr{gf}}$ measurements (see Supplementary Information Sec.~V.E). 
Therefore, $\Ttwo{\mr{gf}}$ will be significantly impacted by low-frequency decoherence processes, such as qubit frequency shifts induced by drifts in gain-pump amplitude. Such noise is expected to be partially refocused in the echo sequence, which contributes to $\Ttwoe{\mr{gf}}>\Ttwo{\mr{gf}}$~\cite{Bylander11}.

Beyond extending the relaxation time of the qubit, we demonstrate coherent control of the $\transgf$ transition while the gain pump is applied. To perform single-qubit gates, we synthesize a two-photon drive that exchanges population between the $\gstate$ and $\fstate$ states without directly populating $\estate$. 
We apply an external drive at $\omegaqubit-\anharm/2$, such that two drive photons excite the $\gstate-\fstate$ transition via a virtual level detuned from $\estate$ (see the inset of Fig.~\ref{fig:four}a)~\cite{Schirk25,Xia25}.
This process can occur simultaneously with the engineered gain.
The conceptual speed limit of this gate is set by the detuning to the nearest unwanted transition. For the $\transgf$ encoding, this is $2/\anharm=\taulimgf$ due to the detuning of the drive from the single-photon $\transge$ and $\transef$ transitions. 
In the experiment, we apply $6\sigma$ Gaussian pulses with $\sigma=\taugfsigmaval$.
Faster gate speeds can be achieved by using derivative removal by adiabatic gate (DRAG) pulse shaping methods~\cite{Motzoi13}.

In Fig.~\ref{fig:four}a, we show the resulting Rabi oscillations on the $\transgf$ transition as a function of the pulse amplitude $A_{\mr{gf}}$ for $\ggain=0$ (for the full range of $\ggain$, see Supplementary Information Sec.~V.B). The oscillations are well-described by \roamp\ $\propto \cos (\pi (A_{\mr{gf}}/A_\pi)^2)$, where $A_\pi$ is the amplitude for a $\pi$ pulse. This dependence is explained by the mixing process involving two drive photons, which leads to a scaling of the oscillation frequency with the drive amplitude squared (see Supplementary Information Sec.~V.B). 

To assess the performance of the gate operations, we use randomized benchmarking to measure the single-qubit gate fidelity $\fid$ in the presence of the engineered gain~\cite{Knill08,Chow09,Magesan11}. 
We ramp on the gain pump to an interaction rate $\ggain$, apply a sequence of $m$ random Clifford gates $U$ to the qubit initialized in $\gstate$ followed by a recovery gate $R$, ramp off the gain pump, and perform dispersive readout (Fig.~\ref{fig:four}b). Ideally, $R$ would return the qubit to $\gstate$. In practice, gate errors accumulate with increasing sequence length, leading to an exponential decay to a completely mixed state.
Fitting each decay curve yields the gate fidelities shown in Fig.~\ref{fig:four}c. The extracted values slightly decrease as we increase $\ggain$ such that we measure $\fid = \gatefiedelitysel$ at $\ggain/2\pi = \ggainvalsel$ ($\kappagain/2\pi = \kappagainsel$).
Note that these gates are calibrated using amplitude-Rabi measurements, such as the one presented Fig~\ref{fig:four}a, but are not otherwise optimized.
Therefore, the numbers reported here are lower bounds of the coherence-limited fidelities. 
Nevertheless, $\fid$ remains above $99.55\%$ across the full range of $\ggain$ explored, demonstrating that the engineered gain is fully compatible with high-fidelity single-qubit operations.

We conclude our characterization by comparing the relaxation and coherence times of the gain-engineered transmon at $\ggain/2\pi = \ggainvalsel$ ($\kappagain/2\pi = \kappagainsel$) with the conventional undriven transmon $\transge$ encoding. The results are summarized in Table~\ref{tab:encoding_comparison}. At this operating point, we enhance the decay time $\Tone{\mr{f}}$ ($\Tone{\mr{g}}$) to a value \TonefRatio\ (\TonegRatio) longer than $\Tone{\mr{e}}$, corresponding to an average relaxation time enhancement of \ToneavgRatio. Here, we have taken the relaxation times of both the $\gstate$ and $\estate$ states to be the same in the $\transge$ encoding.
The increased relaxation time comes at a cost in the echo coherence time $\Ttwoe{\mr{gf}}$, which is only \TtwoeRatioText-times shorter than $\Ttwoe{\mr{ge}}$. 
While $\Ttwo{\mr{gf}}$ for the full four-day averaging duration is much shorter than $\Ttwo{\mr{ge}}$ (\TtwoRatio), averaging over individual four-minute measurement runs yields higher values, with a median of $\Ttwogfmedvalsel$ at this $\ggain$ (see Supplementary Information Sec.~V.E). 
This supports our conclusion that low-frequency noise processes acting over multiple days degrade the measured $\Ttwo{\mr{gf}}$.
This $\Ttwo{\mr{gf}}$ therefore likely underestimates the coherence time relevant for QEC applications, where calibration steps can be interleaved to counteract slow frequency drifts.

\begin{table}
    \centering
    \begin{tabular}{lccc}
        \hline & \multicolumn{2}{c}{Manifold $\ket{\mr{g}} - \ket{i}$} & \\
        \cline{2-3}
         & $\transge$ & $\transgf$ & Ratio \\
        \hline
        $\ggain$ & $0$ & $2\pi \times \ggainvalsel$ & $-$ \\
        $\Tone{i}$ & $\Tonege$ & $\Tonefvalsel$ & \TonefRatio \\
        $\Tone{\mr{g}}$ & $\Tonege$ & $\Tonegvalsel$ & \TonegRatio \\
        $\Ttwo{}$ & $\Ttwoge$ & $\Ttwogfvalsel$ & \TtwoRatio \\
        $\Ttwoe{}$ & $\Ttwogeecho$ & $\Ttwoegfvalsel$ & \TtwoeRatio \\
        \hline
    \end{tabular}
    \caption{\label{tab:encoding_comparison}\textbf{Comparison of the $\transge$ encoding and $\transgf$ encoding at $\ggain/2\pi = \ggainvalsel$.} The comparison metrics are the decay time from $\ket{i}$ ($\gstate$) state, $\Tone{i}$ ($\Tone{\mr{g}}$), Ramsey and echo coherence times $\Ttwo{}$ and $\Ttwoe{}$, for the $\transge$ and $\transgf$ encodings.
    }
\end{table}
 
In summary, our results show that engineered gain can be used to autonomously correct loss errors in a qubit encoded in the $\gstateh$ and $\fstateh$ states of a standard transmon. 
This process can strongly increase the qubit noise-bias while remaining compatible with high gate fidelities.
Note that these figures of merit are moreover \textit{in situ} tuneable via $\ggain$.
For example, smaller $\ggain$ values trade a lower noise bias for higher coherence and gate fidelities. 

Reaching even larger noise bias will require three main improvements in our experimental setup. 
We have identified $\nth{\mr{a}}$ as a significant limitation on $\Tone{\mr{g}}$ and $\Tone{\mr{f}}$ in our current experiment (see Supplementary Information V.D.3). Strategies to reduce this value include improved thermalization and filtering at infrared frequencies~\cite{kerschbaum2026}. We note here that this is not a fundamental constraint of our scheme, as thermal populations as low as $\approx0.1\%$ have been reported in literature~\cite{Jin15}.
A subsequent limitation we anticipate is the finite linewidth of the engineered gain $\kappacav$.
At larger $\ggain$ values, our numerical simulations show that the process will off-resonantly drive the $\transge$ and $\transfh$ transitions leading to a reduction in $\Tone{\mr{g,f}}$ (see Supplementary Information Sec.~V.D.3). This can be mitigated by substituting the readout cavity with a dedicated element such as a multi-stage lossy filter~\cite{Putterman2022,Thorbeck24,Direkci26}, enabling higher $\kappagain$ rates while preserving sharp frequency selectivity. This same strategy could also be used to address the next anticipated limit: the breakdown of the adiabatic elimination regime due to the finite resonator bandwidth $\kappacav$. Finally, the experiment can be improved by including Floquet simulations of the transmon-resonator system in our design process to avoid spurious multi-photon resonances~\cite{Dai2026}.

We envision the gain-engineered transmon as a simple and modular component of larger QEC schemes in two main areas.
Firstly, it can be used as a noise-biased ancilla qubit for fault-tolerant error detection, where the engineered gain can suppress ancilla decay errors from propagating into the logical qubit~\cite{Puri2019,Direkci26} (see Supplementary Information Sec.~VI). 
As an added benefit, the $\transgf$ encoding results in a two-fold increase in the interaction strength as compared to the $\transge$ encoding, reducing the impact of decoherence during error-correction protocols.
Moreover, such decoherence can be further mitigated by introducing echo pulses in some detection schemes~\cite{Eickbusch2022}.
Secondly, the gain-engineered transmon could serve as a data qubit in QEC codes adapted for noise-biased qubits such as the XZZX surface code~\cite{BonillaAtaides2021,Darmawan2021}, which would require a bias-preserving CX gate. While it can be shown that this gate is prohibited for two-level system qubits~\cite{Guillaud19}, no such no-go theorem exists for multi-level systems like the one we use here. Such a gate could therefore be enabled via an additional transmon level, or replaced altogether by multi-qubit quantum-non-demolition measurements~\cite{Vuillot26}.

Finally, the simplicity of the gain-engineered transmon naturally lends itself to several extensions. 
One can, for example, replace the $\fstate$ state with an even higher transmon level $\ket{n}$ in the qubit encoding and use similar dissipative processes to those presented here to engineer gain to $\ket{n}$ and loss to $\gstate$.
Such a scheme is predicted to achieve exponential suppression of bit flips in $n$, while the phase-flip rate increases only linearly~\cite{Direkci26}.
The inclusion of engineered loss to the ground state has the added benefit of mitigating the impact of $\nth{\mr{a}}$ in the experiment. 
One can furthermore consider analogous schemes for the dissipative stabilization of qudit states. 
These examples indicate that the combination of dissipation engineering with the multilevel structure of the transmon showcased in this work constitutes a rich and versatile playground for quantum information processing applications.

\begin{datav}
Numerical simulations were performed using a Python-based open-source software (QuTiP). The data and code that support the findings of this study are available from the corresponding authors upon reasonable request.
\end{datav}

\begin{ack}
We acknowledge useful discussions with Su Direkci, Connor T.~Hann and Yiwen Chu.
We thank the cleanroom operations teams of the Paul Scherrer Institute and of the Binnig and Rohrer Nanotechnology Center for their help and support. This work was supported by the Swiss National Science Foundation Grant No. 200021\_1972551, the Swiss Nanoscience Institute Fellowship Grant No. P2101 and the Swiss State Secretariat for Education, Research and Innovation (SERI).
\end{ack}

\begin{ac}
The original concept of the gain-engineered transmon was proposed by F.A., and developed by F.A. and A.G. with input from all coauthors. The device was fabricated by V.H.K. and A.B.. The experimental setup was assembled by I.Y., F.A., A.B. and D.Z.H.. All measurements were performed by I.Y.. The data analysis was carried out by I.Y. and F.A., with input from D.Z.H. and A.G.. The theoretical description and numerical simulations of the experiment were developed by F.A. and A.B. with input from I.Y., D.Z.H., and A.G.. The theory of fault-tolerant syndrome extraction was developed by F.A. with input from A.G.. The project was supervised by A.G.. The manuscript was written by I.Y., F.A., D.Z.H. and A.G. with input from all authors.
\end{ac}

\begin{sicon}
The accompanying Supplementary Information contains details on: the experimental setup (Section I, Supplementary Figs.~S1,2, Supplementary Table~S1); the full system parameters (Section II, Supplementary Table~S2, Supplementary Figs.~S3,4); the decoherence model (Section III, Supplementary Table~S3); the theoretical model of the engineered gain (Section IV); the experimental measurements and numerical simulations with the engineered gain (Section V, Supplementary Figs.~S5---10) and the use of the gain-engineered transmon as a noise bias ancilla qubit (Section VI, Supplementary Figs.~S11---14, Supplementary Table~S4---6).
\end{sicon}

\bibliography{bibliography}

\setcounter{figure}{0}
\setcounter{table}{0}
\setcounter{equation}{0}

\renewcommand{\thefigure}{S\arabic{figure}}
\renewcommand{\thetable}{S\arabic{table}}
\renewcommand{\theequation}{S\arabic{equation}}

\newpage
\onecolumngrid
\begin{center}
	\textsc{\Large{Supplementary Information}}
\end{center}

\section{Experimental setup}
\subsection{Sample: description, fabrication and design}

\begin{figure}[h]
  \centering
  \includegraphics[width=\linewidth]{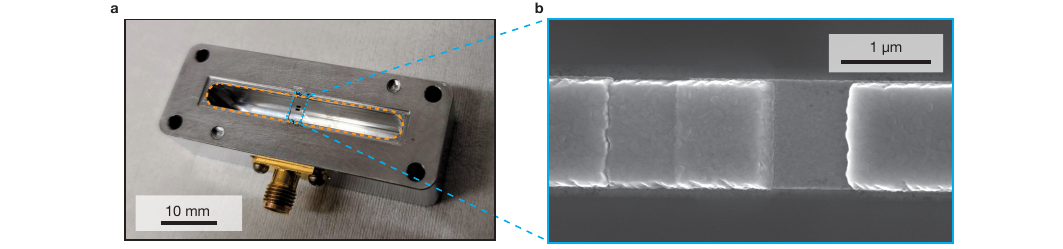}
    \caption{\textbf{Sample pictures.} \textbf{a,} Photograph of the aluminium 3D microwave cavity (dashed orange outline) with the transmon chip mounted inside (dashed blue outline). \textbf{b,} Scanning electron micrograph of the Josephson junction from a representative test sample. }
    \label{fig:heatmon_sample}
\end{figure}

The device comprises a superconducting transmon qubit and a 3D microwave cavity, shown in Fig.~\ref{fig:heatmon_sample}a. The transmon qubit is fabricated on a sapphire substrate and consists of a single Josephson junction connecting two superconducting capacitor pads. The capacitor pads are made of tantalum and the junction is fabricated from aluminium and aluminium oxide. A representative scanning electron microscope image of the junction is shown in Fig.~\ref{fig:heatmon_sample}b.

The device fabrication follows the recipe described in Refs.~\cite{Place21,Crowley23}. We start from a C-plane HEMEX sapphire wafer coated by Star Cryoelectronics with a \SI{200}{\nano\meter} alpha-tantalum film. The supplier cleans the sapphire in piranha solution and oxygen plasma and deposits the alpha-tantalum at \SI{500}{\celsius}. We clean the wafer in a 2:1 H\textsubscript{2}SO\textsubscript{4}/H\textsubscript{2}O\textsubscript{2} piranha solution at \SI{100}{\celsius} for \SI{20}{\minute}, followed by rinsing in deionized water and isopropanol (IPA). We define the capacitor geometry using optical lithography. We spin-coat AZ1518 photoresist to a thickness of \SI{2}{\micro\meter}, pattern it with a Heidelberg DWL66+ laser writer, and develop the resist in AZ~726~MIF for \SI{81}{\second}. We then perform an oxygen plasma descum for \SI{15}{\second} at \SI{125}{\volt} bias in an Oxford ICP-RIE to remove resist residues in the developed areas. We etch the tantalum layer in a 1:1:1 HF/HNO\textsubscript{3}/H\textsubscript{2}O solution (Tantalum Etchant~111, Transene Company) for \SI{21}{\second} at room temperature. We remove the resist by sonicating the wafer in N-methyl-2-pyrrolidone (NMP), acetone, and IPA for \SI{5}{\minute} each, followed by a piranha clean identical to the first step and a \SI{20}{\minute} buffered oxide etch (10:1 H\textsubscript{2}O/HF). We fabricate the Josephson junction using electron-beam lithography following the Dolan-bridge technique~\cite{Dolan77}. We spin-coat a bilayer resist stack of MMA EL13 and PMMA A4 and cover it with a \SI{30}{\nano\meter} gold charge-dissipation layer. After exposure, we remove the gold layer in a potassium-iodide etch and develop the resist in a 3:1 IPA/deionized-water mixture for \SI{2}{\minute} at \SI{6}{\celsius}. We load the wafer into a Plassys MEB550SL3 electron-beam evaporation system and perform argon ion milling at \SI{400}{\volt} and $\pm\SI{45}{\degree}$ to clean the developed areas and remove the native tantalum oxide. We pump the evaporation chamber to a base pressure below \SI{5e-9}{\torr} and define the Josephson junction via double-angle shadow evaporation of aluminium at $\pm\SI{40}{\degree}$, with an intermediate static oxidation step at \SI{30}{\milli\bar} for \SI{10}{\minute} to form the tunnel barrier. The two aluminium layers have thicknesses of \SI{25}{\nano\meter} (bottom) and \SI{35}{\nano\meter} (top). We perform lift off of the resist stack in NMP heated to \SI{90}{\celsius} for more than two hours. Afterwards, we do a solvent cleaning in acetone and IPA, followed by blow-drying with a nitrogen gun.
Before dicing the wafer, we spin a protective layer of AZ1518 optical resist, approximately \SI{5}{\micro\meter} thick, onto the wafer and bake it at \SI{110}{\degreeCelsius} for \SI{2}{min}. The cleanroom team at the Paul Scherrer Institute (PSI) then dices the wafer into individual chips. During the handling process, we use an ionizing air fan to prevent electrostatic discharge from damaging the Josephson junctions. After dicing, we perform a final solvent cleaning on each chip: \SI{45}{min} in NMP at \SI{90}{\celsius}, \SI{2}{min} sonication in acetone, and \SI{2}{min} sonication in IPA, followed by blow-drying with a nitrogen gun.

The fundamental mode of a rectangular aluminium 3D cavity, with frequency $\omegacav$, serves both as the readout resonator and as the lossy buffer mode for the engineered gain. The cavity is machined from aluminium and is coupled to the external transmission line through two ports: an aluminium waveguide connected via a small aperture, and a coaxial pin inserted directly into the cavity. The waveguide acts as a Purcell filter, allowing us to increase the coupling of the cavity to the external transmission line without increasing the single-photon loss rate of the transmon $\kappaqubit$. The waveguide cutoff frequency is designed to lie above the transmon frequency but below the cavity frequency, and a matched coupler connects the waveguide to the readout line. The cavity readout tone at frequency $\omegacav$ and the engineered-gain pump at frequency $\omegagain$ are both delivered through the waveguide port. The coaxial pin is used to drive Rabi oscillations in the transmon. 

\subsection{Wiring Diagram}

\begin{figure}
    \centering
    \includegraphics[angle = 0, width = \figwidthWide]{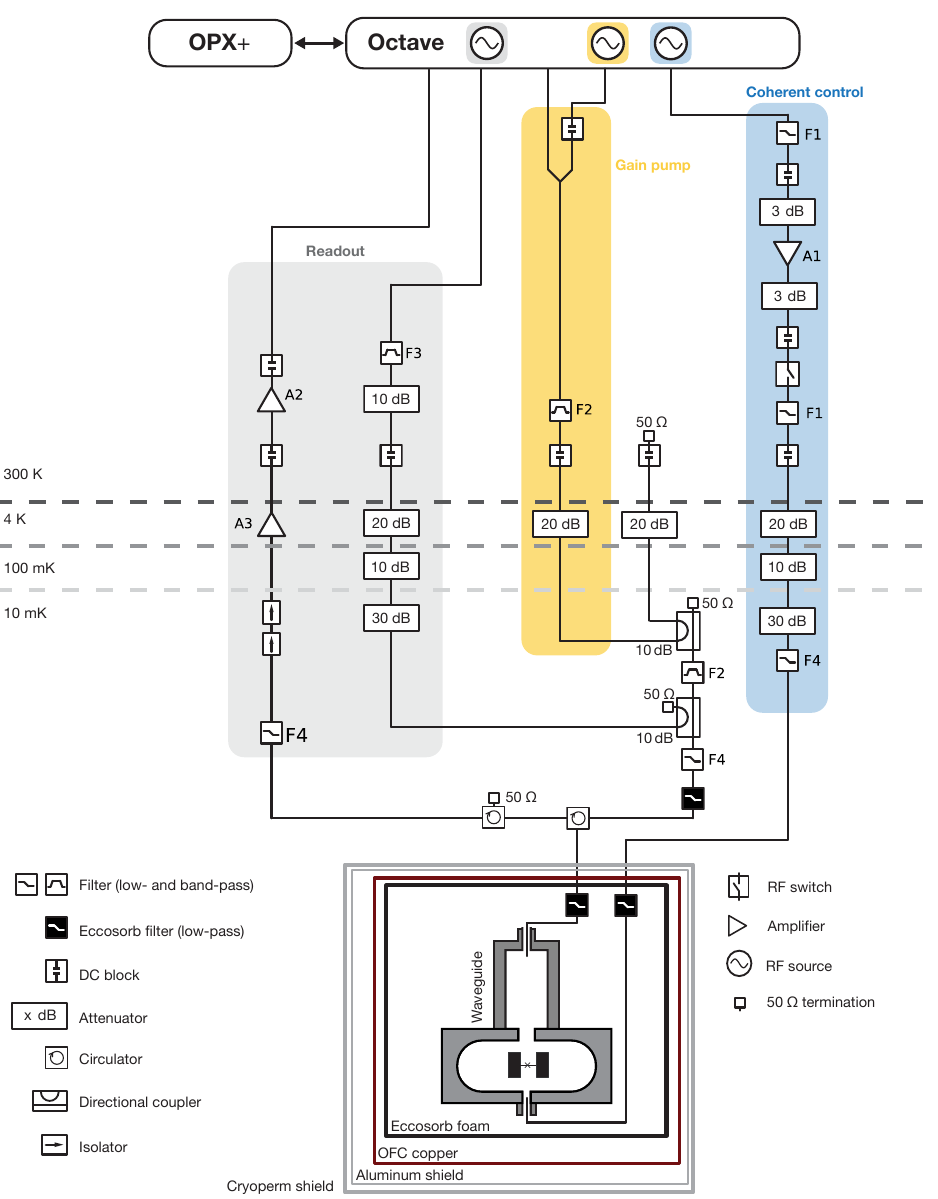}
    \caption{\textbf{Wiring diagram of the experimental setup.} Component labels correspond to those in Table~\ref{tab:wiring_diagram_heatmon}.}
    \label{fig:wiring_diagram}
\end{figure}

The complete wiring diagram of the experiment is shown in Fig.~\ref{fig:wiring_diagram}, with the main microwave and DC components summarized in Table~\ref{tab:wiring_diagram_heatmon}. The device is hosted in a Bluefors BF-LD250 dilution refrigerator (DR), cooled to temperatures below \SI{10}{\milli\kelvin}. 

We generate and demodulate control and readout signals using a Quantum Machines OPX+ combined with an Octave. We use room-temperature filtering to suppress spurious harmonics from the Octave outputs, DC blocks to avoid ground loops, and cryogenic attenuators throughout the DR to suppress thermal noise. Eccosorb filters at the mixing-chamber (MXC) stage suppress infrared radiation. We carefully filter the gain pump line at the MXC stage, providing more than \SI{30}{\decibel} of attenuation at the cavity frequency. From left to right in Fig.~\ref{fig:wiring_diagram}, the three signal lines are the readout input-output line (gray), the gain pump line (orange), and the coherent-control line (blue).

We use the readout line to probe the cavity response and to measure the state of the gain-engineered transmon via dispersive readout. We generate the readout signal, filter and attenuate it at room temperature, and send it into the DR. Inside the DR, the signal is attenuated at successive cryogenic stages, combined with the gain pump tone via a directional coupler, filtered, and delivered to the sample through a circulator. The cavity-reflected output is routed through two circulators at the MXC stage and an isolator, and then amplified by a HEMT amplifier at the \SI{4}{\kelvin} stage. At room temperature, a low-noise amplifier provides additional gain before the signal is routed to the Octave for demodulation. 

The single-photon-gain pump at frequency $\omegagain$ is delivered through the gain-pump line. Two OPX+ output ports are dedicated to this line: one for generating the calibrated $\transefh$-gain pump tone used in the experiments, and the other for diagnostic purposes. The latter was not used in the results presented in the text. 
The signal passes through a band-pass filter and a DC block before entering the DR, where it is attenuated and reaches the MXC stage. The signal enters the \SI{10}{\decibel} port of a directional coupler, and the coupled output is filtered by a sharp band-pass filter providing at least \SI{30}{\decibel} of attenuation at the cavity frequency. The other port is dissipated at the \SI{4}{\kelvin} stage and terminated at room temperature, reducing the heat load at the MXC stage. The pump signal at the MXC is then combined with the readout tone via a second directional coupler and delivered to the sample through a circulator.

We use the coherent-control line for transmon spectroscopy and single-qubit gates. We generate the signal, filter and amplify it at room temperature, and route it through a microwave switch and a filter before it enters the DR. Inside the DR, the signal is attenuated at successive cryogenic stages, further filtered and attenuated at the MXC, and delivered to the sample through a pin connected to the cavity.

The sample is mounted at the MXC stage on an oxygen-free copper bracket and wrapped in Eccosorb foam to suppress stray infrared radiation. The device is enclosed in a three-layer shield consisting of an inner oxygen-free copper layer for thermalization, an aluminium layer and an outer Cryoperm layer to suppress stray electromagnetic fields.

\begin{table}[h]
    \caption{\textbf{Microwave and DC components used in the experimental setup.} Labels correspond to those indicated in Fig.~\ref{fig:wiring_diagram}.}
    \label{tab:wiring_diagram_heatmon}
    \begin{ruledtabular}
    \renewcommand{\arraystretch}{1.2}
    \begin{tabular}{lll}
        \textbf{Label} & \textbf{Component} & \textbf{Manufacturer / Model} \\
        \hline
        \multicolumn{3}{l}{\textbf{Filters}} \\
        F1 & Low-pass filter         & Mini-Circuits ZLSS-4R8G-S+ \\
        F2 & Band-pass filter        & Mini-Circuits ZBSS-4G-S+ \\
        F3 & Band-pass filter        & Mini-Circuits VHF-5500+ \\
        F4 & Low-pass filter         & K\&L 6L250-12000/T26000-OP/O \\
        \hline
        \multicolumn{3}{l}{\textbf{Passive components}} \\
        -- & Cryogenic circulator    & Low Noise Factory LNF-CIC4\_12A \\
        -- & Cryogenic isolator      & Low Noise Factory LNF-ISISC4\_12A \\
        -- & Directional coupler     & Krytar 1822-267 \\
        -- & DC block                & Inmet 8039 \\
        -- & Room-temperature attenuators & Mini-Circuits \\
        -- & Cryogenic attenuators   & Bluefors / XMA \\
        \hline
        \multicolumn{3}{l}{\textbf{Amplifiers}} \\
        A1 & Room-temperature amplifier       & Mini-Circuits ZVA-183-S+ \\
        A2 & Low-noise room-temperature amplifier & Narda-MITEQ LNA-40-04001200-15-10P \\
        A3 & HEMT amplifier                   & Low Noise Factory LNF-LNC4\_16C \\
        -- & Microwave switch                 & Analog Devices HMC-C019 \\
        \hline
        \multicolumn{3}{l}{\textbf{Signal sources and acquisition}} \\
        -- & Control and acquisition & Quantum Machines OPX+ \& Octave \\
        -- & VNA                     & Keysight P9374A \\
        \hline
        \multicolumn{3}{l}{\textbf{Shielding}} \\
        -- & Eccosorb absorber foam  & Laird 301060005 \\
        -- & Cryoperm shield         & Amuneal 32941-01 \\
    \end{tabular}
    \end{ruledtabular}
\end{table}

\newpage
\section{System Parameters}
In this section, we present the parameters of the experimental system. We first describe the characterization experiments and pulse sequences used to extract the device parameters. Then, we show the coherence time measurements for the $\transgeh$ qubit and the $\transgfh$ qubit. The extracted parameters are summarized in Table~\ref{tab:params_heatmon}.

\begin{table}
    \caption{\textbf{System parameters.} Reported uncertainties correspond to 1 standard deviation. }
    \label{tab:params_heatmon}
    \begin{ruledtabular}
    \renewcommand{\arraystretch}{1.2}
    \begin{tabular}{lr}
        \textbf{Parameter} & \textbf{Value} \\
        \hline
        \multicolumn{2}{l}{\textbf{Transmon}} \\
        Josephson junction inductance $\Lj$  & $\approx$ \Ljvalh \\
        Frequency $\omegaqubit/2\pi$ & \fqubith \\
        Anharmonicity $\anharm/2\pi$ & \anharmh \\
        $\transgeh$ decay time $\Tone{\mr{ge}}$ & \bareToneh \\
        $\transgeh$ Ramsey time $\Ttwo{\mr{ge}}$ & \Ttwoge \\
        $\transgeh$ echo time $\Ttwoe{\mr{ge}}$ & \Ttwogeecho \\
        $\transefh$ decay time $\Tone{\mr{ef}}$ & \bareTonehef \\
        $\transgfh$ Ramsey time $\Ttwo{\mr{gf}}$ & \Ttwogf \\
        $\transgfh$ echo time $\Ttwoe{\mr{gf}}$ & \Ttwogfecho \\
        Thermal photon number $\nth{a}$ & $\nthhval$ \\
        \hline
        \multicolumn{2}{l}{\textbf{Readout cavity}} \\
        Frequency $\omegacav/2\pi$ & \fresh \\
        Output coupling $\kappacavcoup/2\pi$ & \kappacawavval \\
        Other losses $\kappacavlosses/2\pi$ & \kappacavintval \\
        Thermal photon number $\nth{b}$ & \nthhcavval \\
        \hline
        \multicolumn{2}{l}{\textbf{Couplings}} \\
        Dispersive shift $\chidisp/2\pi$ & $\approx \dispshifth$ \\
    \end{tabular}
    \end{ruledtabular}
\end{table}

\subsection{Device Characterization Experiments}
\label{subsec:sys_char_heatmon}

We extract the transmon and cavity parameters through room-temperature resistance measurements and cryogenic spectroscopy measurements.
We start by characterizing the readout cavity. We measure the cavity reflection coefficient $S_{11}(\omega_\mr{p})$ as a function of probe frequency $\omega_\mr{p}$ using a VNA. Fitting the response yields the cavity resonance frequency $\omegacav$, the external coupling rate to the readout waveguide $\kappacavcoup$, and the additional loss rate $\kappacavlosses = \kappacavint + \kappacavpin$, which includes the internal cavity losses and losses from the cavity coupling pin, estimated to be $\kappacavpin/2\pi \approx \SI{9}{\kilo\hertz}$.
To extract the transmon frequency $\omegaqubit$ and the anharmonicity $\anharm$~\cite{Schreier08}, we measure the $\transgeh$ and $\transefh$ transitions. We characterize the dispersive shift $\chidisp$ between the transmon and the readout cavity by applying a $\pi$ pulse on the $\transgeh$ transition and measuring the resulting shift in the cavity resonance frequency~\cite{Schuster2007}. We determine the thermal photon numbers $\nth{a}$, following a self-calibrated Rabi-contrast protocol~\cite{Geerlings2013, Jin2015}. The cavity thermal population $\nth{b}$ is similarly measured using selective Rabi oscillations on the qubit frequency peaks split by the cavity photon number.

\subsection{Pulse Sequence Parameters}
\label{subsec:pulse_params}

We perform dispersive readout with a square pulse of duration $\troval$. We choose the readout frequency according to the qubit manifold being measured in order to maximize the readout contrast. For experiments in the $\transgeh$ manifold, we read out at $\omegacav$ or $\omegacav - \chidisp$, corresponding to the cavity frequency when the transmon is in $\gstateh$ or $\estateh$, respectively. For experiments in the $\transgfh$ manifold, we read out at $\omegacav$ or $\omegacav - 2\chidisp$, corresponding to the cavity frequency when the transmon is in $\gstateh$ or $\fstateh$, respectively.

We implement coherent control pulses for the $\transgeh$ ($\transgfh$) qubit as single-photon (two-photon) drives at $\omegaqubit$ ($\omegaqubit - \anharm/2$) with a Gaussian envelope of standard deviation $\sigma = \sigmageval$ ($\sigmagfval$) and total duration $6\sigma = \taugeval$ ($\taugfval)$.
The engineered-gain pulse has a flat-top Gaussian envelope with $3\sigma$ Gaussian rise and fall edges, with $\sigma = \sigmagainval$. After the engineered-gain pulse, we insert a waiting time of $\twaitgain \gg \kappacav^{-1}$, which allows the cavity field to relax to vacuum before we read out the transmon state.

\subsection{Transmon Coherence Times}
\label{subsubsec:coherence_times_heatmon}

In this subsection, we present the coherence time measurements of the $\gstateh$, $\estateh$, and $\fstateh$ states of the transmon. The analytical expressions used to fit the relaxation, Ramsey, and echo coherence experiments are derived in Sec.~\ref{sec:decoherence_model}, where we also specify the fixed and free parameters used in each fit. 

\subsubsection{Measurement of the relaxation and coherence times of the $\transgeh$ qubit}
\label{sec:ge_coherence_times}

\begin{figure}
    \centering
    \includegraphics[width=\linewidth]{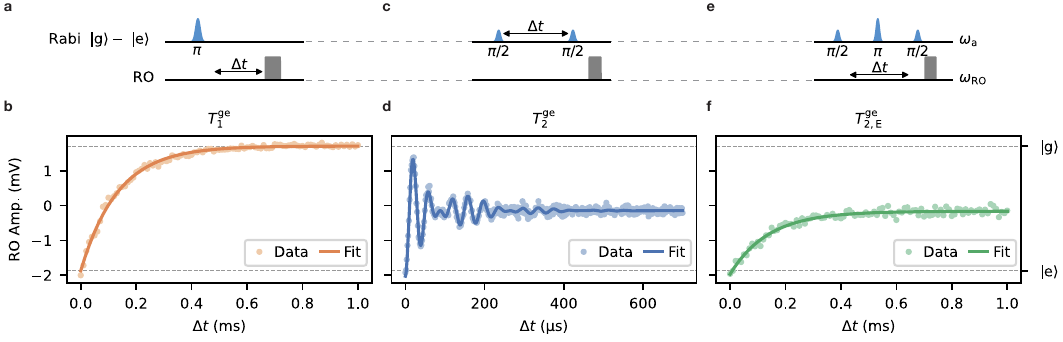}
    \caption{\textbf{Relaxation and coherence time measurements for the $\transgeh$ qubit.} All measurements use single-photon drives at $\omegaqubit$ and dispersive readout at $\omegaro = \omegacav$. \textbf{a,} Pulse sequence and \textbf{b,} data (readout amplitude as a function of delay $\dt$) for the $\Tone{\mr{ge}}$ experiment. Fitting the data to an exponential decay gives $\Tone{\mr{ge}} = \bareToneh$. \textbf{c,} Pulse sequence and \textbf{d,} data for the Ramsey $\Ttwo{\mr{ge}}$ measurement. The measurement shows an oscillatory decay with a beating pattern arising from two frequencies $\delta\omega_1^{\mr{ge}}$ and $\delta\omega_2^{\mr{ge}}$, with a decay time $\Ttwo{\mr{ge}} = \Ttwoge$. \textbf{e,} Pulse sequence and \textbf{f,} data for the echo $\Ttwoe{\mr{ge}}$ measurement. We extract $\Ttwoe{\mr{ge}} = \Ttwogeecho$. Circles are data and solid lines are fits. The dashed lines indicate reference values for $\gstate$ and $\estate$ obtained from Rabi measurements.}
    \label{fig:heatmon_ge_coherence_times}
\end{figure}

Measurements of the relaxation time $\Tone{\mr{ge}}$, the Ramsey coherence time $\Ttwo{\mr{ge}}$, and the echo coherence time $\Ttwoe{\mr{ge}}$ for the $\transgeh$ qubit are shown in Fig.~\ref{fig:heatmon_ge_coherence_times}. The top row shows the corresponding pulse sequences, and the bottom row shows the experimental data alongside the fits. Fitting the data to the respective analytical expressions in Subsec.~\ref{subsec:decoherence_ge} yields $\Tone{\mr{ge}} =1/\kappaone{\mr{ge}} = \bareToneh$, $\Ttwo{\mr{ge}} = 1/\Gammadec{\mr{ge}} = \Ttwoge$ and $\Ttwoe{\mr{ge}} = 1/\Gammaecho{\mr{ge}} = \Ttwogeecho$, with $\Gammadec{\mr{ge}} = \kappaone{\mr{ge}}/2 + \kappadeph{\mr{e}}$ and $\Gammaecho{\mr{ge}} = \kappaone{\mr{ge}}/2 + \kappadephecho{\mr{e}}$.

The Ramsey signal exhibits a beating pattern, which we fit with two oscillation frequencies $\delta\omega_1^{\mr{ge}}$ and $\delta\omega_2^{\mr{ge}}$. We attribute this structure to discrete frequency jumps of the transmon between two values separated by $\delta\omegaqubit = \delta\omega_2^{\mr{ge}} - \delta\omega_1^{\mr{ge}} \approx \omegajumph$. Since the amplitude of the frequency jump is much smaller than the linewidth of the engineered gain $\delta\omegaqubit \ll \kappacav$, this behavior has a negligible effect on the engineered gain rate $\kappagain$ (see Subsec.~\ref{subsec:frq_sel_gain}). We therefore do not expect this effect to influence the results of the gain-engineered experiments presented in the main text. Note that a similar beating pattern was observed over multiple cooldown cycles.

\subsubsection{Measurement of the coherence times of the $\transgfh$ qubit}

\begin{figure}
    \centering
    \includegraphics[width=\linewidth]{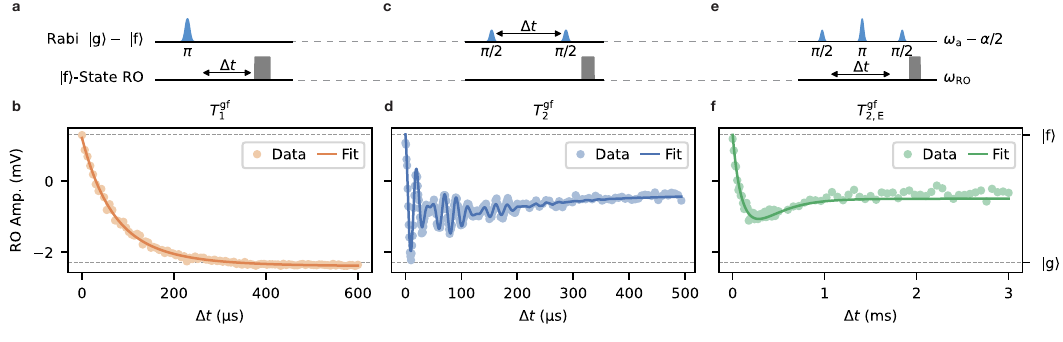}
    \caption{\textbf{Relaxation and coherence time measurements for the $\transgfh$ qubit.} All measurements use the two-photon drive at $\omegaqubit - \anharm/2$ and $\fstateh$-state readout at $\omegaro = \omegacav - 2\chidisp$. \textbf{a,} Pulse sequence and \textbf{b,} data (readout amplitude as a function of delay $\dt$) for the $\Tone{\mr{f}}$ measurement. We fit the measurement to an exponential decay with $\Tone{\mr{ef}} = \bareTonehef$. \textbf{c,} Pulse sequence and \textbf{d,} data for the Ramsey $\Ttwo{\mr{gf}}$ measurement. The oscillatory decay has a beating pattern arising from two frequencies $\delta\omega_1^{\mr{gf}}$ and $\delta\omega_2^{\mr{gf}}$, and a decay constant $\Ttwo{\mr{gf}} = \Ttwogf$. \textbf{e,} Pulse sequence and \textbf{f,} data for the echo $\Ttwoe{\mr{gf}}$ measurement. Fitting the data yields $\Ttwoe{\mr{gf}} = \Ttwogfecho$. Circles are data and solid lines are fits. The dashed lines indicate reference values for $\gstate$, $\fstate$ obtained from Rabi measurements.}
    \label{fig:coherence_times_heatmon}
\end{figure}

Figure~\ref{fig:coherence_times_heatmon} shows measurements of the relaxation time $\Tone{\mr{f}}$, the Ramsey coherence time $\Ttwo{\mr{gf}}$, and the echo coherence time $\Ttwoe{\mr{gf}}$ for the $\transgfh$ qubit. The corresponding pulse sequences (top row) and experimental data (bottom row) are shown alongside the fits. The relaxation measurement is fitted using Eq.~\eqref{eq:tone_gf} and we extract $\Tone{\mr{ef}} =1/\kappaone{\mr{ef}} = \Toneef$. For the coherence time measurements, we fit the data to Eq.~\eqref{eq:T2_gf_ramsey} and obtain $\Ttwo{\mr{gf}} = 1/\Gammadec{\mr{ef}} = \Ttwogf$ and $\Ttwoe{\mr{gf}} = 1/\Gammaecho{\mr{ef}} = \Ttwogfecho$, with $\Gammadec{\mr{ef}} = \kappaone{\mr{ef}}/2 + \kappadeph{\mr{f}}$ and $\Gammaecho{\mr{ef}} = \kappaone{\mr{ef}}/2 + \kappadephecho{\mr{f}}$.

As in the $\transgeh$ case, the Ramsey signal exhibits a beating pattern that we fit with two oscillation frequencies $\delta\omega_1^{\mr{gf}}$ and $\delta\omega_2^{\mr{gf}}$, which we attribute to discrete transmon frequency jumps of amplitude $\delta\omegaqubit$. For the $\transgfh$ two-photon transition, each transmon frequency jump shifts the transition frequency by $2\delta\omegaqubit$, giving a frequency difference $|\delta\omega_1^{\mr{gf}} - \delta\omega_2^{\mr{gf}}| = 2\delta\omegaqubit = \SI{12}{\kilo\hertz}$, confirmed by our fit.

\newpage
\section{Decoherence model}
\label{sec:decoherence_model}
In this section, we review the standard theoretical treatment of relaxation and coherence experiments in the $\transgeh$ and $\transgfh$ manifolds. These expressions are well known in the literature~\cite{Krantz19}. We derive them here for completeness and to introduce the basic decoherence model, which we then extend to include the engineered single-photon-gain dissipator that modifies the population dynamics in the $\transgfh$ manifold. We model the transmon as an anharmonic oscillator. In a rotating frame at a reference frequency $\omegar$, the Hamiltonian is
\begin{equation}
    \hat{H}/\hbar = -\frac{\anharm}{2}\hat{a}^{\dagger 2}\aop^2 +(\omegaqubit - \omegar)\aod\aop,
    \label{eq:hamiltonian_qubit_decoherence}
\end{equation}
where $\anharm$ is the anharmonicity and $\omegaqubit$ is the $\transgeh$ transition frequency.

\subsection{Modeling decoherence in the $\transgeh$ manifold}
\label{subsec:decoherence_ge}

We model the relaxation $\Tone{\mr{ge}}$, Ramsey $\Ttwo{\mr{ge}}$ and echo $\Ttwoe{\mr{ge}}$ experiments in the $\transgeh$ manifold. We use the Hamiltonian defined in Eq.~\eqref{eq:hamiltonian_qubit_decoherence} and set $\omega_r = \omegaqubit - \delta\omega_1$, where $\delta\omega_1$ is a small detuning introduced in the Ramsey experiment. We introduce the dissipators
\begin{equation}
    \kappaone{\mr{ge}}\diss{\gstateh\estatehb}, \quad 2\kappadephgen{\mr{e}}\diss{\estateh\estatehb}.
    \label{eq:diss_ge}
\end{equation}
where $\kappaone{\mr{ge}} = 1/\Tone{\mr{ge}}$ is the relaxation rate for the $\transgeh$ transition and $\kappadephgen{\mr{e}}$ is the dephasing rate of the $\estateh$ state. The label in parentheses indicates which sequence samples the dephasing. A Ramsey (echo) sequence samples $\kappadeph{\mr{e}} = 1/\Tphi{\mr{ge}}$ ($\kappadephecho{\mr{e}} = 1/\Tphie{\mr{ge}}$). The echo rate is, in general, smaller because the refocusing pulse cancels the low-frequency part of the dephasing noise~\cite{Bylander11}. Within this Markovian model, both cases take the same form, so we carry the rate as $\kappadephgen{\mr{e}}$ and insert the value appropriate to each experiment when fitting.
The Lindblad master equation for the two-dimensional $\{\gstateh, \estateh\}$ subspace is
\begin{equation}
   \frac{d \dm(t)}{dt} = \left[\begin{matrix}\kappaone{\mr{ge}} \rho_{\mr{ee}}(t) & \left(i \delta\omega_1 -\Gammagen{\mr{ge}}\right)\rho_{\mr{ge}}(t) \\ \left(- i \delta\omega_1 - \Gammagen{\mr{ge}} \right)\rho_{\mr{eg}}(t) & - \kappaone{\mr{ge}} \rho_{\mr{ee}}(t)\end{matrix}\right]
   \label{eq:lindblad1}
\end{equation}
with $\dm$ the density matrix of the system, $\rho_{ij} = \bra{i}\dm\ket{j}$ and $\Gammagen{\mr{ge}} =\kappaone{\mr{ge}}/2 + \kappadephgen{\mr{e}}$.

\subsubsection{$\Tone{\mr{ge}}$ experiment}

We begin with modeling the $\Tone{\mr{ge}}$ experiment, which describes the decay of population from $\estateh$ to $\gstateh$. The initial state is
\begin{equation}
    \dm_0 = \estateh\estatehb.
\end{equation}
Solving Eq.~\eqref{eq:lindblad1} yields
\begin{equation}
    \rho_{\mr{ee}}(t) = e^{-\kappaone{\mr{ge}} t}, \quad \rho_{\mr{gg}}(t) = 1- e^{-\kappaone{\mr{ge}} t}, \quad \rho_{\mr{ge}}(t) = \rho_{\mr{eg}}(t) = 0.
\end{equation}
The population $\rho_{\mr{ee}}$ decays exponentially with rate $\kappaone{\mr{ge}}$. The readout signal is
\begin{equation}
    \msig{\Tone{\mr{ge}}}(t) = \rcon{\mr{g}} \left(1- e^{-\kappaone{\mr{ge}} t}\right) + \rcon{\mr{e}} e^{-\kappaone{\mr{ge}} t},
    \label{eq:tone_ge}
\end{equation}
where $\rcon{i} \propto \bra{i}\aod\aop\ket{i}$ is the readout contrast of state $\ket{i}$, proportional to its average photon number as expected for a dispersive readout. In our data analysis, we fit Eq.~\eqref{eq:tone_ge} to the measured signal with the readout contrasts $\rcon{\mr{g}}$ and $\rcon{\mr{e}}$ and the relaxation rate $\kappaone{\mr{ge}}$ as free parameters.

\subsubsection{$\Ttwo{\mr{ge}}$ Ramsey and $\Ttwoe{\mr{ge}}$ Echo experiment}

The Ramsey and echo experiments follow a similar derivation. We first model the $\Ttwo{\mr{ge}}$ Ramsey experiment, which describes the decoherence of a $\transgeh$ superposition. The initial state is
\begin{equation}
    \dm_0 = \frac{1}{2} \left( \gstateh + \estateh \right) \left( \gstatehb + \estatehb \right).
\end{equation}
Solving Eq.~\eqref{eq:lindblad1} gives
\begin{equation}
    \rho_{\mr{ee}}(t) = \frac{1}{2}e^{-\kappaone{\mr{ge}} t}, \quad \rho_{\mr{gg}}(t) = 1- \frac{1}{2} e^{-\kappaone{\mr{ge}} t},
\end{equation}
\begin{equation}
    \rho_{\mr{ge}}(t) = \frac{1}{2} e^{+i\delta\omega_1 t - \Gammadec{\mr{ge}}t}, \quad \rho_{\mr{eg}}(t) = \frac{1}{2} e^{-i\delta\omega_1 t -\Gammadec{\mr{ge}}t}.
\end{equation}
After the free evolution, a final $\pi/2$ pulse is applied. The corresponding rotation operator is
\begin{equation}
    \rot{\frac{\pi}{2}}{\mr{ge}} = \frac{1}{\sqrt{2}} \left(
    \begin{matrix}
        1 & -1 \\
        1  & 1
    \end{matrix}\right),
\end{equation}
which maps the coherences $\rho_{\mr{ge}}$, $\rho_{\mr{eg}}$ onto the populations $\rho_{\mr{gg}}$, $\rho_{\mr{ee}}$. The readout signal is
\begin{equation}
    \msig{\Ttwo{\mr{ge}}}(t) = \frac{\rcon{\mr{g}} + \rcon{\mr{e}}}{2} + \frac{\rcon{\mr{e}} - \rcon{\mr{g}}}{2} \cos(\delta\omega_1\, t)\,e^{-\Gammadec{\mr{ge}}t}.
\end{equation}

As discussed in Subsec.~\ref{subsubsec:coherence_times_heatmon}, the transmon frequency does not remain constant during the experiment. Instead, it alternates between $\omegaqubit$ and $\omegaqubit + \delta\omegaqubit$ on a time scale which is longer than the duration of a single Ramsey trace ($\Ttrace \approx \Ttraceval$). The measured signal is therefore an ensemble average over traces acquired at each of the two frequencies. Including this effect, the measurement signal becomes
\begin{equation}
    \msig{\Ttwo{\mr{ge}}}^\mathrm{switch}(t) = \frac{\rcon{\mr{g}} + \rcon{\mr{e}}}{2}\left(\rho_{\mr{gg}}(t)+\rho_{\mr{ee}}(t)\right) + \frac{\rcon{\mr{e}} - \rcon{\mr{g}}}{2} \left( p \cos(\delta\omega_1\, t) + \left(1-p\right) \cos\!\left(\delta\omega_2\, t\right)\right)e^{-\Gammadec{\mr{ge}}t},
    \label{eq:T2_ge_ramsey}
\end{equation}
where $\delta\omega_2 = \delta\omega_1 + \delta\omegaqubit$ and $p$ is the probability that the transmon frequency is $\omegaqubit$ during a given trace. We fit Eq.~\eqref{eq:T2_ge_ramsey} with the readout contrasts $\rcon{\mr{g}}$ and $\rcon{\mr{e}}$, the dephasing rate $\kappadeph{\mr{e}}$, the two detunings $\delta\omega_1$ and $\delta\omega_2$, and the probability $p$ as free parameters. We fix the relaxation rate $\kappaone{\mr{ge}}$ to the value obtained from the $\Tone{\mr{ge}}$ fit.

The expressions for the $\Ttwoe{\mr{ge}}$ can be similarly derived. This is done by inserting a refocusing $\pi$ pulse at $t/2$, which cancels the oscillating phase and replaces $\Gammadec{\mr{ge}}$ by $\Gammaecho{\mr{ge}} = \kappaone{\mr{ge}}/2 + \kappadephecho{\mr{e}}$ in the decay envelope. In this case the free parameters are $\rcon{\mr{g}}$, $\rcon{\mr{e}}$, and the echo dephasing rate $\kappadephecho{\mr{e}}$, again with $\kappaone{\mr{ge}}$ fixed to its $\Tone{\mr{ge}}$ value.

\resumetoc
\subsection{Modeling decoherence in the $\transgfh$ manifold}
\label{subsec:decoherence_gf}

In this subsection, we model the relaxation $\Tone{\mr{f}}$, Ramsey $\Ttwo{\mr{gf}}$ and echo $\Ttwoe{\mr{gf}}$ experiments in the $\transgfh$ qubit encoding. We use the Hamiltonian defined in Eq.~\eqref{eq:hamiltonian_qubit_decoherence} and set $\omega_r = \omegaqubit - \anharm - \delta\omega_1/2$, so that $\delta\omega_1$ is the detuning of the $\transgfh$ two-photon transition from the rotating frame. Together with the dissipators introduced in Eq.~\eqref{eq:diss_ge}, we introduce the additional dissipators
\begin{equation}
    \kappaone{\mr{ef}}\diss{\estateh\fstatehb}, \quad 2\kappadephgen{\mr{f}}\diss{\fstateh\fstatehb}.
    \label{eq:diss_ef}
\end{equation}
where $\kappaone{\mr{ef}} = 1/\Tone{\mr{ef}}$ is the relaxation rate for the $\transefh$ transition and $\kappadephgen{\mr{f}}$ is the pure dephasing rate of the $\fstateh$ state. As in the $\transgeh$ manifold, a Ramsey (echo) sequence samples $\kappadeph{\mr{f}} = 1/\Tphi{\mr{ef}}$ ($\kappadephecho{\mr{f}}$).
We define $\Gammagen{\mr{ef}} = \kappaone{\mr{ef}}/2 + \kappadephgen{\mr{f}}$. The Lindblad master equation for the three-dimensional $\{\gstateh, \estateh, \fstateh\}$ subspace is
\begin{equation}
   \frac{d \dm(t)}{dt} = \resizebox{0.95\linewidth}{!}{$\displaystyle\left[\begin{matrix}\kappaone{\mr{ge}} \rho_{\mr{ee}}(t) &  \left(-\Gammagen{\mr{ge}} + i \left(\anharm + \delta\omega_1\right)\right)\rho_{\mr{ge}}(t) &  \left(i \delta\omega_1 -\Gammagen{\mr{ef}}\right) \rho_{\mr{gf}}(t) \\ \left(-\Gammagen{\mr{ge}} - i \left(\anharm + \delta\omega_1\right)\right) \rho_{\mr{eg}}(t) & - \kappaone{\mr{ge}} \rho_{\mr{ee}}(t) + \kappaone{\mr{ef}} \rho_{\mr{ff}}(t) &  \left(-\Gammagen{\mr{ge}} -\Gammagen{\mr{ef}} + i \left(- \anharm + \delta\omega_1\right)\right) \rho_{\mr{ef}}(t) \\  \left(- i \delta\omega_1 - \Gammagen{\mr{ef}}\right) \rho_{\mr{fg}}(t) &  \left(-\Gammagen{\mr{ge}} -\Gammagen{\mr{ef}}+ i \left(\anharm - \delta\omega_1\right)\right) \rho_{\mr{fe}}(t) & - \kappaone{\mr{ef}} \rho_{\mr{ff}} (t) \end{matrix}\right]$}
   \label{eq:lindblad2}
\end{equation}
\subsubsection{$\Tone{\mr{f}}$ experiment}

We model the $\Tone{\mr{f}}$ experiment, which describes the population decay from $\fstateh$ to $\gstateh$ through the single-photon loss processes $\fstateh \to \estateh \to \gstateh$. The initial state is
\begin{equation}
    \dm_0 = \fstateh\fstatehb.
\end{equation}
Solving Eq.~\eqref{eq:lindblad2} yields
\begin{align*}
    \rho_{\mr{ff}}(t) &= e^{-\kappaone{\mr{ef}}t}, & \rho_{\mr{ee}}(t) &= \frac{\kappaone{\mr{ef}}}{\kappaone{\mr{ef}} - \kappaone{\mr{ge}}}(e^{-\kappaone{\mr{ge}}t} - e^{-\kappaone{\mr{ef}}t}), \\
    \rho_{\mr{gg}}(t) &= 1-\rho_{\mr{ff}}(t) - \rho_{\mr{ee}}(t), & \rho_{xy}(t) &= 0,
\end{align*}
where $x,y \in \{\mr{g,e,f}\}$ denote all index combinations not listed explicitly above. The readout signal is
\begin{equation}
    \msig{\Tone{\mr{f}}}(t) = (\rcon{\mr{f}} -\rcon{\mr{g}})e^{-\kappaone{\mr{ef}}t} + (\rcon{\mr{e}}-\rcon{\mr{g}}) \frac{\kappaone{\mr{ef}}}{\kappaone{\mr{ef}} - \kappaone{\mr{ge}}}\left(e^{-\kappaone{\mr{ge}}t} - e^{-\kappaone{\mr{ef}}t}\right) + \rcon{\mr{g}}.
    \label{eq:tone_gf}
\end{equation}
The readout signal exhibits a multi-exponential decay at two distinct rates: $\kappaone{\mr{ef}}$ for the $\transefh$ decay and $\kappaone{\mr{ge}}$ for the $\transgeh$ decay. This same behavior appears in the $\Ttwo{\mr{gf}}$ Ramsey experiment, which we describe next. In our data analysis, we fit Eq.~\eqref{eq:tone_gf} with the readout contrasts $\rcon{\mr{g}}$, $\rcon{\mr{e}}$, and $\rcon{\mr{f}}$ and the relaxation rate $\kappaone{\mr{ef}}$ as free parameters, and we fix $\kappaone{\mr{ge}}$ to the value obtained from the $\Tone{\mr{ge}}$ fit. 

\subsubsection{$\Ttwo{\mr{gf}}$ Ramsey experiment}

We model the $\Ttwo{\mr{gf}}$ Ramsey experiment, which describes the decoherence of a $\gstateh$--$\fstateh$ superposition. The initial state is
\begin{equation}
    \dm_0 = \frac{1}{2}\left( \gstateh + \fstateh \right) \left( \gstatehb + \fstatehb\right).
\end{equation}
Solving Eq.~\eqref{eq:lindblad2} yields
\begin{equation}
\begin{aligned}
    \rho_{\mr{ff}}(t) &= \frac{1}{2}e^{-\kappaone{\mr{ef}}t}, & \quad \rho_{\mr{ee}}(t) &= \frac{\kappaone{\mr{ef}}}{2(\kappaone{\mr{ef}} - \kappaone{\mr{ge}})}(e^{-\kappaone{\mr{ge}}t} - e^{-\kappaone{\mr{ef}}t}), \\
    \rho_{\mr{gg}}(t) &= 1- \rho_{\mr{ee}}(t) - \rho_{\mr{ff}}(t), & \rho_{\mr{gf}}(t) &= \frac{1}{2}e^{i\delta \omega_1 t -\Gammadec{\mr{ef}}t},\\
    \rho_{\mr{fg}}(t) &= \frac{1}{2}e^{-i\delta \omega_1 t -\Gammadec{\mr{ef}}t}, & \rho_{xy}(t) &= 0,
    \label{eq:T2_02}
    \end{aligned}
\end{equation}
where $x,y \in \{\mr{g,e,f}\}$ denote all index combinations not listed explicitly above. After the free evolution, a final $\pi/2$ pulse is applied. The corresponding rotation operator is
\begin{equation}
    \rot{\frac{\pi}{2}}{\mr{gf}} = \frac{1}{\sqrt{2}} \left(
    \begin{matrix}
        1 & 0 & -1 \\
        0 & \sqrt{2} & 0 \\
        1 & 0 & 1
    \end{matrix}\right),
    \label{eq:pi_2_gf}
\end{equation}
which maps the coherences $\rho_{\mr{gf}}$, $\rho_{\mr{fg}}$ onto the populations $\rho_{\mr{gg}}$, $\rho_{\mr{ff}}$. 

The resulting readout signal is
\begin{equation}
    \msig{\Ttwo{\mr{gf}}}(t) = \rcon{\mr{e}}\rho_{\mr{ee}}(t) + \frac{\rcon{\mr{g}} + \rcon{\mr{f}}}{2}\left(\rho_{\mr{gg}}(t)+\rho_{\mr{ff}}(t)\right) + \frac{\rcon{\mr{f}} - \rcon{\mr{g}}}{2} \cos(\delta\omega_1\, t)\,e^{-\Gammadec{\mr{ef}}t}.
\end{equation}
In the experiment, the transmon frequency alternates between $\omegaqubit$ and $\omegaqubit + \delta\omegaqubit$ on the time scale $\Twchange \gg \Ttrace$. Including this effect, the ensemble-averaged measurement signal becomes
\begin{equation}
    \msig{\Ttwo{\mr{gf}}}^\mathrm{switch}(t) = \rcon{\mr{e}}\rho_{\mr{ee}}(t) + \frac{\rcon{\mr{g}} + \rcon{\mr{f}}}{2}\left(\rho_{\mr{gg}}(t)+\rho_{\mr{ff}}(t)\right) + \frac{\rcon{\mr{f}} - \rcon{\mr{g}}}{2} \left( p \cos(\delta\omega_1\, t) + \left(1-p\right) \cos\!\left(\delta\omega_2\, t\right)\right)e^{-\Gammadec{\mr{ef}}t},
    \label{eq:T2_gf_ramsey}
\end{equation}
where $\delta \omega_2 = \delta\omega_1 + 2\delta\omegaqubit$ and $p$ is the probability that the transmon frequency is $\omegaqubit$ during a given trace. We fit our experimental results to Eq.~\eqref{eq:T2_gf_ramsey} with the readout contrasts $\rcon{\mr{g}}$, $\rcon{\mr{e}}$, and $\rcon{\mr{f}}$, the dephasing rate $\kappadeph{\mr{f}}$, the two detunings $\delta\omega_1$ and $\delta\omega_2$, and the probability $p$ as free parameters. We fix the relaxation rates $\kappaone{\mr{ge}}$ and $\kappaone{\mr{ef}}$ to the values obtained from the $\Tone{\mr{ge}}$ and $\Tone{\mr{f}}$ fits.

\subsubsection{$\Ttwoe{\mr{gf}}$ echo experiment}

Similarly, we derive the model of the $\Ttwoe{\mr{gf}}$ echo experiment, in which a refocusing $\pi$ pulse is applied in the middle of the free evolution to cancel low-frequency dephasing noise. The initial state is
\begin{equation}
    \dm_0 = \frac{1}{2}\left( \gstateh + \fstateh \right) \left( \gstatehb + \fstatehb\right).
\end{equation}
Following a free evolution for a duration $t/2$, the resulting density matrix elements are given by Eq.~\eqref{eq:T2_02} evaluated at $t/2$, with $\Gammadec{\mr{ef}}$ replaced by the echo rate $\Gammaecho{\mr{ef}} = \kappaone{\mr{ef}}/2 + \kappadephecho{\mr{f}}$. This sequence samples $\kappadephecho{\mr{f}}$ rather than $\kappadeph{\mr{f}}$ because the refocusing pulse cancels the low-frequency part of the dephasing noise.

We then apply a $\pi$ rotation defined as
\begin{equation}
    \rot{\pi}{\mr{gf}} = \left(
    \begin{matrix}
        0 & 0 & 1 \\
        0 & 1 & 0 \\
        1 & 0 & 0
    \end{matrix}\right),
    \label{eq:pi_gf}
\end{equation}
which acts as $\sigma_x$ in the $\{\gstateh, \fstateh\}$ subspace, swapping $\gstateh$ and $\fstateh$ while leaving $\estateh$ unchanged. When applied to the density matrix, $\rot{\pi}{\mr{gf}} \dm \rot{\pi}{\mr{gf}}^\dagger$ swaps the populations $\rho_{\mr{gg}} \leftrightarrow \rho_{\mr{ff}}$ and maps the coherence $\rho_{\mr{gf}\,(\mr{fg})} \to \rho_{\mr{fg}\,(\mr{gf})}$.

We then evolve for the second half of the sequence for a duration $t/2$, solving the Lindblad master equation of Eq.~\eqref{eq:lindblad2} with the initial conditions set by the first half evolution followed by the $\pi$ pulse. The density matrix elements after the second $t/2$ evolution are
\begin{equation}
\begin{aligned}
    \rho_{\mr{ff}}^{(2)}(t/2) &= \rho_{\mr{gg}}^{(1)}(t/2)\,e^{-\kappaone{\mr{ef}}t/2}, \\
    \rho_{\mr{ee}}^{(2)}(t/2) &= \rho_{\mr{ee}}^{(1)}(t/2)\,e^{-\kappaone{\mr{ge}}t/2} + \rho_{\mr{gg}}^{(1)}(t/2) \frac{\kappaone{\mr{ef}}}{\kappaone{\mr{ef}} - \kappaone{\mr{ge}}}(e^{-\kappaone{\mr{ge}}t/2} - e^{-\kappaone{\mr{ef}}t/2}), \\
    \rho_{\mr{gg}}^{(2)}(t/2) &= 1- \rho_{\mr{ee}}^{(2)}(t/2) - \rho_{\mr{ff}}^{(2)}(t/2), \\
    \rho_{\mr{gf}}^{(2)}(t/2) &= \rho_{\mr{fg}}^{(1)}(t/2)\,e^{i\delta \omega_1 t/2 -\Gammaecho{\mr{ef}}t/2},\\
    \rho_{\mr{fg}}^{(2)}(t/2) &= \rho_{\mr{gf}}^{(1)}(t/2)\,e^{-i\delta \omega_1 t/2 -\Gammaecho{\mr{ef}}t/2}, \\
    \rho_{xy}^{(2)}(t/2) &= 0.
\end{aligned}
\end{equation}
where $x,y \in \{\mr{g,e,f}\}$ denote all index combinations not explicitly listed. Note that, because of the $\pi$ pulse, the oscillating phase in the coherences $\rho_{\mr{gf}}$ and $\rho_{\mr{fg}}$ is perfectly cancelled after the second free evolution, and only the incoherent decay remains. After the second free evolution, we apply a final $\pi/2$ rotation (Eq.~\eqref{eq:pi_2_gf}) and obtain the measurement signal
\begin{equation}
    \msig{\Ttwoe{\mr{gf}}}(t) = \rcon{\mr{e}}\rho_{\mr{ee}}^{(2)}(t/2) + \frac{\rcon{\mr{g}} + \rcon{\mr{f}}}{2} \left(\rho_{\mr{gg}}^{(2)}(t/2)+\rho_{\mr{ff}}^{(2)}(t/2)\right) + \frac{\rcon{\mr{f}} - \rcon{\mr{g}}}{2}\,e^{-\Gammaecho{\mr{ef}}\,t}.
    \label{eq:echo_signal}
\end{equation}
In our analysis, we fit Eq.~\eqref{eq:echo_signal} with the readout contrasts $\rcon{\mr{g}}$ and $\rcon{\mr{e}}$ and the echo dephasing rate $\kappadephecho{\mr{f}}$ as free parameters. We fix the relaxation rates $\kappaone{\mr{ge}}$ and $\kappaone{\mr{ef}}$ to the values obtained from the $\Tone{\mr{ge}}$ and $\Tone{\mr{f}}$ fits.

\resumetoc
\subsection{Modeling decoherence in the $\transgfh$ manifold with engineered single-photon gain}
\label{subsec:decoherence_gf_gain}

In this subsection, we extend the model of Subsec.~\ref{subsec:decoherence_gf} to include the engineered-gain dissipator. 
We use the same Hamiltonian and dissipators as before, and additionally introduce the single-photon-gain dissipator
\begin{equation}
    \kappag{} \diss{\fstateh\estatehb},
\end{equation}
which drives the $\transefh$ transition at rate $\kappag{}$. Including this term, the Lindblad master equation for the $\transgfh$ manifold becomes
\begin{equation}
   \frac{d \dm(t)}{dt} = \resizebox{0.95\linewidth}{!}{$\displaystyle\left[\begin{matrix}\kappaone{\mr{ge}} \rho_{\mr{ee}}(t) &  \left(-\Gammagen{\mr{ge}} - \kappag{}/2 + i \left(\anharm + \delta\omega_1\right)\right)\rho_{\mr{ge}}(t) &  \left(i \delta\omega_1 -\Gammagen{\mr{ef}}\right) \rho_{\mr{gf}}(t) \\ \left(-\Gammagen{\mr{ge}} - \kappag{}/2 - i \left(\anharm + \delta\omega_1\right)\right) \rho_{\mr{eg}}(t) & - (\kappaone{\mr{ge}} + \kappag{}) \rho_{\mr{ee}}(t) + \kappaone{\mr{ef}} \rho_{\mr{ff}}(t) &  \left(-\Gammagen{\mr{ge}} -\Gammagen{\mr{ef}} - \kappag{}/2 + i \left(- \anharm + \delta\omega_1\right)\right) \rho_{\mr{ef}}(t) \\  \left(- i \delta\omega_1 - \Gammagen{\mr{ef}}\right) \rho_{\mr{fg}}(t) &  \left(-\Gammagen{\mr{ge}} -\Gammagen{\mr{ef}} - \kappag{}/2 + i \left(\anharm - \delta\omega_1\right)\right) \rho_{\mr{fe}}(t) & - \kappaone{\mr{ef}} \rho_{\mr{ff}}(t)+\kappag{} \rho_{\mr{ee}}(t)\end{matrix}\right]$}
   \label{eq:lindblad3}
\end{equation}
Here, the engineered gain opens an additional excitation channel $\estateh \to \fstateh$ at rate $\kappag{}$, so every coherence involving the $\estateh$ state decays at an extra rate $\kappag{}/2$. We write this contribution explicitly in Eq.~\eqref{eq:lindblad3} and keep the rates $\Gammagen{\mr{ge}} $ and $\Gammagen{\mr{ef}} $ as defined in Subsec.~\ref{subsec:decoherence_gf}.

\subsubsection{$\Tone{\mr{f}}$ experiment}

We first model the $\Tone{\mr{f}}$ experiment in the presence of single-photon gain. The initial state is
\begin{equation}
    \dm_0 = \fstateh\fstatehb.
\end{equation}
Solving Eq.~\eqref{eq:lindblad3} yields
\begin{align*}
    \rho_{\mr{ff}}(t) &= \frac{1}{\lambda^+-\lambda^-} \left( (\lambda^+ +\kappaone{\mr{ef}}) e^{\lambda^- t} - (\lambda^- +\kappaone{\mr{ef}}) e^{\lambda^+ t}\right),  & \rho_{\mr{ee}}(t) &= \frac{\kappaone{\mr{ef}}}{\lambda^+ - \lambda^-}(e^{\lambda^+ t} - e^{\lambda^- t}),\\
    \rho_{\mr{gg}}(t) &= 1-\rho_{\mr{ff}}(t) - \rho_{\mr{ee}}(t), & \rho_{xy}(t) &= 0,
\end{align*}
where $x,y \in \{\mr{g,e,f}\}$ denote all index combinations not explicitly listed, and
\begin{equation}
    \lambda^\pm = \frac{1}{2}\left( -\Sigma \pm \sqrt{\Sigma^2 -4\kappaone{\mr{ge}}\kappaone{\mr{ef}}} \right),
    \label{eq:lambda_pm_gain}
\end{equation}
with
\begin{equation}
    \Sigma = \kappaone{\mr{ge}} + \kappaone{\mr{ef}} + \kappag{}.
    \label{eq:sigma_gain}
\end{equation}
The readout signal is then
\begin{equation}
    \msig{\Tone{\mr{f}}}^\mathrm{gain}(t) = \rcon{\mr{f}}\rho_{\mr{ff}}(t) + \rcon{\mr{e}}\rho_{\mr{ee}}(t) + \rcon{\mr{g}}\rho_{\mr{gg}}(t).
    \label{eq:T1_gf_gain_signal}
\end{equation}
As before, the readout signal is described exactly by a bi-exponential decay, now with rates $\lambda^+$ and $\lambda^-$ modified by the single-photon-gain rate $\kappag{}$. The same applies to the $\Ttwo{\mr{gf}}$ Ramsey experiment, which we describe next. We discuss the conditions under which this bi-exponential decay is well approximated by a single exponential in Subsec.~\ref{subsec:biexp_singleexp_gf}.

\subsubsection{$\Ttwo{\mr{gf}}$ Ramsey experiment}

For the $\Ttwo{\mr{gf}}$ Ramsey experiment in the presence of single-photon gain, the initial state is
\begin{equation}
    \dm_0 = \frac{1}{2}\left( \gstateh + \fstateh \right) \left( \gstatehb + \fstatehb\right).
\end{equation}
Solving Eq.~\eqref{eq:lindblad3} yields the density matrix elements
\begin{equation}
\begin{aligned}
    \rho_{\mr{ff}}(t) &= \frac{1}{2(\lambda^+-\lambda^-)} \left( (\lambda^+ +\kappaone{\mr{ef}}) e^{\lambda^- t} - (\lambda^- +\kappaone{\mr{ef}}) e^{\lambda^+ t}\right),  & \rho_{\mr{ee}}(t) &= \frac{\kappaone{\mr{ef}}}{2 (\lambda^+ - \lambda^-)}(e^{\lambda^+ t} - e^{\lambda^- t}),\\
    \rho_{\mr{gg}}(t) &= 1-\rho_{\mr{ff}}(t) - \rho_{\mr{ee}}(t),& \rho_{\mr{gf}}(t) &= \frac{1}{2}e^{i\delta \omega_1 t -\Gammadec{\mr{ef}}t},\\
    \rho_{\mr{fg}}(t) &= \frac{1}{2}e^{-i\delta \omega_1 t -\Gammadec{\mr{ef}}t}, & \rho_{xy}(t) &= 0.
\end{aligned}
\label{eq:T2gf_free_evolution}
\end{equation}
Note that the coherences $\rho_{\mr{gf}}$ and $\rho_{\mr{fg}}$ are unaffected by the single-photon-gain dissipator, which acts only on the populations and does not introduce additional dephasing. The reason is that the gain dissipator only drives the $\estateh \to \fstateh$ transition. Starting from a $\gstateh$--$\fstateh$ superposition, the system populates $\estateh$ only after a single-photon loss event on the $\transefh$ transition. The engineered gain can therefore act only once a loss has already occurred, so the loss event alone sets the collapse of the $\transgfh$ coherence. After the free evolution, a final $\pi/2$ pulse (Eq.~\eqref{eq:pi_2_gf}) is applied as described in Subsec.~\ref{subsec:decoherence_gf}. Including the effect of the qubit frequency jumps (see Subsec.~\ref{subsec:decoherence_ge}), the readout signal evolves as
\begin{align}
\msig{\Ttwo{\mr{gf}}}^\mathrm{gain}(t) &= \rcon{\mr{e}}\rho_{\mr{ee}}(t)
+ \frac{\rcon{\mr{g}} + \rcon{\mr{f}}}{2}\left(\rho_{\mr{gg}}(t)+\rho_{\mr{ff}}(t)\right)\label{eq:T2_gf_ramsey_gain} \\
\nonumber
&+ \frac{\rcon{\mr{f}} - \rcon{\mr{g}}}{2}
\left( p \cos(\delta\omega_1\, t) + \left(1-p\right) \cos\!\left(\delta \omega_2 \, t\right)\right)
e^{-\Gammadec{\mr{ef}}t}.
\end{align}
When analysing the measurement results, we fit Eq.~\eqref{eq:T2_gf_ramsey_gain} with the readout contrasts $\rcon{\mr{g}}$, $\rcon{\mr{e}}$, and $\rcon{\mr{f}}$, the dephasing rate $\kappadeph{\mr{f}}$, the two detunings $\delta\omega_1$ and $\delta\omega_2$, and the probability $p$ as free parameters. We fix the relaxation rates $\kappaone{\mr{ge}}$ and $\kappaone{\mr{ef}}$ to the values obtained from the $\Tone{\mr{ge}}$ and $\Tone{\mr{f}}$ fits, and the gain rate $\kappag{}$ to its calibrated value.

\subsubsection{$\Ttwoe{\mr{gf}}$ echo experiment}

We model the $\Ttwoe{\mr{gf}}$ echo experiment in the presence of engineered single-photon gain. The sequence is the same as in Subsec.~\ref{subsec:decoherence_gf}, with the addition of the engineered-gain dissipator of Eq.~\eqref{eq:lindblad3}. The initial state is
\begin{equation}
    \dm_0 = \frac{1}{2}\left( \gstateh + \fstateh \right) \left( \gstatehb + \fstatehb\right).
\end{equation}

Following a free evolution for a duration $t/2$, the density matrix elements are given by Eq.~\eqref{eq:T2gf_free_evolution} evaluated at $t/2$, again with $\Gammadec{\mr{ef}}$ replaced by the echo rate $\Gammaecho{\mr{ef}}$.
We then apply a $\pi$ rotation (Eq.~\eqref{eq:pi_gf}), which swaps the populations $\rho_{\mr{gg}} \leftrightarrow \rho_{\mr{ff}}$ and maps the coherence $\rho_{\mr{gf}\,(\mr{fg})} \to \rho_{\mr{fg}\,(\mr{gf})}$.

For the second half of the sequence, we evolve under the Lindblad master equation (Eq.~\eqref{eq:lindblad3}) for a duration $t/2$, with the initial conditions set by the first half evolution and the $\pi$ pulse. For arbitrary initial conditions $\rho_{\mr{ff}}(0) = a$ and $\rho_{\mr{ee}}(0) = b$, the solutions for the populations are
\begin{align}
    \rho_{\mr{ff}}(t) &= \frac{1}{\lambda^+ - \lambda^-}\left[ \left(a(\lambda^+ + \kappaone{\mr{ge}} + \kappag{}) + b\,\kappag{}\right) e^{\lambda^+ t} - \left(a(\lambda^- + \kappaone{\mr{ge}} + \kappag{}) + b\,\kappag{}\right) e^{\lambda^- t}\right], \nonumber \\
    \rho_{\mr{ee}}(t) &= \frac{1}{\lambda^+ - \lambda^-}\left[ \left(b(\lambda^+ + \kappaone{\mr{ef}}) + a\,\kappaone{\mr{ef}}\right) e^{\lambda^+ t} - \left(b(\lambda^- + \kappaone{\mr{ef}}) + a\,\kappaone{\mr{ef}}\right) e^{\lambda^- t}\right]. \label{eq:pop_general_gain}
\end{align}
Here, $a$ and $b$ denote the $\fstateh$ and $\estateh$ populations at the start of the second free evolution. The $\pi$ pulse swaps the $\gstateh$ and $\fstateh$ populations, so the $\fstateh$ population entering the second half is $\rho_{\mr{gg}}^{(1)}(t/2)$, while the $\estateh$ population is unchanged at $\rho_{\mr{ee}}^{(1)}(t/2)$. Substituting these as $a$ and $b$ in Eq.~\eqref{eq:pop_general_gain} and evaluating at the elapsed time $t/2$, the density matrix elements after the second half are
\begin{equation}
\begin{aligned}
    \rho_{\mr{ff}}^{(2)}(t/2) &= \rho_{\mr{ff}}(t/2)\Big|_{a = \rho_{\mr{gg}}^{(1)}(t/2),\, b = \rho_{\mr{ee}}^{(1)}(t/2)}, \\
    \rho_{\mr{ee}}^{(2)}(t/2) &= \rho_{\mr{ee}}(t/2)\Big|_{a = \rho_{\mr{gg}}^{(1)}(t/2),\, b = \rho_{\mr{ee}}^{(1)}(t/2)}, \\
    \rho_{\mr{gg}}^{(2)}(t/2) &= 1- \rho_{\mr{ee}}^{(2)}(t/2) - \rho_{\mr{ff}}^{(2)}(t/2), \\
    \rho_{\mr{gf}}^{(2)}(t/2) &= \rho_{\mr{fg}}^{(1)}(t/2)\,e^{i\delta \omega_1 t/2 -\Gammaecho{\mr{ef}}t/2},\\
    \rho_{\mr{fg}}^{(2)}(t/2) &= \rho_{\mr{gf}}^{(1)}(t/2)\,e^{-i\delta \omega_1 t/2 -\Gammaecho{\mr{ef}}t/2}, \\
    \rho_{xy}^{(2)}(t/2) &= 0.
\end{aligned}
\end{equation}
Note that, because of the $\pi$ pulse, the oscillating phase in the coherence terms $\rho_{\mr{gf}}$ and $\rho_{\mr{fg}}$ is perfectly canceled after the second free evolution, and only the incoherent decay remains.

After the second free evolution, we apply a final $\pi/2$ rotation (Eq.~\eqref{eq:pi_2_gf}) and obtain the measurement signal
\begin{equation}
    \msig{\Ttwoe{\mr{gf}}}^{\mathrm{gain}}(t) = \rcon{\mr{e}}\rho_{\mr{ee}}^{(2)}(t/2) + \frac{\rcon{\mr{g}} + \rcon{\mr{f}}}{2} \left(\rho_{\mr{gg}}^{(2)}(t/2)+\rho_{\mr{ff}}^{(2)}(t/2)\right) + \frac{\rcon{\mr{f}} - \rcon{\mr{g}}}{2}\,e^{-\Gammaecho{\mr{ef}}\,t}.
    \label{eq:echo_signal_gain}
\end{equation}
We fit our measurement results to Eq.~\eqref{eq:echo_signal_gain} with the readout contrasts $\rcon{\mr{g}}$ and $\rcon{\mr{e}}$ and the echo dephasing rate $\kappadephecho{\mr{f}}$ as free parameters. We fix the relaxation rates $\kappaone{\mr{ge}}$ and $\kappaone{\mr{ef}}$ to the values obtained from the $\Tone{\mr{ge}}$ and $\Tone{\mr{f}}$ fits, and the gain rate $\kappag{}$ to its calibrated value.

\resumetoc
\subsection{From bi-exponential to single-exponential decay in the $\transgfh$ manifold}
\label{subsec:biexp_singleexp_gf}
All relaxation and coherence experiments in the $\transgfh$ manifold produce bi-exponential readout signals. When we fit the relaxation time $\Tone{\mr{f}}$ in the presence of gain, we nevertheless describe decay with a single exponential for $\kappagain \gg \kappaone{\mr{ge}},\, \kappaone{\mr{ef}}$. We justify this simplification here.

The $\Tone{\mr{f}}$ relaxation follows the two-step cascade $\fstateh \to \estateh \to \gstateh$, since $\fstateh$ reaches $\gstateh$ only through the intermediate level $\estateh$. The resulting signal is a sum of two exponentials whose rates encode the two stages of cascaded decay. Without engineered gain, these are simply the bare transition rates $\kappaone{\mr{ef}}$ and $\kappaone{\mr{ge}}$ (Subsec.~\ref{subsec:decoherence_gf}). The engineered gain returns population from $\estateh$ to $\fstateh$ at rate $\kappag{}$, coupling the two populations and replacing the decay rates by the eigenvalues $\lambda^\pm$ of Eq.~\eqref{eq:lambda_pm_gain}.

In the strong-gain regime $\kappag{} \gg \kappaone{\mr{ge}}, \kappaone{\mr{ef}}$, expanding the two eigenvalues $\lambda^\pm$ to leading order, we get,
\begin{equation}
    \lambda^+ \approx -\frac{\kappaone{\mr{ge}}\kappaone{\mr{ef}}}{\kappag{}}, \qquad \lambda^- \approx -\kappag{}.
    \label{eq:lambda_pm_asymptotic}
\end{equation}
Their ratio $|\lambda^-|/|\lambda^+| \approx (\kappag{})^2/(\kappaone{\mr{ge}}\kappaone{\mr{ef}}) \gg 1$ shows that the two rates are widely separated, so the fast exponential, associated to $\lambda^-$, survives only as a short transient at early times $t \simeq 1/\kappag{}$, while the subsequent decay is a single exponential with a larger relaxation time.

To make this explicit, we insert the asymptotic rates of Eq.~\eqref{eq:lambda_pm_asymptotic} into the exact $\Tone{\mr{f}}$ solutions of Subsec.~\ref{subsec:decoherence_gf_gain}. The density-matrix elements reduce to
\begin{align}
    \rho_{\mr{ff}}(t) &\approx e^{\lambda^+ t} + \frac{\kappaone{\mr{ef}}}{\kappag{}}\,\left(e^{\lambda^- t} - e^{\lambda^+ t}\right), &
    \rho_{\mr{ee}}(t) &\approx \frac{\kappaone{\mr{ef}}}{\kappag{}}\left(e^{\lambda^+ t} - e^{\lambda^- t}\right), \nonumber \\
    \rho_{\mr{gg}}(t) &\approx 1 - e^{\lambda^+ t}, & \rho_{xy}(t) &= 0,
    \label{eq:t1gf_asymptotic}
\end{align}
where $x,y \in \{\mr{g,e,f}\}$ again denote all index combinations not written explicitly. The intermediate population $\rho_{\mr{ee}}$ never exceeds the small bound $\kappaone{\mr{ef}}/\kappag{} \ll 1$. The fast $e^{\lambda^- t}$ transients in $\rho_{\mr{ff}}$ and $\rho_{\mr{ee}}$ cancel in $\rho_{\mr{gg}}$, which therefore rises as a single exponential.

Dropping the $O(\kappaone{\mr{ef}}/\kappag{})$ corrections, the readout signal of Eq.~\eqref{eq:T1_gf_gain_signal} collapses to
\begin{equation}
    \msig{\Tone{\mr{f}}}^\mathrm{gain}(t) \approx \rcon{\mr{g}} + (\rcon{\mr{f}} - \rcon{\mr{g}})\,e^{-t/\Tone{\mr{f}}},
    \label{eq:t1gf_signal_single}
\end{equation}
a single exponential with
\begin{equation}
    \Tone{\mr{f}} \approx \frac{1}{|\lambda^+|} \approx \frac{\kappag{}}{\kappaone{\mr{ge}}\kappaone{\mr{ef}}}.
    \label{eq:t1gf_enhanced}
\end{equation}
The signal therefore decays from $\rcon{\mr{f}}$ to $\rcon{\mr{g}}$ with an enhanced relaxation time that grows in direct proportion to $\kappagain$.

The enhancement of $\Tone{\mr{f}}$ has a simple physical picture. A photon-loss event takes $\fstateh \to \estateh$ at rate $\kappaone{\mr{ef}}$. From $\estateh$ the engineered gain returns the population to $\fstateh$ at rate $\kappag{}$, while the residual relaxation $\estateh \to \gstateh$ removes it to the ground state at rate $\kappaone{\mr{ge}}$. When $\kappag{} \gg \kappaone{\mr{ge}}$, the population is almost always pumped back before it can relax to $\gstateh$, so the net relaxation rate is suppressed to $\kappaone{\mr{ef}}\,\kappaone{\mr{ge}}/\kappag{}$, exactly the slow rate $|\lambda^+|$.

For smaller values of $\kappagain$, the single-exponential fit still provides a reasonable measure of the qubit relaxation time. To evaluate this, we compare fitted exponential decay times with the 1/e time across the first four $\kappagain$ values explored (see Table~\ref{tab:single_exponential_vs_1/e}). If the bi-exponential is prominent, we would see a large difference between the values extracted via these two methods. Instead, we find that there is only a small difference in the extracted values, justifying the use of a single-exponential fit for the full range of $\ggain$ in Fig.~\ref{fig:three}c of the main text.

\begin{table}[h]
    \caption{\textbf{Comparison between exponential fits and the $1/e$ characteristic time.} Reported uncertainties correspond to 1 standard deviation. The first four rows refer to a $\Tone{\mr{f}}$ measurement, while the final row corresponds to a $\Tone{\mr{g}}$ measurement.}
    \label{tab:single_exponential_vs_1/e}
    \begin{ruledtabular}
    \renewcommand{\arraystretch}{1.2}
    \begin{tabular}{llrr}
        \textbf{$\ggain$} & \textbf{$\kappagain$} & \textbf{Exponential fit} & \textbf{$1/e$ time} \\
        \hline
        \SI{0}{\kilo\hertz} & \SI{0}{\kilo\hertz}  & \SI{0.077\pm0.001}{\milli\second}& \SI{0.073\pm0.001}{\milli\second} \\
        \SI{5.345}{\kilo\hertz} & \SI{0.3}{\kilo\hertz}  & \SI{0.085\pm0.001}{\milli\second}& \SI{0.078\pm0.001}{\milli\second} \\
        \SI{21.35}{\kilo\hertz} & \SI{4.2}{\kilo\hertz}  & \SI{0.279\pm0.008}{\milli\second}& \SI{0.202\pm0.010}{\milli\second} \\
        \SI{48.03}{\kilo\hertz} & \SI{21.5}{\kilo\hertz}  & \SI{0.995\pm0.022}{\milli\second}& \SI{0.916\pm0.012}{\milli\second} \\
        \hline
        ($\Tone{\mr{g}}$) \SI{48.03}{\kilo\hertz}  & \SI{21.5}{\kilo\hertz}   & \SI{0.966\pm0.34}{\milli\second}& \SI{0.619\pm0.091}{\milli\second} \\
    \end{tabular}
    \end{ruledtabular}
\end{table}

\newpage

\section{Engineered Gain (Theory)} 

\subsection{System Hamiltonian}
\label{subsec:sys_ham}
Here, we introduce the full Hamiltonian of the transmon-cavity system, including the microwave drives that activate the gain interaction and the $\transgf$ two-photon drive used to perform single-qubit gates on the gain-engineered transmon.

The system Hamiltonian is
\begin{equation}
\hat{H}_{0} = \hat{H}_{\mathrm{a,0}} + \hat{H}_{\mathrm{b,0}} + \hat{H}_{\mathrm{i}} + \hat{H}_{\mathrm{d}}.
\end{equation}
The transmon Hamiltonian is $\hat{H}_{\mathrm{a,0}}/\hbar = \omegaqubit^0 \aod_0 \aop_0 + \gfour(\aod_0 + \aop_0)^{4}$, with bare-mode annihilation operator $\aop_0$, bare frequency $\omegaqubit^0$, and fourth-order nonlinearity $\gfour$~\cite{Frattini2017,frattini2021phd}. The readout cavity Hamiltonian is $\hat{H}_{\mathrm{b,0}}/\hbar = \omegacav^0 \bod_0 \bop_0$, with bare frequency $\omegacav^0$ and annihilation operator $\bop_0$. The transmon-cavity interaction is $\hat{H}_{\mathrm{i}}/\hbar = g(\aod_0-\aop_0)(\bod_0-\bop_0)$, with coupling strength $g$. The drive Hamiltonian is $\hat{H}_{\mathrm{d}}/\hbar = 2 \left[ \epsgain \, \mathrm{Re}(e^{i \omegagain t})  \right] (\aod_0 + \aop_0)$, where the slowly-varying amplitude $\epsgain$ drives the gain interaction at frequency $\omegagain$.
We set the frequency of the parametric process to be
\begin{align}
    \omegagain &= (\omegaqubit -\anharm + \omegacav -2\chidisp)/2 - \anharm\left( |\xigain|^2 \right), \label{eq:omegagain}\\
    \xigain &= \epsgain\left( \frac{1}{\omegagain-\omegaqubit} - \frac{1}{\omegagain+\omegaqubit} \right), \label{eq:xigain}
\end{align}
with $\chidisp = 2\anharm\lambda^2$ the dispersive-shift interaction strength, $\lambda = g/\Deltaab$, $\anharm = -12\gfour$ the anharmonicity of the transmon, and $\xigain$ the effective displacement generated by the gain pump. The final term in Eq.~\eqref{eq:omegagain} compensates for the Stark shift from the parametric drive on the transmon. 

We apply a sequence of transformations to $\hat{H}_{0}$ following standard techniques for parametric processes in circuit QED~\cite{Grimm2020,Frattini2024}, which we briefly summarize here. We first apply a first-order rotating-wave approximation (RWA) to $\hat{H}_{\mathrm{i}}$~\cite{Blais2021}. The interaction contains the excitation-conserving terms $\aod_0 \bop_0$ and $\aop_0 \bod_0$, which oscillate at the detuning $\Deltaab = \omegaqubit^0 - \omegacav^0$ in the frame rotating at the bare mode frequencies, together with the counter-rotating terms $\aod_0 \bod_0$ and $\aop_0 \bop_0$, which oscillate at $\omegaqubit^0 + \omegacav^0$. We drop the counter-rotating terms under the conditions $g, |\Deltaab|\ll \omegaqubit^0 + \omegacav^0,$
which make these terms average to zero, leaving
\begin{equation}
    \hat{H}_{\mathrm{i}}^{\mr{RWA}}/\hbar = -g(\aod_0 \bop_0 + \aop_0 \bod_0)
\end{equation}
We work in the dispersive regime $g \ll |\Deltaab|$, and we can therefore diagonalize the linear coupling with the Bogoliubov transformation~\cite{Blais2021}
\begin{equation}
    \hat{U}_\lambda = e^{\lambda(\aop \bod - \aod \bop)}, \qquad \lambda = g/\Deltaab \ll 1,
    \label{eq:bogoliubov}
\end{equation}
which substitutes
\begin{align}
    \aop_0 &\to \aop + \lambda \bop + \Obig{\lambda^2}, \label{eq:sub_a_dressed}\\
    \bop_0 &\to \bop - \lambda \aop + \Obig{\lambda^2}, \label{eq:sub_b_dressed}
\end{align}
with $\aop$ and $\bop$ the dressed-mode operators. This substitution cancels the linear coupling at first order in $\lambda$ and yields the dressed-mode frequencies \( \omegaqubit = \omegaqubit^0 + 2g^{2}/\Deltaab\) and \( \omegacav = \omegacav^0 - 2g^{2}/\Deltaab \).
We then apply a time-dependent displacement transformation
to eliminate the linear pump drive $\hat{H}_{\mathrm{d}}$~\cite{Grimm2020, frattini2021phd}, obtaining
\begin{align}
    \aop &\to \aop + \xigain e^{-i\omegagain t}, \label{eq:sub_a_displaced}\\
    \aod &\to \aod + \xigain^* e^{i\omegagain t}, \label{eq:sub_adag_displaced}
\end{align}
with $\xigain$ the effective displacement of Eq.~\eqref{eq:xigain}. 
We finally move into rotating frames~\cite{Blais2021} at $\wRa = \omegaqubit - \anharm/2$ and $\wRb = \omegacav - \anharm/2 -2\chidisp$. Expanding $\gfour(\aod_0 + \aop_0)^4$ in this displaced, rotating frame generates a sum of terms of the form
\begin{equation}
    \Omega_{jkmn}^{pq}\, \aop^{\dagger j} \aop^{k}\, \bop^{\dagger m} \bop^{n}\,
    e^{i\left[(j-k)\wRa + (m-n)\wRb + (q-p)\omegagain\right]t},
    \label{eq:rwa_terms}
\end{equation}
with $j+k+m+n+p+q = 4$ and amplitudes $\Omega_{jkmn}^{pq} \propto \gfour\, \lambda^{m+n}\, \xigain^{p} (\xigain^{*})^{q}$. We perform a RWA by retaining the static terms, for which the bracket in Eq.~\eqref{eq:rwa_terms} vanishes, and dropping every other term under the condition
\begin{equation}
    \left| \Omega_{jkmn}^{pq} \right| \ll \left| (j-k)\wRa + (m-n)\wRb + (q-p)\omegagain \right|,
    \label{eq:rwa_condition_full}
\end{equation}
which makes it average to zero over the timescale of the experiment dynamics. 
We finally get
\begin{equation}
    \hat{H} = \Ha + \Hgain + \Hb + \Hchi
    \label{eq:ham_full}
\end{equation}
with
\begin{align}
    \Ha/\hbar &= \frac{\anharm}{2}\aod \aop - \frac{\anharm}{2}\aop^{2\dagger} \aop^2, \label{eq:Ha}\\
    \Hgain/\hbar &= \ggain(\aod \bod + \aop \bop), \label{eq:Hgain}\\
    \Hb/\hbar &= (\anharm/2+2\chi)\bod\bop, \label{eq:Hb}\\
    \Hchi/\hbar &= -\chi \aod\aop \bod\bop. \label{eq:Hchi}
\end{align}
In contrast to Eq.~\eqref{eq:hamiltonian_main} of the main text, the rotating frames chosen here ($\wRa,\wRb$) allow both $\Hgain$ to be a time-independent interaction and the computational states $\ket{\mr{g},0}$ and $\ket{\mr{f},0}$ to be energy-degenerate.
The gain interaction, $\Hgain$, is a two-mode squeezing term that resonantly couples $\ket{\mr{e}, 0}$ and $\ket{\mr{f}, 1}$ of the transmon-cavity system with rate $\ggain = \anharm\lambda|\xigain|^2$. The residual cavity Hamiltonian in the rotating frame is given by $\Hb$. Finally, $\Hchi$ is the dispersive coupling between the transmon and the cavity, enabling dispersive readout of the transmon state. The parametric drive also imposes a Stark shift on the transmon frequency,
\begin{equation}
    \Hstark/\hbar = -2\anharm \left( |\xigain|^2  \right) \aod \aop,
    \label{eq:Hstark}
\end{equation}
which we have already compensated through the choice of pump frequency $\omegagain$ above, so it does not appear in $\hat{H}$.

\subsection{Frequency-selective engineered gain}
\label{subsec:frq_sel_gain}

Here, we show how the coherent two-mode-squeezing interaction $\Hgain$ (Eq.~\eqref{eq:Hgain}), combined with the single-photon loss of the cavity, produces a frequency-selective single-photon gain acting on the $\transefh$ transition of the transmon. The derivation adiabatically eliminates the cavity mode and yields the effective gain rate $\kappagain = 8\ggain^2/\kappacav$ quoted in the main text, together with the frequency selectivity that suppresses the gain in neighboring transmon transitions.
We model the cavity as being coupled to a zero-temperature bath ($\nth{b} = 0$) at rate $\kappacav$, described by the dissipator $\kappacav \diss{\bop}$. Throughout this subsection we label the transmon levels using Fock-state notation $\ket{n}$, $n = 0,1,2,\dots$, corresponding to $\gstateh, \estateh, \fstateh, \hstateh, \dots$.

\subsubsection{Decomposition into independent interaction channels}
In this system, the cavity frequency depends on the transmon Fock state $\ket{n}$ through the dispersive shift $\chidisp$. To make this dependence explicit, we insert the resolution of the identity $\sum_n \Pi(n) = \hat{I}$, with $\Pi(n) = \ket{n}\bra{n}$, on either side of the transmon operators in the gain interaction,
\begin{equation}
   \Hgain/\hbar = \ggain \sum_{n,m}\left(\Pi(n)\,\aod\,\Pi(m)\,\bod + \Pi(n)\,\aop\,\Pi(m)\,\bop\right).
    \label{eq:gain_interaction_decomposed}
\end{equation}
The interaction Hamiltonian $\Hgain$ describes a two-mode-squeezing process: each term simultaneously creates (or annihilates) one excitation in the transmon and one photon in the cavity. For a given final transmon state $\ket{n}$, the cavity term of Eq.~\eqref{eq:ham_full} reads
\begin{equation}
     \Hb^{n}/\hbar + \Hchi^{n}/\hbar = \left(-\chidisp_n + \anharm/2 + 2\chidisp\right)\bod\bop \equiv \Delta_n\,\bod\bop,
    \label{eq:Delta_n}
\end{equation}
with $\chidisp_n = n\chidisp$ the dispersive shift of the cavity when the transmon occupies $\ket{n}$.
In the adiabatic-elimination limit $\ggain \ll \kappacav$, the cavity responds independently to each transmon transition, so we replace the single cavity mode $\bop$ by a set of independent effective bath modes $\bop_n$, one per final state $\ket{n}$, each with detuning $\Delta_n$ and decay rate $\kappacav$. Because $\aod$ raises the transmon by a single quantum, $\Pi(n+1)\,\aod\,\Pi(n) = \sqrt{n+1}\,\ket{n+1}\bra{n}$, so only the adjacent transitions $\ket{n}\to\ket{n+1}$ contribute and the double sum over $\{m,n\}$ collapses to a single sum over $n$. Absorbing the Fock-state matrix element $\sqrt{n+1}$ into a transition-dependent coupling $\sqrt{n+1}\,\ggain$, the Hamiltonian becomes
\begin{equation}
    \hat{H}/\hbar = \Ha/\hbar + \sum_{n}\left[\Delta_{n+1}\,\bod_{n+1}\bop_{n+1} + \sqrt{n+1}\,\ggain\left(\ket{n+1}\bra{n}\,\bod_{n+1} + \ket{n}\bra{n+1}\,\bop_{n+1}\right)\right].
    \label{eq:gain_hami_decomposed}
\end{equation}

\subsubsection{Adiabatic elimination and transition-resolved gain rates}
We now eliminate each bath mode $\bop_{n+1}$ independently. In the limit $\ggain \ll \kappacav$, the transmon is effectively static on the timescale $1/\kappacav$ over which the cavity relaxes, and we can neglect the final term in Eq.~\eqref{eq:gain_hami_decomposed}. The Heisenberg equation of motion for $\bop_{n+1}$ reads
\begin{equation}
    \frac{\mr{d}}{\mr{d}t}\bop_{n+1}(t) = -\left(\frac{\kappacav}{2} + i\Delta_{n+1}\right)\bop_{n+1}(t) + \sqrt{\kappacav}\,\bop_\mr{in}(t),
\end{equation}
with Fourier-domain solution
\begin{equation}
    \bop_{n+1}[\omega] = \frac{\sqrt{\kappacav}\,\bop_\mr{in}[\omega]}{\dfrac{\kappacav}{2} + i(\Delta_{n+1} - \omega)}.
    \label{eq:bn_fourier}
\end{equation}
This susceptibility is a Lorentzian peaked at $\omega = \Delta_{n+1}$ with width $\kappacav$, and its resonant structure is what makes the engineered gain frequency selective. Due to the two-mode-squeezing nature of the interaction, the transmon transition $\ket{n}\to\ket{n+1}$ is accompanied by the emission of a cavity photon, so the associated single-photon-gain rate follows from the bath correlation function~\cite{Breuer2007},
\begin{equation}
    \gamma(n+1,n) = (n+1)\,\ggain^2 \int_{-\infty}^{\infty} \mr{d}\tau\, e^{i\tilde{\omega}_{n,n+1}\tau}\,\langle \bod_{n+1}(\tau)\,\bop_{n+1}(0)\rangle,
\end{equation}
where $\tilde{\omega}_{n,n+1} = \tilde{\omega}_n - \tilde{\omega}_{n+1}$ is the transmon transition frequency, with $\tilde{\omega}_n = \bra{n}\Ha\ket{n}/\hbar$. For the zero-temperature bath, the input-field correlators reduce to $\langle \bop_\mr{in}[\omega]\bod_\mr{in}[\nu]\rangle = 2\pi\,\delta(\omega - \nu)$ and $\langle \bod_\mr{in}[\omega]\bop_\mr{in}[\nu]\rangle = 0$. Substituting Eq.~\eqref{eq:bn_fourier} yields the Lorentzian single-photon-gain rate
\begin{equation}
    \gamma(n+1,n) = \frac{(n+1)\,\kappacav\,\ggain^2}{\dfrac{\kappacav^2}{4} + \left(\Delta_{n+1} - \tilde{\omega}_{n,n+1}\right)^2}.
    \label{eq:gamma_nm}
\end{equation}
Collecting all loss channels, the reduced master equation for the transmon is
\begin{equation}
    \frac{\mr{d}}{\mr{d}t}\rhoa = -\frac{i}{\hbar}\left[\Ha,\rhoa\right] + \diss{\aod_\mr{eff}}\rhoa,
\end{equation}
with the effective dissipator~\cite{Breuer2007}
\begin{equation}
    \diss{\aod_\mr{eff}}\rhoa = \sum_{n}\gamma(n+1,n)\left(\ket{n+1}\bra{n}\,\rhoa\,\ket{n}\bra{n+1} - \frac{1}{2}\left\{\ket{n}\bra{n},\,\rhoa\right\}\right).
    \label{eq:diss_transmon_explicit}
\end{equation}
Unlike a generic white single-photon gain channel, here the transition rate $\gamma(n+1,n)$ depends on the Fock state $\ket{n}$. We will see now how this dependence produces a transition-selective single-photon gain channel acting on the $\transgf$.

\subsubsection{Frequency selectivity of the single-photon gain}
To build intuition, we write out the transition rates explicitly for the lowest transmon transitions $\{\transgeh,\transefh,\transfhh\}$, with jump operators $\estateh\bra{\mr{g}}$, $\fstateh\bra{\mr{e}}$, and $\hstateh\bra{\mr{f}}$, using $\tilde{\omega}_n = \anharm n - \frac{\anharm}{2}n^2$ from $\Ha$ (Eq.~\eqref{eq:Ha}),
\begin{alignat}{3}
    \gamma(\mr{e},\mr{g}) &= \frac{\kappacav\,\ggain^2}{\dfrac{\kappacav^2}{4} + (\anharm+\chi)^2}, \quad &\gamma(\mr{f},\mr{e}) &= \frac{8\ggain^2}{\kappacav}, \quad &\gamma(\mr{h},\mr{f}) &= \frac{3\,\kappacav\,\ggain^2}{\dfrac{\kappacav^2}{4} + (\anharm+\chi)^2}.
\end{alignat}
The $\transefh$ transition is resonant with the engineered bath, $\Delta_\mr{f} - \tilde{\omega}_\mr{ef} = 0$, while the $\transgeh$ and $\transfhh$ transitions are detuned by $\anharm + \chidisp$. In the regime $\kappacav \ll \anharm$, we therefore have $\gamma(\mr{f},\mr{e}) \gg \gamma(\mr{e},\mr{g}),\, \gamma(\mr{h},\mr{f}) \approx 0$. The engineered dissipation thus predominantly drives the $\transefh$ transition and leaves the $\transgeh$ and $\transfhh$ transitions unaffected. 

This frequency selectivity relies on three properties of the device. The transmon anharmonicity and dispersive shift set the scale $\anharm + \chidisp$ by which the neighboring transitions are detuned from the engineered bath. The cavity susceptibility of Eq.~\eqref{eq:bn_fourier} sets the bandwidth of the gain process to the cavity linewidth $\kappacav$, so unwanted transitions fall outside this bandwidth whenever $\kappacav \ll \anharm$. The effective dissipator then reduces to a single transition-selective single-photon-gain channel,
\begin{align}
    \diss{\aod_\mr{eff}}\rhoa &\xrightarrow{\kappacav \ll \anharm} \kappagain\,\diss{\fstateh\bra{\mr{e}}}\rhoa, \\
    \kappagain &\equiv \gamma(\mr{f},\mr{e}) = \frac{8\ggain^2}{\kappacav}.
    \label{eq:kappagain}
\end{align}
In Subsec.~\ref{subsec:t1_vs_detuning}, we present measurements of the $\Tone{\mr{f}}$ as a function of frequency detuning of the engineered gain pump. 

\newpage
\section{Engineered Gain (Experimental Implementation and Simulations)}
\subsection{Calibration of effective displacement $\xigain$}
\label{subsec:xigain_calib}

In this section, we describe the calibration of the effective gain-drive displacement $\xigain$, which sets the engineered-gain rate $\ggain$ used throughout the main text. We first show that the pump-induced Stark shift of the transmon serves as a direct probe of $\xigain$. We then use a Floquet analysis to identify the $\transge$ transition as the one that best tracks the analytic Stark-shift relation, and we perform a spectroscopy measurement of the $\transge$ shift to extract the conversion between the pump amplitude $\Ap$ and $\xigain$. This direct measurement is necessary because the shift of the gain resonance $\omegagain(\Ap)$, mapped in the main text (Fig.~\ref{fig:two}b), follows the $\transef$ transition and therefore does not track $\xigain$ reliably.

The engineered-gain rate follows from the effective drive amplitude through $\ggain = \anharm\lambda|\xigain|^2$ (Eq.~\eqref{eq:Hgain}), so calibrating $\ggain$ reduces to determining the effective displacement $\xigain$ that the gain pump produces at each pump amplitude $\Ap$. We extract $\xigain$ from the pump-induced Stark shift of the transmon. Equation~\eqref{eq:Hstark} identifies the shift in the $n$-th transmon level as $-2\anharm|\xigain|^2\,n$, so to leading order both the $\transge$ and $\transef$ transitions shift by the same amount $\domegageac = -2\anharm|\xigain|^2$. Although either transition could serve as a probe of $\xigain$, the equivalence only holds in the perturbative limit, and the two transition frequencies differ as the pump amplitude grows.

\begin{figure}[h]
    \centering
    \includegraphics[width=\linewidth]{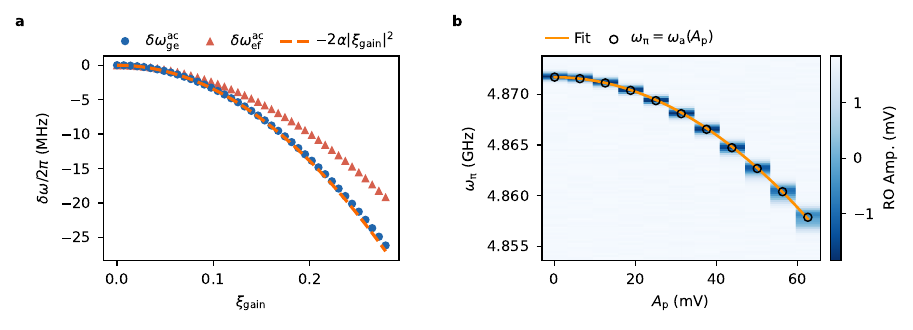}
    \caption{\textbf{Calibration of the effective gain-drive amplitude $\xigain$ from the Stark shift.} \textbf{a,} Theoretical Stark shift $\delta\omega/2\pi$ of the $\transge$ (blue circles) and $\transef$ (red triangles) transitions as a function of $\xigain$, obtained from a Floquet diagonalization of the driven transmon. The dashed orange line is the perturbative prediction $-2\anharm|\xigain|^2$. The $\transge$ shift follows this prediction over the full range, whereas the $\transef$ shift deviates at large $\xigain$, so we calibrate $\xigain$ from the $\transge$ transition. \textbf{b,} Measured \roamp\ of $\transge$ spectroscopy as a function of pump amplitude $\Ap$, with the pump-shifted resonance frequency $\wpi = \omegaqubit(\Ap)$ extracted from using Gaussian fits (open circles). The solid orange line is a fit to $\wpi(\Ap) = \omegaqubit - 2\anharm|\xigain|^2$ with $\xigain = c\,\Ap$, which calibrates the conversion $c = \xigainconv$ from pump amplitude to effective displacement.}
    \label{fig:starkshift_calib}
\end{figure}

To identify the transition that best tracks the analytic relation, we compute the exact Stark shift of the driven transmon with a Floquet analysis. We model the transmon under the gain pump with the time-periodic Hamiltonian~\cite{Koch07}
\begin{equation}
    \hat{H}(t) = 4 \Ec \hat{n}_\mr{c}^2 - \Ej \cos\hat{\varphi} + 2\hbar \, \epsgain  \left(\aop + \aod\right)\cos\!\left(\omegagain t\right),
    \label{eq:driven_transmon_floquet}
\end{equation}
where $\Ec$ and $\Ej$ are the charging and Josephson energies taken from the device measurement and $\Ej \gg \Ec$. We express the charge operator $\hat{n}_\mr{c} = i\,n_\mr{zpf}\left(\aod - \aop\right)$ and the superconducting phase operator $\hat{\varphi} = \varphi_\mr{zpf}\left(\aop + \aod\right)$ through the transmon ladder operators $\aop$, $\aod$, with zero-point fluctuations $n_\mr{zpf} = \left(\Ej/32 \Ec\right)^{1/4}$ and $\varphi_\mr{zpf} = \left(2 \Ec/\Ej\right)^{1/4}$~\cite{Koch07}. The final term is the gain pump of amplitude $\epsgain$ at frequency $\omegagain$. For this calculation, we truncate the transmon Hilbert space to $N = \Nfloquet$ levels. We diagonalize the Floquet Hamiltonian associated with Eq.~\eqref{eq:driven_transmon_floquet} over one drive period $T = 2\pi/\omegagain$, and track the quasienergies of the Floquet modes adiabatically connected to $\gstateh$, $\estateh$, and $\fstateh$.

The extracted shift of the $\transge$ and $\transef$ transitions as a function of $\xigain$ is shown in Fig.~\ref{fig:starkshift_calib}a. The $\transge$ shift follows the perturbative $-2\anharm|\xigain|^2$ scaling across the full range explored, while the $\transef$ shift departs from it at larger $\xigain$. Therefore, we calibrate $\xigain$ from the $\transge$ Stark shift, for which the analytic relation $\domegageac = -2\anharm|\xigain|^2$ remains accurate over the range of pump amplitudes used. This result also explains why we do not extract $\xigain$ from the pump-induced shift of the gain resonance $\omegagain(\Ap)$ shown in Fig.~\ref{fig:two}b of the main text. The resonance peaks mark the $\estateh \to \fstateh$ gain condition and therefore inherit the $\transef$ Stark shift, which differs from $-2\anharm|\xigain|^2$ at large pump amplitude and would bias the extracted $\xigain$. Measuring the $\transge$ Stark shift directly avoids this problem and provides a reliable calibration.  We verify this with the Floquet simulation. Feeding the $\xigain$ extracted from each Stark shift into $\ggain = \anharm\lambda|\xigain|^2$ and comparing with the $\ggain$ obtained from the Floquet quasienergy splitting of the driven transmon-cavity system, the prediction based on the $\transge$ shift is closer to the Floquet value than the one based on the $\transef$ shift.

We measure the $\transge$ Stark shift by spectroscopy of the transmon while the gain pump is applied. For each pump amplitude $\Ap$, we apply the pump at the frequency $\omegagain(\Ap)$ (see Fig.~2b of the main text) and then apply a  $\pi_\mr{ge}$ pulse of frequency $\wpi$ on the $\transge$ transition. When $\wpi$ is on resonance with the shifted $\transge$ transition frequency $\omegaqubit(\Ap)$, the $\pi_\mr{ge}$ pulse excites the transmon to $\estateh$, which the gain then excites to $\fstateh$, generating a shift in the readout amplitude. For each $\Ap$, we sweep $\wpi$ across the $\transge$ transition and fit the resonance to a Gaussian peak to extract the pump-shifted transition frequency (Fig.~\ref{fig:starkshift_calib}b). The resonance moves down in frequency as $\Ap$ increases, following $\wpi(\Ap) = \omegaqubit - 2\anharm|\xigain|^2$ with $\xigain = c\,\Ap$. A least-squares fit to this model, using the independently measured anharmonicity $\anharm$, yields the conversion $c = \xigain/\Ap = \xigainconv$, which maps the pump amplitude onto the effective displacement. Combined with $\ggain = \anharm\lambda|\xigain|^2$, equivalently $\ggain = \tfrac{1}{2}\lambda|\domegageac|$, this calibration fixes the engineered-gain rate reported throughout the main text. Note that this analysis supposes that $\xigain$ is only dependent on $\epsgain$ and independent of the frequency $\omegagain$ (see Subsec.~\ref{subsec:sys_ham}). This relies on two conditions. First, that the drive power $\epsgain$ delivered to the sample is independent of the frequency, meaning that the transmission response to the sample is flat over the considered pump frequency range. Second, that $\delta \omegagain = \omegagain(\Ap) - \omegagain(\Ap = 0) \ll \omegagain \pm \omegaqubit$, so that $\omegagain$ varies negligibly over the range of pump amplitudes used, and $\xigain$ can be treated as frequency independent to good approximation. In our experiment, we have a maximum deviation of  $\delta \omegagain \approx \SI{20}{\mega\hertz}$ with a detuning of $\omegagain - \omegaqubit \approx\SI{1.2}{\giga\hertz}$, resulting in a negligible change in $\xigain$.

\subsection{Rabi oscillations with engineered gain}
\label{subsec:rabi_gf_gain}

\begin{figure}[h]
    \centering
    \includegraphics[width=\linewidth]{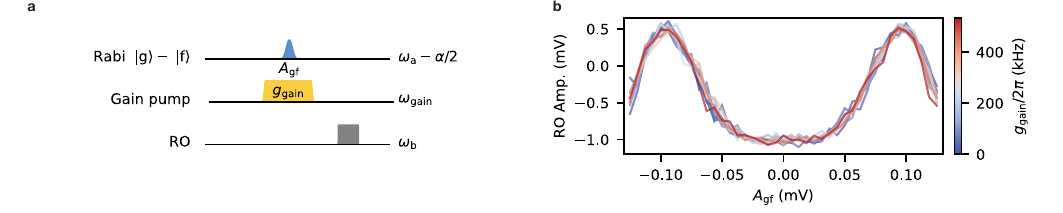}
    \caption{\textbf{Rabi oscillations on the $\transgf$ transition in the presence of the engineered gain.} \textbf{a,} Pulse sequence. We ramp on the engineered gain with a variable interaction rate $\ggain$, apply a variable-amplitude $A_\mr{gf}$ Rabi pulse on the $\transgf$ transition, ramp down the engineered-gain pulse, and perform a dispersive-shift readout. \textbf{b,} Measured \roamp\ as a function of the drive amplitude $A_{\mr{gf}}$ for increasing interaction rates $\ggain$ (color scale). The Rabi traces overlap for all $\ggain$, showing that the engineered gain leaves the coherent $\transgf$ drive unchanged.}
    \label{fig:rabi_gf_gain}
\end{figure}

In this section, we demonstrate that we can drive Rabi oscillations on the $\transgf$ transition in the presence of engineered gain, confirming that single-qubit control is compatible with the single-photon engineered gain. To show this, we perform the pulse sequence shown in Fig.~\ref{fig:rabi_gf_gain}a. We ramp on the engineered gain with a variable interaction rate $\ggain$, we wait $\dt = \SI{3}{\micro\second}$, and we apply a Rabi pulse of variable amplitude $A_\mr{gf}$, the peak output voltage of the OPX+. We then ramp down the engineered gain and perform a dispersive-shift readout.

The result of the measurement is shown in Fig.~\ref{fig:rabi_gf_gain}b. As discussed in the main text, we observe Rabi oscillations that depend on the squared drive amplitude $A_\mr{gf}^2$. Regardless of the value of the engineered gain, the Rabi oscillation amplitude and period remain the same, showing that the engineered gain does not affect the $\transgf$ transition. This is expected because the engineered gain is frequency selective and only affects the $\transef$ transition, while the two-photon drive directly induces $\transgf$ oscillations.

We now derive the Rabi oscillation contrast as a function of $A_\mr{gf}$ and show it has a quadratic dependence. Here, for simplicity, we suppose that the Rabi pulse has a square shape of duration $\tau_\mr{gf}$, but a similar analysis extends to the Gaussian pulse shape used in the experiment. 

The two-photon drive adds an additional term to the system Hamiltonian, Eq.~\eqref{eq:ham_full},
\begin{equation}
    \Htwo/\hbar = \etwo\,\aop^{\dagger 2} + \etwo^*\,\aop^2,
    \label{eq:Htwo}
\end{equation}
where $\etwo$ is the amplitude of the two-photon process. The effective drive amplitude is given by $\etwo = \anharm\,|\xitwo|^2$, where $\xitwo = \epstwo\left(\frac{1}{\omegatwo-\omegaqubit} - \frac{1}{\omegatwo+\omegaqubit}\right)$ is the effective displacement generated by a two-photon parametric pump, which has amplitude $\epstwo$ and frequency $\omegatwo=\omegaqubit - \anharm/2$.

Projecting $\Htwo$ onto the $\transgf$ subspace gives an effective coupling between $\gstateh$ and $\fstateh$,
\begin{equation}
    \Htwo^\mr{gf}/\hbar = \frac{\Omegagf}{2}\,\sigmaxgf,
    \qquad
    \sigmaxgf = \gstateh\bra{\mr{f}} + \fstateh\bra{\mr{g}},
    \label{eq:H2_rabi_gf}
\end{equation}
with two-photon Rabi rate $\Omegagf = 2\sqrt{2}\,\etwo$. Since $\Htwo$ arises from a process that is second order in the single-photon drive $\epstwo$, i.e. $\etwo\propto\epstwo^2$, the Rabi rate inherits a quadratic dependence on the applied pump amplitude:
\begin{equation}
    \Omegagf \;\propto\; \etwo \;\propto\; \epstwo^2 \;\propto\; A_\mr{gf}^2,
    \label{eq:omega_gf_scaling}
\end{equation}
where $A_\mr{gf}$ is the amplitude of the two-photon-drive pump, given by the peak output voltage of the OPX+.

Starting from the transmon in $\gstateh$ and evolving it under $\Htwo^\mr{gf}$, the Hamiltonian generates rotation about the $x$ axis of the $\transgf$ Bloch sphere. The state after the pulse time $\tau_{\mr{gf}}$ is
\begin{equation}
    \ket{\psi(\tau_{\mr{gf}})} = \cos\!\left(\frac{\Omegagf\,\tau_{\mr{gf}}}{2}\right)\gstateh - i\sin\!\left(\frac{\Omegagf\,\tau_{\mr{gf}}}{2}\right)\fstateh,
    \label{eq:rabi_gf_state}
\end{equation}
so the readout contrast after the evolution is 
\begin{equation}
   \msig{\mr{Rabi}}^\mr{gf}(\tau_{\mr{gf}}) = \frac{\rcon{\mr{g}}+\rcon{\mr{f}}}{2} + \frac{\rcon{\mr{g}}-\rcon{\mr{f}}}{2}\,\cos(\Omegagf\,\tau_{\mr{gf}}).
    \label{eq:rabi_gf_contrast}
\end{equation}
Using the dependence of $\Omegagf$ on $A_\mr{gf}^2$, we can rewrite the contrast as a function of the drive amplitude as
\begin{equation}
    \msig{\mr{Rabi}}^\mr{gf}(A_\mr{gf}) = \frac{\rcon{\mr{g}}+\rcon{\mr{f}}}{2} + \frac{\rcon{\mr{g}}-\rcon{\mr{f}}}{2}\,\cos\!\left(\pi\,\frac{A_\mr{gf}^2}{A_\pi^2}\right)\,
    \label{eq:rabi_gf_fit}
\end{equation}
where $A_\pi$ is the drive amplitude that produces a $\pi$ rotation on the $\transgf$ Bloch sphere after a pulse time $\tau_{\mr{gf}}$. The contrast oscillates with $A_\mr{gf}^2$, as observed in Figs.~4a and~\ref{fig:rabi_gf_gain}b.

\subsection{Relaxation time versus engineered-gain frequency}
\label{subsec:t1_vs_detuning}

In this section, we measure the relaxation time $\Tone{\mr{f}}$ of the gain-engineered transmon as a function of the gain-pump frequency $\omegap$. We show that the enhancement follows a Lorentzian dependence on $\omegap$, whose linewidth is comparable to the cavity decay rate $\kappacav$. This provides experimental support for the adiabatic-elimination description of Subsec.~\ref{subsec:frq_sel_gain}, in which the frequency selectivity of the engineered gain derives from the cavity susceptibility.

The lineshape of the relaxation time follows from the frequency selectivity derived in Subsec.~\ref{subsec:frq_sel_gain}. The single-photon-gain rate on the $\transefh$ transition inherits the Lorentzian cavity susceptibility of Eq.~\eqref{eq:bn_fourier}, so a pump detuning $\omegap - \omegagain$ suppresses the gain rate as
\begin{equation}
    \kappagain(\omegap) = \frac{8\ggain^2}{\kappacav}\,\frac{\kappacav^2/4}{\kappacav^2/4 + \left(\omegap - \omegagain\right)^2},
    \label{eq:kappagain_detuning}
\end{equation}
a Lorentzian centered at $\omegap = \omegagain$ with a full width at half maximum of $\kappacav$ . 

We perform the $\Tone{\mr{f}}$ pulse sequence of Fig.~\ref{fig:three}a of the main text while detuning the gain pump from its resonant value $\omegagain$. For each pump frequency $\omegap$, we ramp on the engineered gain at fixed interaction rate $\ggain/2\pi = \ggaindwval$ ($\kappagain/2\pi = \kappagaindwval$) and frequency $\omegap$, prepare the transmon in $\fstateh$, wait a variable delay $\dt$, ramp down the gain, and readout at $\omegacav-2\chidisp$. We extract $\Tone{\mr{f}}(\omegap)$ as the $1/e$ decay time of each trace rather than from an exponential fit, which remains a reliable characteristic timescale over the range of $\kappagain$ explored here (Subsec.~\ref{subsec:biexp_singleexp_gf}).

\begin{figure}[h]
    \centering
    \includegraphics[width=\linewidth]{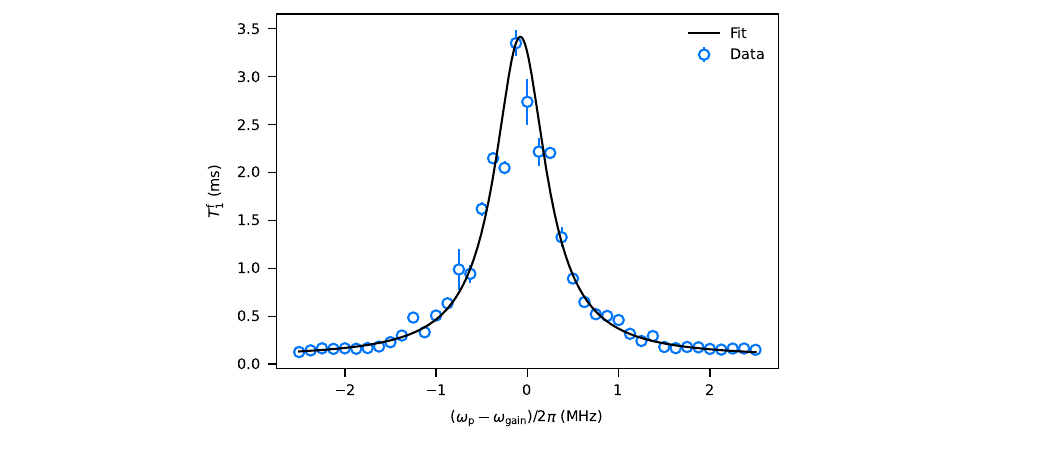}
    \caption{\textbf{Relaxation time of the gain-engineered transmon versus gain-pump frequency.}  Measured $\Tone{\mr{f}}$ as a function of the gain-pump detuning $\omegap - \omegagain$ from the resonant gain condition (open circles, with error bars denoting the uncertainty on the extracted $1/e$ decay time) together with a Lorentzian fit (solid line) at interaction rate $\ggain/2\pi = \ggaindwval$. The relaxation time peaks at $\omegap \approx \omegagain$. The fitted full width at half maximum $\kappalorfit$ matches the cavity decay rate $\kappacav/2\pi = \kappacavtotal$.}
    \label{fig:t1_vs_detuning}
\end{figure}

The result is shown in Fig.~\ref{fig:t1_vs_detuning}. The relaxation time peaks around the resonant gain condition $\omegap \approx \omegagain$, where it reaches $\approx \SI{3.3\pm0.1}{\milli\second}$. The relaxation time drops once the pump is detuned by more than a cavity linewidth. Unlike the experiments presented in the main text, this measurement was not interleaved and the resonance condition has a small detuning in the calibration. This explains the different relaxation times in Fig.~\ref{fig:t1_vs_detuning} compared with those of the main text. If the relaxation time were proportional to the gain rate, as given by Eq.~\eqref{eq:t1gf_enhanced}), we would expect $\Tone{\mr{f}}(\omegap)$ to follow the same Lorentzian dependence on the pump detuning as $\kappagain$. In the regime explored here, this proportionality does not hold exactly. This is due to residual transmon thermal population which caps $\Tone{\mr{f}}$ (Sec.~\ref{sec:ggain_sim}). The enhancement nonetheless grows monotonically with $\kappagain$, so detuning the pump suppresses it and produces a peak in $\Tone{\mr{f}}(\omegap)$ centered at $\omegap = \omegagain$.

We fit the measured $\Tone{\mr{f}}(\omegap)$ to a Lorentzian and extract a full width at half maximum of $\kappalorfit$, in excellent agreement with the independently measured cavity decay rate $\kappacav/2\pi = \kappacavtotal$. This agreement indicates that the engineered gain arises from the adiabatic elimination of the cavity mode, with the cavity linewidth setting both the frequency selectivity of the gain and the bandwidth over which the relaxation time is enhanced.

\subsection{Time-dependent simulation of engineered gain}
\label{sec:ggain_sim}

In this section, we simulate the full transmon-cavity Hamiltonian to confirm the calibration of $\ggain$, and to reproduce the enhancement of the relaxation $\Tone{\mr{f}}$ and coherence $\Ttwo{\mr{gf}}$ times as a function of $\ggain$. The simulations identify the mechanisms that limit $\Tone{\mr{f}}$, namely the transmon thermal population over the experimentally accessed range and, at larger interaction rates, an off-resonant excitation of the $\transgeh$ transition together with the breakdown of the adiabatic elimination set by the cavity linewidth $\kappacav$.

\subsubsection{Master equation}

We simulate the full transmon-cavity system using the system Hamiltonian defined in Eq.~\eqref{eq:ham_full}, and we model the open-system dynamics with the Lindblad master equation
\begin{align}
    \frac{\mr{d}}{\mr{d}t} \dm(t) &= -\frac{i}{\hbar}\left[ \hat{H} + \delta \omega_\mr{a,sim}\aod\aop +\delta \omega_\mr{b,sim}\bod\bop, \, \dm(t)\right] \nonumber \\
    &+ \kappaone{\mr{ge}}\left(1+\nth{a}\right)\diss{\gstateh\estatehb }\,\dm(t) + \kappaone{\mr{ef}}\left(1+\nth{a}\right)\diss{\estateh\fstatehb }\,\dm(t) \nonumber \\
    &+ \kappaone{\mr{ge}}\,\nth{a}\,\diss{\estateh\gstatehb }\,\dm(t) + \kappaone{\mr{ef}}\,\nth{a}\,\diss{\fstateh\estatehb}\,\dm(t) \nonumber \\
    &+ 2\kappadeph{\mr{e}}\diss{\estateh\estatehb }\,\dm(t) + 2\kappadeph{\mr{f}}\diss{\fstateh\fstatehb }\,\dm(t) \nonumber \\
    &+ \kappacav\diss{\bop}\,\dm(t),
    \label{eq:lindblad_full}
\end{align}
where $\delta \omega_\mr{a,sim}$ ($\delta \omega_\mr{b,sim}$) is a frequency detuning for the transmon (cavity) mode. The transmon detuning $\delta \omega_\mr{a,sim}$ is used in the coherence time ($\Ttwo{\mr{gf}}$) simulations (see Subsec.~\ref{subsec:cohtime_sim}), and the cavity detuning $\delta \omega_\mr{b,sim}$ is used to account for a small frequency mismatch on $\omegagain$ during the calibration of $\ggain$. The dissipators account for the single-photon relaxation rates $\kappaone{\mr{ge}}$ and $\kappaone{\mr{ef}}$ of the $\transgeh$ and $\transefh$ transitions, the corresponding dephasing rates $\kappadeph{\mr{e}}$ and $\kappadeph{\mr{f}}$, and the single-photon loss rate $\kappacav$ of the cavity. We include the thermal excitation of the transmon using $\nth{a}$. We model the cavity as coupled to a zero-temperature bath because the measured cavity thermal photon number $\nth{b}$ is sufficiently small that the simulated dynamics remain unchanged. All parameters are taken from the measured values reported in Table~\ref{tab:params_heatmon}, unless stated otherwise. 

To reduce the computational workload, we truncate the Hilbert space of the simulations. For the cavity Hilbert space, we use the first two Fock states. For the calibration simulation, we truncate the transmon to the first three Fock states, which is sufficient to capture the dynamics of the $\transgf$ subspace under the engineered gain. For the relaxation and coherence time simulations, we extend the transmon Hilbert space to the first four Fock states, adding the level $\hstateh$, so that the simulation also captures the population leaving $\fstateh$ toward $\hstateh$. In this case, we add the following dissipators to Eq.~\eqref{eq:lindblad_full}:
\begin{equation}
    \kappaone{\mr{fh}}\left(1+\nth{a}\right)\diss{\fstateh\hstatehb}, \quad \kappaone{\mr{fh}}\,\nth{a}\,\diss{\hstateh\fstatehb},
    \label{eq:lindblad_h}
\end{equation}
where $\kappaone{\mr{fh}}$ is the single-photon relaxation rate of the $\transfhh$ transition. Here, we set $\kappaone{\mr{fh}} = 3\kappaone{\mr{ge}}$ following the predicted scaling of the matrix elements~\cite{Blais2021}.

\subsubsection{Calibration simulation}

\begin{figure}
    \centering
    \includegraphics[width=\linewidth]{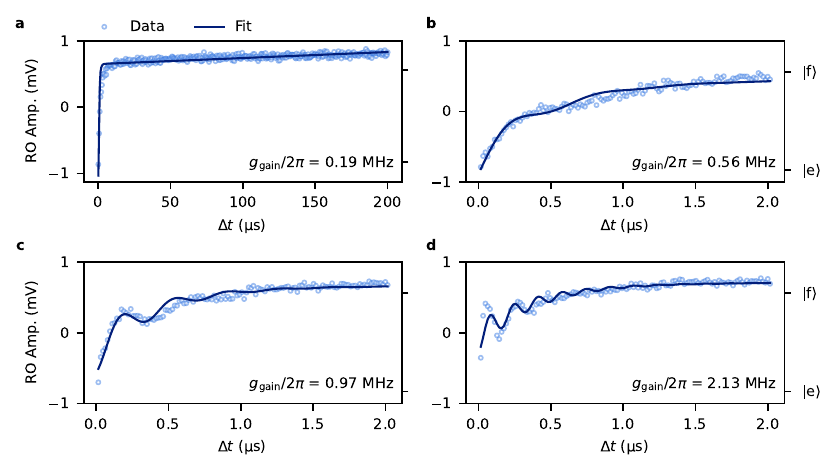}
    \caption{\textbf{Calibration traces at large engineered-interaction rates.} Measured \roamp\ as a function of the pump duration $\dt$ (open circles) together with the master equation simulation of Eq.~\eqref{eq:lindblad_full} (solid lines), for $\ggain/2\pi = \ggaintracea$ (\textbf{a}), $\ggaintraceb$ (\textbf{b}), $\ggaintracec$ (\textbf{c}), and $\ggaintraced$ (\textbf{d}). Each simulation uses the $\ggain$ predicted from the measured $\transge$ Stark shift, with the readout contrasts $\rcon{\mr{g,e,f}}$. The right axes mark the readout values corresponding to the transmon populations in $\estate$ and $\fstate$. The oscillations reflect the breaking of the adiabatic elimination limit at these rates, and the simulations track them reasonably well.}
    \label{fig:ggain_traces_SI}
\end{figure}

To simulate the time-domain calibration experiment shown in Fig.~\ref{fig:two} of the main text, we set $\delta \omega_\mr{a,sim} = 0$ and $\nth{a} = 0$ and initialize the system in $\dm(0) = \ket{\mr{e},0}\bra{\mr{e},0}$, where the first index denotes the transmon state and the second the Fock state of the cavity. We evolve the system for a variable duration $\dt$ and compute the readout signal as
\begin{equation}
    \msig{}(\dt) = \rcon{\mr{f}}\fstatehb \rhoa(\dt) \fstateh + \rcon{\mr{e}}\estatehb \rhoa(\dt) \estateh + \rcon{\mr{g}}\gstatehb \rhoa(\dt) \gstateh,
    \label{eq:readout_sim_gain}
\end{equation}
where $\rcon{\mr{g,e,f}}$ are the readout contrasts of the $\gstateh$, $\estateh$, and $\fstateh$ states, respectively, and $\rhoa$ is the reduced density matrix of the transmon obtained by tracing out the cavity. For each simulated trace, we use the measured system parameters and the value of $\ggain$ predicted from the measured $\transge$ Stark shift $\domegageac$ (see main text), leaving the readout contrasts $\rcon{\mr{g,e,f}}$ as free parameters. To account for a small residual detuning between the experimental pump $\omegap$ and the gain resonance $\omegagain$, we fix $\delta \omega_\mr{b,sim}$ to the measured value at each $\ggain$.
The agreement between simulation and experiment, shown in Figs.~3c,d, validates the predicted $\ggain$.

For completeness, we show the results of the simulations at values of $\ggain$ that we do not report in the main text. Figure~\ref{fig:ggain_traces_SI} shows the measured and simulated calibration traces at $\ggain/2\pi = \ggaintracea$, $\ggaintraceb$, $\ggaintracec$, and $\ggaintraced$. For reference, we include the readout signals corresponding to $\estate$- and $\fstate$-state populations obtained from independent Rabi measurements. As for the data presented in Fig.~\ref{fig:two} of the main text, the readout signal increases from close to the $\estate$-value towards the $\fstate$-value for increasing pump duration $\dt$. Note that the initial readout contrast for higher $\ggain$ does not start at $\estate$ due to a finite ramp time for the engineered gain pulse. For the highest $\ggain$ values, the transmon reaches $\fstateh$ within a few hundred nanoseconds and the traces develop coherent oscillations with a frequency that increases with $\ggain$. These oscillations are due to the breaking of the adiabatic elimination limit. The simulation reproduces the oscillation frequency with the $\ggain$ predicted from the measured $\transge$ Stark shift. 
As discussed in the main text, we attribute the saturation in measured readout contrast above the $\fstate$-value to leakage to higher transmon levels. In the simulations, higher transmon levels are not explicitly included. However, the simulation closely follows the experimental result with the readout contrast as a free fit parameter, demonstrating that this leakage does not have a significant impact on the two-mode squeezing interaction dynamics as a function of pump duration. 

\subsubsection{Relaxation time simulation}
To simulate the enhancement of the relaxation times of the two computational states as a function of $\ggain$, we set $\delta \omega_\mr{a,sim} = \delta \omega_\mr{b,sim} = 0$ and evolve the system for a variable duration $\dt$. Here, we include the fourth transmon level $\hstateh$ together with the relaxation and thermal excitation dissipators of the $\transfhh$ transition (Eq.~\eqref{eq:lindblad_h}), so that the simulation accounts for the population that leaves the $\transgf$ subspace. We initialize the system in $\dm(0) = \ket{\mr{f},0}\bra{\mr{f},0}$ ($\ket{\mr{g},0}\bra{\mr{g},0}$) and compute the $\fstateh$- ($\gstateh$-) state population $\fstatehb \rhoa(\dt) \fstateh$ ($\gstatehb \rhoa(\dt) \gstateh$) to obtain $\Tone{\mr{f,sim}}$ ($\Tone{\mr{g,sim}}$).
In both cases, we extract the decay time constant by fitting to a simple exponential decay (see Subsec.~\ref{subsec:biexp_singleexp_gf}). We repeat this procedure over a range of $\ggain$ values.

\begin{figure}
    \centering
    \includegraphics[width=\linewidth]{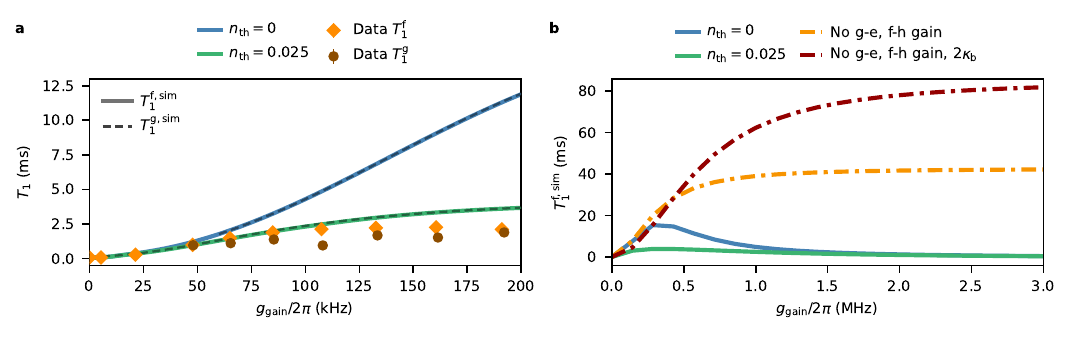}
    \caption{\textbf{Simulated relaxation time of the gain-engineered transmon as a function of the engineered interaction rate $\ggain$.} \textbf{a,} Simulated relaxation times $\Tone{\mr{f,sim}}$ (thick solid lines) and $\Tone{\mr{g,sim}}$ (thin dashed lines) over the interaction rates accessed in the experiment. The blue lines are the simulation for a zero transmon thermal population $\nth{a} = 0$ and the green lines use the measured $\nth{a} = \nthhvalsim$. Orange diamonds and brown circles are the measured $\Tone{\mr{f}}$ and $\Tone{\mr{g}}$. \textbf{b,} The same simulation extended to interaction rates well beyond the experimental range, showing $\Tone{\mr{f,sim}}$ only. The full-gain simulations (blue, $\nth{a} = 0$, green, measured $\nth{a}$) reach a maximum and then decrease, due to the off-resonant excitation of the $\transgeh$ and $\transfhh$ transitions. Keeping only the resonant $\transefh$ term of the gain Hamiltonian (Eq.~\eqref{eq:gain_ham_ef_only}, orange dash-dotted, $\nth{a} = 0$) suppresses this turnover, and $\Tone{\mr{f,sim}}$ saturates at the adiabatic-elimination ceiling. Doubling the cavity linewidth to $2\kappacav$ (dark red dash-dotted) doubles the saturated $\Tone{\mr{f,sim}}$.}
    \label{fig:sim_T1_vs_ggain}
\end{figure}

The results are shown in Fig.~\ref{fig:sim_T1_vs_ggain}a. We find that $\Tone{\mr{g,sim}}$ matches $\Tone{\mr{f,sim}}$ for $\ggain/2\pi > \ggainsimonetwolevel$, both for the measured transmon thermal photon number $\nth{a} =\nthhvalsim$ and for $\nth{a} = 0$. For $\nth{a} = 0$ and $\ggain/2\pi < \ggainsimonetwolevel$, we cannot extract $\Tone{\mr{g,sim}}$ because the off-resonant excitation of the $\transgeh$ transition is negligible at these interaction rates and the system remains in $\gstateh$ for the entire simulation time. 
At low interaction rates, the simulation reproduces the measured rise of $\Tone{\mr{f}}$ and $\Tone{\mr{g}}$ with no free parameters. At higher interaction rates, the simulation with $\nth{a} = 0$ yields much longer relaxation times than the measured ones. The simulated relaxation time with the measured $\nth{a}$ more closely follows the measured values. This indicates that the transmon thermal population is the dominant limitation on the measured relaxation times. The agreement is better for $\Tone{\mr{f}}$, while the measured $\Tone{\mr{g}}$ falls below the simulation in this range. Toward higher $\ggain$, however, the simulations still overestimate both relaxation times and do not capture the measured saturation. We attribute this discrepancy to an increase in excitations away from the computational states $\{\gstate,\fstate\}$ with increasing $\ggain$, which could be due to pump-induced thermal heating or spurious multiphoton transitions in the cavity-transmon system~\cite{Dai2026}. 

We see that $\Tone{\mr{f}}\approx\Tone{\mr{g}}$ in the simulation. We understand this from a competition between the engineered gain rate and the intrinsic decay rates. For $\kappagain\gg\kappaone{\mr{ef}}>\kappaone{\mr{ge}}$, the engineered gain evacuates any population in $\estate$ arising from single-photon decay from $\fstate$, as well as excitations from $\gstate$ arising due to thermal fluctuations or off-resonant driving from the gain pump. As a result, there is negligible population in $\estate$ such that the dynamics can be described by transitions in an effective two-level subspace $\{\gstate,\fstate\}$. Hence, we expect $\Tone{\mr{g}}\approx\Tone{\mr{f}}$.

To identify the mechanisms that ultimately bound $\Tone{\mr{f}}$, we extend the simulation to interaction rates well beyond the experimental range, up to $\ggain/2\pi = \ggainsimonemax$ (Fig.~\ref{fig:sim_T1_vs_ggain}b). Even for $\nth{a} = 0$, $\Tone{\mr{f,sim}}$ does not follow the ideal scaling indefinitely but reaches a maximum of $\Tonefsimpeak$ near $\ggain/2\pi = \ggainsimonepeak$ and then decreases. This turnover arises because the transmon is a multilevel system, in which the gain operator $\aod\bod$ drives not only the resonant $\transefh$ gain but also the off-resonant $\transgeh$ and $\transfhh$ transitions, which move population out of the computational states $\gstateh$, $\fstateh$. Restricting the transmon to the first four levels, the gain interaction of Eq.~\eqref{eq:gain_interaction_decomposed} can be written as
\begin{equation}
    \Hgain/\hbar = \ggain\left[\left(\estateh\gstatehb + \sqrt{2}\,\fstateh\estatehb + \sqrt{3}\,\hstateh\fstatehb\right)\bod + \mr{h.c.}\right].
    \label{eq:gain_ham_levels}
\end{equation}
To isolate the effect of the resonant $\transefh$ term and identify the limit set by the finite cavity loss rate, we repeat the simulation with the two off-resonant terms removed from $\Hgain$, leaving a Hamiltonian that acts only on the $\transefh$ transition,
\begin{equation}
    \Hgain^{\mr{ef}}/\hbar = \sqrt{2}\,\ggain\left(\fstateh\estatehb\,\bod + \estateh\fstatehb\,\bop\right),
    \label{eq:gain_ham_ef_only}
\end{equation}
while all dissipators, including those of the $\transfhh$ transition (Eq.~\eqref{eq:lindblad_h}), are left unchanged. With this Hamiltonian (dash-dotted lines in Fig.~\ref{fig:sim_T1_vs_ggain}b), the decrease of $\Tone{\mr{f,sim}}$ disappears and $\Tone{\mr{f,sim}}$ instead saturates around $\Tonefsimceil$. This remaining ceiling comes from the breakdown of the adiabatic elimination when $\ggain \gtrsim \kappacav$. Doubling the cavity linewidth to $2\kappacav$ raises the ceiling to $\Tonefsimceiltwokappa$, indicating that the ultimate limit on $\Tone{\mr{f}}$ is set by the cavity linewidth. Experimentally, a large frequency-selective engineered gain rate can be implemented by using multi-stage lossy filter~\cite{Putterman2022,Thorbeck24,Direkci26},
which will keep the engineered gain acting only on the $\transef$ transition while still enabling large $\kappagain$.

\subsubsection{Coherence time simulation.}
\label{subsec:cohtime_sim}
To simulate the coherence time $\Ttwo{\mr{gf,sim}}$, we set $\delta \omega_\mr{a,sim}/2\pi = \SI{10}{\kilo\hertz}$ and $\delta \omega_\mr{b,sim} = 0$ in the master equation (Eqs.~\eqref{eq:lindblad_full} and~\eqref{eq:lindblad_h}). Further, we set $\kappadeph{\mr{e,f}}$ in Eq.~\eqref{eq:lindblad_full} to the measured value at $\ggain=0$. As for the relaxation time simulation, we include the fourth transmon level $\hstateh$ together with the dissipators of the $\transfhh$ transition (Eq.~\eqref{eq:lindblad_h}). We initialize the system in $\dm(0) = (\ket{\mr{f},0} + \ket{\mr{g},0})(\bra{\mr{f},0}+\bra{\mr{g},0})/2$ and evolve it for a variable duration $\dt$. We then apply a $\rot{\frac{\pi}{2}}{\mr{gf}}$ (Eq.~\eqref{eq:pi_2_gf}) pulse following the protocol described in Subsec.~\ref{subsec:decoherence_gf_gain}, and compute $\fstatehb \rot{\frac{\pi}{2}}{\mr{gf}} \rhoa(\dt)\rot{\frac{\pi}{2}}{\mr{\dagger,gf}} \fstateh$. We extract $\Ttwo{\mr{gf,sim}}$ by fitting the resulting signal using Eq.~\eqref{eq:T2_gf_ramsey_gain}, with $\kappagain$ computed from $\ggain$ (Eq.~\eqref{eq:kappagain}) and the readout contrasts and $\kappadeph{\mr{f}}$ as free fit parameters. This is the same procedure that is applied to the experimental data, such that $\kappadeph{\mr{f}}$ is extracted as a function of $\ggain$ from the simulated traces.
We repeat the simulation from $\ggain /2\pi = 0$ up to $\ggain/2\pi = \ggainsimtwomax$, which covers the whole range of interaction rates we use in the experiment for $\Ttwo{\mr{gf}}$.

We find from the simulation that $\Ttwo{\mr{gf,sim}}$ is set by $\kappadeph{\mr{f}}$ at $\ggain=0$  and does not depend on $\ggain$. This is consistent with the model discussed in Subsec.~\ref{subsec:frq_sel_gain}, which we briefly recap below. The engineered gain acts exclusively on the $\estateh \to \fstateh$ transition. To illustrate this mechanism, consider a system initialized in a superposition of $\gstateh$ and $\fstateh$. A single photon loss event collapses this superposition into $\estateh$, and the engineered single-photon gain restores the population to $\fstateh$ without introducing additional dephasing. This prediction is in stark contrast with the experimentally measured $\Ttwo{\mr{gf}}$, which decreases substantially with increasing $\ggain$. The discrepancy indicates that the single-photon-gain pump introduces a dephasing mechanism not captured by the model. 
As discussed in the main text, this dephasing mechanism likely has a significant low-frequency component since the measured $\Ttwo{\mr{gf}}$ is lower than $\Ttwoe{\mr{gf}}$.

\subsection{Interleaved coherence time measurements}
\label{sec:interleaved coherence time measurements}

In this section, we compare the results of the interleaved Ramsey coherence measurements presented in Fig.~\ref{fig:three}e of the main text to those obtained from individual runs of the interleaved measurement. As stated in the main text, these results were acquired in an interleaved manner over a continuous four-day period, with a total of 84 individual runs of approximately four minutes each. 

\begin{figure}
    \centering
    \includegraphics[width=\linewidth]{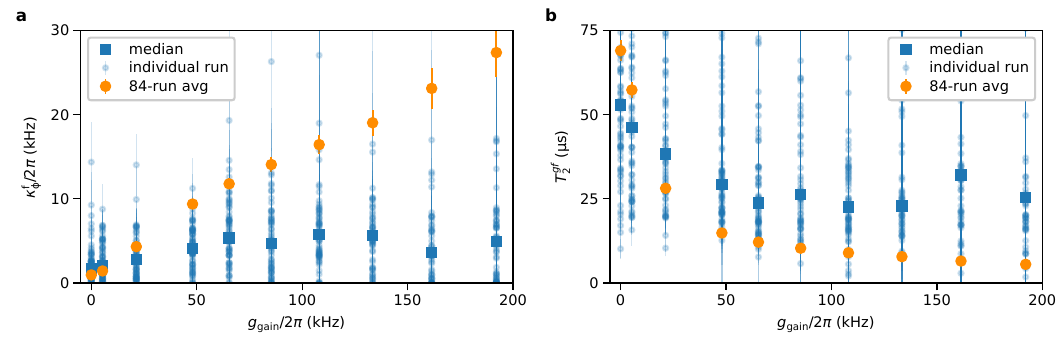}
    \caption{\textbf{Comparison of averaged and individual coherence measurements.}
    Dephasing rate $\kappadeph{\mr{f}}$ (\textbf{a}) and coherence time $\Ttwo{\mr{gf}}$ (\textbf{b}) as a function of $\ggain$. Yellow (blue) circles are the fitted results from the measurement across all (individual) runs. The blue square is the median of the individual runs for each $\ggain$. Error bars indicate the uncertainty on the fitted result.
    }
    \label{fig:dephasing_median}
\end{figure}

We fit the individual traces from each run to Eq.~\eqref{eq:T2_gf_ramsey_gain} and extract $\kappadeph{\mr{f}}$, shown as light blue circles in Fig.~\ref{fig:dephasing_median}a. The median $\kappadeph{\mr{f}}$ from the distribution at each $\ggain$ is plotted as the blue squares. The individual traces have a low signal-to-noise ratio, leading to a high uncertainty on each $\kappadeph{\mr{f}}$. Nevertheless, almost all individual values, including the median, lie below the $\kappadeph{\mr{f}}$ obtained from the signal averaged across all 84 runs (orange circles). The corresponding $\Ttwo{\mr{gf}}$ values are shown in Fig.~\ref{fig:dephasing_median}b (orange circles identical to Fig.~\ref{fig:three}e of the main text).

The significant difference in $\kappadeph{\mr{f}}$ obtained from individual runs compared with the average across the full dataset suggests noise processes acting on timescales longer than an individual run. Additionally, the linear scaling of the averaged $\kappadeph{\mr{f}}$ also indicates low-frequency noise induced by the gain pump. This could be related to long-time drifts in the input power, which change the qubit Stark shift. Such frequency drifts would manifest as a larger extracted $\kappadeph{\mr{f}}$ when the data is averaged across all measurement runs. Echo coherence measurements, instead, are not sensitive to such low-frequency noise acting between runs. This effect might explain the significant difference between $\Ttwoe{\mr{gf}}$ and $\Ttwo{\mr{gf}}$ presented in the main text.

Finally, repeating the same analysis procedure for relaxation times, echo coherence times, and gate fidelities yields no significant difference between the median and 84-run averaged result. 

\newpage
\section{Gain-Engineered Transmon as a noise-biased ancilla qubit}
\label{sec:heatmon_ancilla}

In this section, we explore the use of the gain-engineered transmon as a noise-biased ancilla qubit for fault-tolerant syndrome extraction. We first present the theoretical framework that establishes how the engineered gain suppresses the propagation of ancilla loss errors to a logical qubit. We then support this analysis with numerical simulations, which show that increasing the engineered gain rate $\kappagain$ progressively reduces the ancilla backaction. Our treatment follows the framework introduced in Ref.~\cite{Puri2019} for the Kerr-cat syndrome detector.

Throughout this section, we use $\bop$ to indicate a high-quality-factor storage mode that holds the logical qubit~\cite{Blais2021}. The readout cavity of Sec.~\ref{subsec:sys_ham} does not appear explicitly, because we model the engineered gain as the effective dissipator $\kappagain\diss{\fstateh\estatehb}$ derived in Subsec.~\ref{subsec:frq_sel_gain}. We also describe the relaxation of the ancilla with the single collapse operator $\kappaqubit\diss{\aop}$ acting on the transmon mode, rather than with the transition-resolved rates $\kappaone{\mr{ge}}$ and $\kappaone{\mr{ef}}$ of Sec.~\ref{sec:decoherence_model}. The matrix element of $\aop$ on the $\transefh$ transition is $\sqrt{2}$, so this choice fixes $\kappaone{\mr{ef}} = 2\kappaone{\mr{ge}} = 2\kappaqubit$.

\subsection{Fault-tolerant syndrome extraction}
\label{subsec:heatmon_syndrome_extraction}

Here, we explain how the gain-engineered transmon suppresses the propagation of loss errors to the logical qubit. We first analyze the undriven $\transgeh$ encoding, and then extend the analysis to the gain-engineered $\transgfh$ encoding.

To see how an error propagates from a transmon ancilla to the logical qubit during syndrome extraction, we consider a general interaction between the two,
\begin{equation}
    \Vint/\hbar = \etaint \hat{O}\aod \aop,
    \label{eq:Vint_ancilla}
\end{equation}
where $\etaint$ is the interaction strength, $\hat{O}$ is an operator acting on the logical qubit, and $\aop$ is the transmon annihilation operator. In the undriven encoding, we store the ancilla in the $\gstateh$ and $\estateh$ transmon states, and we use $\Vint$ to synthesize a syndrome measurement on the logical qubit~\cite{Puri2019}. An error on the transmon during the interaction can propagate to the logical qubit. Dephasing errors, represented by the action of $\aod \aop$, commute with the interaction, $[\aod \aop, \Vint] = 0$, and therefore do not propagate to the logical qubit. In contrast, energy loss errors, represented by the action of $\aop$, do not commute with the interaction, $[\aop, \Vint] \neq 0$. A single-photon loss event on the transmon therefore corrupts the information stored in the logical qubit.

We derive the condition under which the engineered gain suppresses this propagation. In the gain-engineered transmon, the ancilla states are $\gstateh$ and $\fstateh$, and the engineered gain pumps from $\estateh$ to $\fstateh$ at rate $\kappagain$. The ancilla and the logical qubit interact for a total time $T$ to synthesize the syndrome measurement~\cite{Puri2019}. Suppose a loss event occurs at time $\tau$ and maps $\fstateh \rightarrow \estateh$. The gain-engineered transmon then dwells in $\estateh$ for a time $\taudwell$ before the engineered gain restores it to $\fstateh$. The resulting evolution operator, up to a normalization constant, is
\begin{equation}
    U(T-\tau-\taudwell)\, \hat{C}\, U(\taudwell)\,\aop\, U(\tau),
    \label{eq:ancilla_evolution_loss}
\end{equation}
where $\hat{C} = \hat{I}_\mr{b} \otimes \fstateh\estatehb_\mr{a}$ is the correction operator applied by the engineered gain and $U(t) = e^{-i \Vint t/\hbar }$ is the unitary generated by $\Vint$. Equation~\eqref{eq:ancilla_evolution_loss} describes a single trajectory, in which the loss event occurs at a given time $\tau$ and the engineered gain restores the ancilla after a given dwell time $\taudwell$. During the dwell time, the ancilla sits outside the computational subspace, in $\estateh$, while the logical qubit keeps evolving under $\Vint$. The logical qubit therefore acquires a spurious rotation of the form $\Rrot = e^{-i\thetaerr \hat{O}}$, generated by the same operator $\hat{O}$ that synthesizes the syndrome measurement, with an amplitude $\thetaerr \propto \etaint\taudwell$. This rotation is the error that a loss event propagates from the ancilla to the logical qubit.

Both $\tau$ and $\taudwell$ are random variables, set by the stochastic jump processes of the two dissipators. The engineered gain returns the ancilla to $\fstateh$ at rate $\kappagain$, such that the average dwell time is $\taudwellavg = 1/\kappagain$. The mean error amplitude therefore scales as $\thetaerravg \propto \etaint/\kappagain$. In the regime $\kappagain \gg \etaint$, the dwell time is short compared to the timescale on which $\Vint$ acts for every trajectory, and $\thetaerr$ goes to zero. The ancilla then becomes transparent to single-photon loss, which renders the stabilizer measurement fault-tolerant with respect to ancilla loss errors.

This analysis sets the required hierarchy $\kappagain \gg \etaint,\,\kappaqubit$. The engineered gain must be faster than the rate $\etaint$ at which errors transfer from the ancilla to the logical qubit, and faster than the single-photon loss rate $\kappaqubit$ of the transmon.

\subsection{Simulations of syndrome extraction}
\label{subsec:heatmon_ancilla_sims}

In this subsection, we confirm the fault-tolerance property of the gain-engineered transmon with numerical simulations. We first introduce the simulation framework and the two metrics we use to quantify the fault tolerance of the syndrome extraction, the logical-qubit backaction $\backact$ and the ancilla error-detection infidelity $\detinf$. We then apply this framework to three logical-qubit encodings.

We consider a logical qubit encoded in the harmonic mode of a high-quality-factor storage resonator, dispersively coupled to the gain-engineered transmon. The system Hamiltonian in a rotating frame reads
\begin{equation}
    \hat{H}/\hbar = \frac{\anharm}{2}\aod\aop - \frac{\anharm}{2}\aop^{2\dagger}\aop^{2} + \Vint/\hbar,
    \label{eq:ancilla_hamiltonian}
\end{equation}
where $\anharm$ is the transmon anharmonicity, so that the first two terms are the transmon Hamiltonian $\Ha$ of Eq.~\eqref{eq:Ha}. The rotating frame is $\omegaqubit - \anharm/2$ for the transmon mode $\aop$ and $\omegacav$ for the storage mode $\bop$, with $\omegacav$ the frequency of the storage resonator. The operator $\hat{O}$ synthesizes the target stabilizer $\Sstab$ on the logical qubit. The ancilla dissipators are
\begin{equation}
    \kappaqubit\diss{\aop}, \qquad \kappagain\diss{\fstateh\estatehb_\mr{a}},
    \label{eq:ancilla_diss}
\end{equation}
with $\kappaqubit$ the single-photon loss rate and $\kappagain$ the engineered gain rate of the ancilla. We assume no storage decoherence, so that we can attribute any error on the logical qubit to ancilla errors. The master equation is then
\begin{equation}
    \frac{\mr{d}}{\mr{d}t}\dm(t)= -\frac{i}{\hbar}[\hat{H}, \dm(t)] + \kappaqubit\diss{\aop}\dm(t) + \kappagain\diss{\fstateh\estatehb_\mr{a}}\dm(t).
    \label{eq:ancilla_me}
\end{equation}

Using this master equation, we simulate a single round of error detection, in which we check a stabilizer $\Sstab$ of the logical qubit. We initialize the system in
\begin{equation}
    \dm(0) = \ket{\psi_0} \bra{\psi_0}, \quad \ket{\psi_0} = \ket{\psi_+}_\mr{a} \logstate_\mr{b},
\end{equation}
where $\logstate_\mr{b}$ is a logical state that is a $+1$ eigenstate of the stabilizer $\Sstab$, and $\ket{\psi_+}_\mr{a} = \frac{1}{\sqrt{2}}(\gstateh_\mr{a} + \fstateh_\mr{a})$ is the ancilla $+X$ eigenstate. We then evolve the system for an interaction time $T$ chosen so that the ideal unitary is~\cite{Puri2019}
\begin{equation}
    U(T) = \hat{I}_\mr{a}\,\frac{1+\Sstab}{2} + \hat{\sigma}_\mr{z,a}\,\frac{1-\Sstab}{2},
    \label{eq:ancilla_ideal_unitary}
\end{equation}
where $\hat{\sigma}_\mr{z,a} = \gstateh\gstatehb_\mr{a} - \fstateh\fstatehb_\mr{a}$ and $\hat{I}_\mr{a} = \gstateh\gstatehb_\mr{a} + \fstateh\fstatehb_\mr{a}$ are defined on the computational subspace of the gain-engineered transmon. An error on the logical qubit flips the ancilla from $\ket{\psi_+}_\mr{a}$ to $\ket{\psi_-}_\mr{a} = \frac{1}{\sqrt{2}}(\gstateh_\mr{a} - \fstateh_\mr{a})$. From the final state $\dm(T)$ we extract the logical-qubit backaction and the ancilla error-detection infidelity,
\begin{equation}
    \backact = 1 - \bra{\Psi_\mr{L}}\rhob(T)\logstate, \qquad \detinf = 1 - \bra{\psi_+}\rhoa(T)\ket{\psi_+},
    \label{eq:ancilla_metrics}
\end{equation}
where $\rhoa$ and $\rhob$ are the reduced density matrices of the ancilla and of the logical qubit. The backaction $\backact$ quantifies the disturbance of the logical state caused by ancilla errors propagating through $\Vint$, and it vanishes for a perfectly fault-tolerant interaction. The error-detection infidelity $\detinf$ quantifies the deviation of the ancilla from its initial state. Since we initialize the storage mode in a logical state, the ancilla should remain in $\ket{\psi_+}_\mr{a}$, and $\detinf$ therefore reflects the ancilla errors that would lead to an incorrect stabilizer outcome.

We now apply this framework to three logical-qubit encodings and assess how well the gain-engineered transmon suppresses the backaction and the error-detection infidelity. We report the simulation parameters of each encoding in the corresponding subsection.

\subsubsection{First-order binomial code}
\label{subsubsec:ancilla_binomial}

The first encoding we consider is the first-order binomial code~\cite{Michael2016,Hu2019,Ni2023}, a bosonic code designed to correct single-photon loss errors. The logical codewords are
\begin{equation}
    \zerol_\mr{b} = \frac{1}{\sqrt{2}}(\ket{0}_\mr{b} + \ket{4}_\mr{b}), \quad \onel_\mr{b} = \ket{2}_\mr{b},
    \label{eq:binomial_codewords}
\end{equation}
which consist exclusively of even Fock states and therefore have even photon-number parity. A single-photon loss event maps the qubit into the error subspace spanned by
\begin{equation}
    \zeroe_\mr{b} = \ket{3}_\mr{b}, \quad \onee_\mr{b} = \ket{1}_\mr{b},
    \label{eq:binomial_errorwords}
\end{equation}
which have odd photon-number parity. We can therefore detect errors by monitoring the photon-number parity of the storage mode. Moreover, both logical states share the same average photon number $\nbar = 2$, so a single-photon loss event carries no information about which logical state was occupied. The error does not collapse the logical superposition, and it remains fully correctable once detected.

We realize the parity measurement through a dispersive interaction between the storage mode and the ancilla transmon of the form of Eq.~\eqref{eq:Vint_ancilla}, namely
\begin{equation}
    \Vint/\hbar = -\chidisp \aod \aop \bod \bop,
    \label{eq:ancilla_interaction_parity}
\end{equation}
where $\chidisp$ is the dispersive coupling between the transmon and the storage mode, so that $\etaint = \chidisp$ in the hierarchy derived in Subsec.~\ref{subsec:heatmon_syndrome_extraction}. Applying this interaction for a duration $T = \pi/2\chidisp$, we obtain the unitary
\begin{equation}
    U(T) = e^{-i \Vint T/\hbar} = \frac{1+\Pstab}{2} \hat{I}_\mr{a} + \frac{1-\Pstab}{2} \hat{\sigma}_\mr{z,a},
    \label{eq:ancilla_interaction_parity_unitary}
\end{equation}
where $\Pstab = e^{i\pi \bod \bop}$ is the photon-number parity operator of the storage mode. This duration is half the value required for a $\transgeh$ ancilla~\cite{Sun14}, because the $\fstateh$ state of the $\transgfh$ encoding carries twice the dispersive shift of $\estateh$ and therefore doubles the effective interaction rate. For even logical parity, $\langle \Pstab\rangle = 1$ and the ancilla state is unchanged. For odd parity, $\langle \Pstab\rangle = -1$ and the ancilla acquires a $\hat{\sigma}_\mr{z,a}$ rotation, which reveals the single-photon loss error.

\begin{figure}
    \centering
    \includegraphics[width=\linewidth]{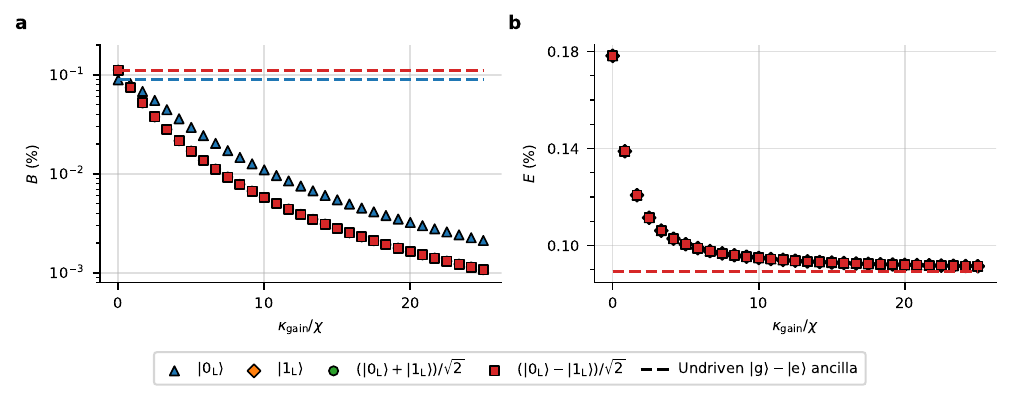}
    \caption{\textbf{First-order binomial code.} \textbf{a,} Measurement backaction $\backact$ as a function of $\kappagain$. The backaction for the states $\left( \zerol_\mr{b} \pm \onel_\mr{b} \right) / \sqrt{2}$ is identical and the markers overlap, while the backaction for the state $\onel_\mr{b}$ is zero. \textbf{b,} Error-detection infidelity $\detinf$ as a function of $\kappagain$. The infidelity is identical for all initial logical states and the markers overlap. Dashed lines are $\backact$ and $\detinf$ for an undriven $\transgeh$ transmon ancilla. Simulation parameters are listed in Table~\ref{tab:ancilla_sim_binomial}.}
    \label{fig:ancilla_binomial}
\end{figure}

We simulate this parity-measurement protocol using the master equation of Eq.~\eqref{eq:ancilla_me}, with the parameters listed in Table~\ref{tab:ancilla_sim_binomial}. Figure~\ref{fig:ancilla_binomial}a shows the measurement backaction $\backact$ and Fig.~\ref{fig:ancilla_binomial}b the error-detection infidelity $\detinf$, both as functions of $\kappagain$, together with the values obtained for an undriven $\transgeh$ ancilla (dashed lines). In the limit $\kappagain \gg \chidisp$, the backaction on the logical qubit drops well below that of the undriven ancilla, and the error-detection infidelity $\detinf$ converges to that of the undriven transmon.

The convergence of $\detinf$ comes from a compensation between two effects. The gain-engineered transmon performs the parity measurement twice as fast as the undriven ancilla, because $\fstateh$ carries twice the dispersive shift of $\estateh$. Its Ramsey coherence time is however twice as short, because $\fstateh$ decays at twice the rate of $\estateh$. The probability $\ploss$ that the ancilla undergoes a single-photon loss event during one parity measurement is therefore the same for the two encodings. The difference in $\detinf$ that we observe at small $\kappagain$ instead comes from the fact that such a single-photon loss event causes the ancilla state to leave the computational subspace $\{\gstate,\fstate\}$.

We intuitively explain the effect of this leakage. We work in the regime $\chidisp \gg \kappaqubit$, so that at most one loss event occurs during a single measurement. Defining $\ket{\psiloss}$ as the ancilla state right after such an event, we can expand the error-detection infidelity of Eq.~\eqref{eq:ancilla_metrics} to first order in $\ploss$ as
\begin{equation}
    \detinf = 1 - \left[ (1-\ploss) + \ploss \left|\braket{\psi_+}{\psiloss}\right|^2 \right] = \ploss \left( 1 - \left|\braket{\psi_+}{\psiloss}\right|^2 \right),
    \label{eq:detinf_single_loss}
\end{equation}
where $\ket{\psi_+}_\mr{a}$ is $(\gstateh_\mr{a} + \estateh_\mr{a})/\sqrt{2}$ for the undriven $\transgeh$ transmon and $(\gstateh_\mr{a} + \fstateh_\mr{a})/\sqrt{2}$ for the $\transgfh$ encoding.
For the undriven ancilla, a loss event maps $\estateh$ onto $\gstateh$, so that $\ket{\psiloss} = \gstateh_\mr{a}$, $\left|\braket{\psi_+}{\psiloss}\right|^2 = 1/2$ and $\detinf = \ploss/2$. For the $\transgfh$ encoding without gain, a loss event on the $\transefh$ transition takes the ancilla to $\ket{\psiloss} = \estateh_\mr{a}$, which lies outside the computational subspace spanned by $\gstateh$ and $\fstateh$. The overlap vanishes and $\detinf = \ploss$, twice the undriven value. A single loss event is therefore twice as damaging for the $\transgfh$ encoding, not because it dephases the ancilla but because it causes the ancilla to leak out of the computational subspace. Once $\kappagain \gg \kappaqubit$, the engineered gain returns this population to $\fstateh$ before the end of the measurement, so that $\ket{\psiloss} = \fstateh_\mr{a}$ and $\left|\braket{\psi_+}{\psiloss}\right|^2 = 1/2$ again. The gain does not restore the superposition destroyed by the loss event, but only returns the ancilla to the computational subspace. In an actual experiment, we can also mitigate the residual leakage at small $\kappagain$ by assigning half of the leaked population to $\gstateh$ and half to $\fstateh$, or by flagging it as an erasure error~\cite{Kubica23,Putterman2025}.

The backaction in Fig.~\ref{fig:ancilla_binomial}a depends on the initial logical state. As shown in Subsec.~\ref{subsec:heatmon_syndrome_extraction}, a single-photon loss event on the ancilla does not commute with the dispersive interaction and imprints a random phase rotation on the logical qubit, which produces a logical error that the code cannot correct. During the dwell in $\estateh$ that follows a loss event, the residual interaction of Eq.~\eqref{eq:ancilla_interaction_parity} rotates each Fock state $\ket{n}_\mr{b}$ of the storage mode by an angle proportional to $n\chidisp\taudwell$. For $\onel_\mr{b} = \ket{2}_\mr{b}$, only a single Fock state is occupied, so this rotation is a global phase and the backaction vanishes. For $\zerol_\mr{b} = (\ket{0}_\mr{b} + \ket{4}_\mr{b})/\sqrt{2}$, the two components acquire a relative rotation proportional to $4\chidisp\taudwell$, which makes this state the most strongly affected. The states $(\zerol_\mr{b} \pm \onel_\mr{b})/\sqrt{2}$ are superpositions of $\ket{0}_\mr{b}$, $\ket{2}_\mr{b}$, and $\ket{4}_\mr{b}$ whose components rotate by $0$, $2\chidisp\taudwell$ and $4\chidisp\taudwell$ respectively. Their backaction is on average smaller than that of $\zerol_\mr{b}$ and independent of the sign of the superposition, so the corresponding markers overlap.

\begin{table}
    \caption{\textbf{Simulation parameters for the first-order binomial code.} $N_\mr{a}$ and $N_\mr{b}$ are the Fock-space truncation dimensions of the ancilla transmon and of the storage mode, $\anharm$ is the transmon anharmonicity of Eq.~\eqref{eq:ancilla_hamiltonian}, $\chidisp$ the dispersive coupling of Eq.~\eqref{eq:ancilla_interaction_parity}, $\kappaqubit$ the single-photon loss rate of the ancilla, and $T$ the interaction time of one parity measurement.}
    \label{tab:ancilla_sim_binomial}
    \begin{ruledtabular}
    \renewcommand{\arraystretch}{1.2}
    \begin{tabular}{lr}
        \textbf{Parameter} & \textbf{Value} \\
        \hline
        Ancilla truncation $N_\mr{a}$ & \Nasim \\
        Storage truncation $N_\mr{b}$ & \Nbsimbin \\
        Anharmonicity $\anharm/2\pi$ & \anharmsimval \\
        Dispersive coupling $\chidisp/2\pi$ & \chisimval \\
        Ancilla loss rate $\kappaqubit$ & \kappaonesimval \\
        Interaction time $T$ & $\pi/2\chidisp$ \\
    \end{tabular}
    \end{ruledtabular}
\end{table}

\begin{figure}
    \centering
    \includegraphics[width=\linewidth]{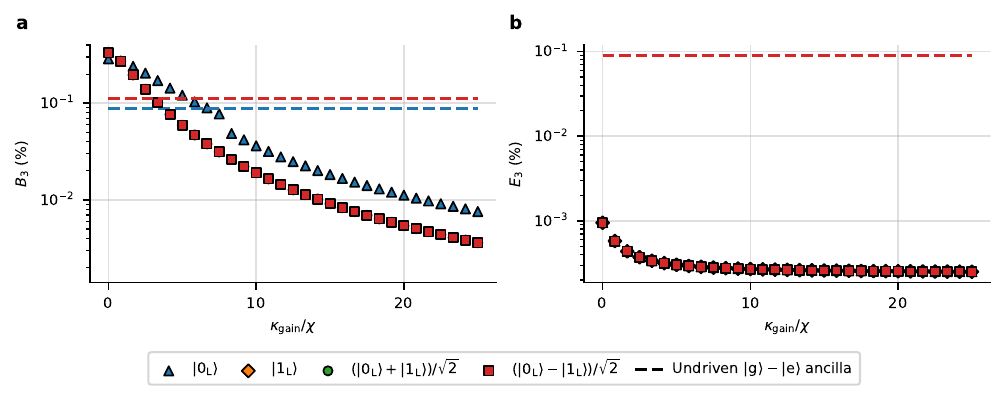}
    \caption{\textbf{First-order binomial code with three repeated measurements.} \textbf{a,} Measurement backaction $\backact_3$ as a function of $\kappagain$ for three repeated measurements with a majority vote. The backaction for the states $\left( \zerol_\mr{b} \pm \onel_\mr{b} \right) / \sqrt{2}$ is identical and the markers overlap, while the backaction for the state $\onel_\mr{b}$ is zero. \textbf{b,} Error-detection infidelity $\detinf_3$ as a function of $\kappagain$. The infidelity is identical for all initial logical states and the markers overlap. Dashed lines are $\backact$ and $\detinf$ for an undriven $\transgeh$ transmon ancilla with a single measurement. Simulation parameters are listed in Table~\ref{tab:ancilla_sim_binomial}.}
    \label{fig:ancilla_binomial_maj}
\end{figure}

We can overcome the elevated error-detection infidelity by repeating the ancilla measurement and taking a majority vote over the outcomes~\cite{Puri2019}. For three repeated measurements, the combined infidelity is
\begin{equation}
    \detinf_3 = 3\detinf^2(1-\detinf) + \detinf^3.
    \label{eq:detec_inf_repeated}
\end{equation}
Figures~\ref{fig:ancilla_binomial_maj}a,b show the backaction $\backact_3$ and the error-detection infidelity $\detinf_3$ for the repeated-measurement case. The total simulation time grows to $3T$, which increases the backaction of the ancilla on the logical qubit. Nevertheless, this repetition scheme reduces both $\backact_3$ and $\detinf_3$ below the values of a single stabilizer measurement with an undriven ancilla transmon, which demonstrates a clear advantage of the gain-stabilized ancilla. This advantage requires $\kappagain \gg \chidisp$, a regime that calls for a larger engineered gain rate than we reach here, as we discuss in Subsec.~\ref{subsubsec:ancilla_discussion}.

\subsubsection{Four-legged cat code}
\label{subsubsec:ancilla_cat}

The four-legged cat code is built from superpositions of coherent states~\cite{Mirrahimi2014,Ofek2016}. Following Ref.~\cite{Ofek2016}, the logical codewords are
\begin{equation}
    \zerol_\mr{b} = \catp_\mr{b}, \quad \onel_\mr{b} = \catpj_\mr{b},
    \label{eq:cat_codewords}
\end{equation}
which are cat states of amplitude $\alphacat$ with even photon-number parity. Here, $\alphacat$ denotes the cat amplitude, which we distinguish from the transmon anharmonicity $\anharm$ used elsewhere in this work. A single-photon loss event maps the logical states into the odd-parity subspace spanned by
\begin{equation}
    \zeroe_\mr{b} = \catm_\mr{b}, \quad \onee_\mr{b} = \catmj_\mr{b}.
    \label{eq:cat_errorwords}
\end{equation}
Each single-photon loss event maps the cat qubit into an orthogonal subspace and adds a $\pi/2$ rotation around the $Z$ axis of the logical Bloch sphere~\cite{Mirrahimi2014,Ofek2016}. Therefore, by monitoring the parity, we enable full correction of single-photon loss events.

\begin{figure}[h]
    \centering
    \includegraphics[width=\linewidth]{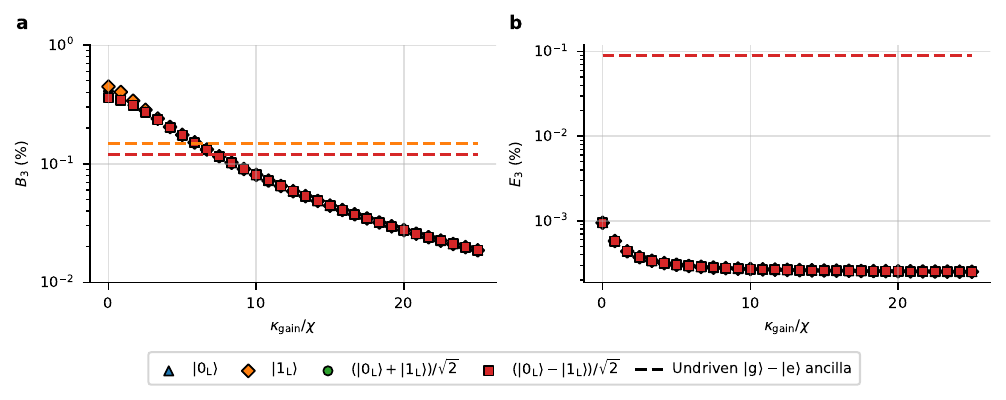}
    \caption{\textbf{Four-legged cat code with three repeated measurements.} \textbf{a,} Measurement backaction $\backact_3$ as a function of $\kappagain/\chidisp$. \textbf{b,} Error-detection infidelity $\detinf_3$ as a function of $\kappagain/\chidisp$. In panel \textbf{b}, $\detinf_3$ is identical for all initial logical states and the markers overlap. In panel \textbf{a}, the backaction is state dependent at small $\kappagain/\chidisp$, where the codewords $\zerol_\mr{b}$ and $\onel_\mr{b}$ show a larger backaction than their superpositions $(\zerol_\mr{b} \pm \onel_\mr{b})/\sqrt{2}$, and becomes identical for all states once $\kappagain \gg \chidisp$. Results are shown for the gain-engineered transmon ancilla with three repeated measurements and a majority vote. Dashed lines are $\backact$ and $\detinf$ for an undriven $\transgeh$ transmon ancilla with a single measurement. Simulation parameters are listed in Table~\ref{tab:ancilla_sim_cat}.}
    \label{fig:ancilla_cat}
\end{figure}

Similar to the protocol used in binomial code case, the logical parity unitary Eq.~\eqref{eq:ancilla_interaction_parity_unitary} is realised by evolving the dispersive interaction Eq.~\eqref{eq:ancilla_interaction_parity} between the ancilla and the logical qubit, for a time $T = \pi/2\chidisp$. We repeat the simulations performed for the binomial code; we apply three repeated checks with a majority vote and compute the resulting backaction and error-detection infidelity (Eq.~\eqref{eq:detec_inf_repeated}) for all logical states of the four-legged cat qubit. The results and simulation parameters are shown in Fig.~\ref{fig:ancilla_cat} and Table~\ref{tab:ancilla_sim_cat}. This scheme reduces both $\backact_3$ and $\detinf_3$ below the values of a single stabilizer measurement with an undriven ancilla transmon, which demonstrates an advantage of the gain-engineered transmon ancilla. As above, $\detinf_3$ lies below the undriven value over the whole simulated range, while $\backact_3$ requires $\kappagain \gg \chidisp$.

While the error-detection infidelity is identical for all initial states, the backaction in Fig.~\ref{fig:ancilla_cat}a depends on the initial logical state at small $\kappagain/\chidisp$. A loss event on the ancilla imprints a random rotation in phase space on the storage mode, with $\thetaerr \propto \chidisp\taudwell$ and a dwell time $\taudwell \propto 1/\kappagain$. At small $\kappagain$, the dwell time is long and the ancilla can decay further to $\gstateh$, so the random rotations are large. The codewords $\zerol_\mr{b}$ and $\onel_\mr{b}$ are more sensitive to this rotation than their superpositions $(\zerol_\mr{b} \pm \onel_\mr{b})/\sqrt{2}$, so their backaction is higher. As $\kappagain$ increases, $\thetaerr \to 0$, the random rotations become negligible, and the backaction becomes the same for all logical states.

\begin{table}[h]
    \caption{\textbf{Simulation parameters for the four-legged cat code.} $N_\mr{a}$ and $N_\mr{b}$ are the Fock-space truncation dimensions of the ancilla transmon and of the storage mode, $\anharm$ is the transmon anharmonicity of Eq.~\eqref{eq:ancilla_hamiltonian}, $\chidisp$ the dispersive coupling of Eq.~\eqref{eq:ancilla_interaction_parity}, $\alphacat$ the amplitude of the cat codewords of Eq.~\eqref{eq:cat_codewords}, $\kappaqubit$ the single-photon loss rate of the ancilla, and $T$ the interaction time of one parity measurement.}
    \label{tab:ancilla_sim_cat}
    \begin{ruledtabular}
    \renewcommand{\arraystretch}{1.2}
    \begin{tabular}{lr}
        \textbf{Parameter} & \textbf{Value} \\
        \hline
        Ancilla truncation $N_\mr{a}$ & \Nasim \\
        Storage truncation $N_\mr{b}$ & \Nbsimcat \\
        Anharmonicity $\anharm/2\pi$ & \anharmsimval \\
        Dispersive coupling $\chidisp/2\pi$ & \chisimval \\
        Cat amplitude $\alphacat$ & \alphacatval \\
        Ancilla loss rate $\kappaqubit$ & \kappaonesimval \\
        Interaction time $T$ & $\pi/2\chidisp$ \\
    \end{tabular}
    \end{ruledtabular}
\end{table}

\subsubsection{Gottesman-Kitaev-Preskill code}
\label{subsubsec:ancilla_gkp}

The Gottesman-Kitaev-Preskill (GKP) code~\cite{Gottesman2001} encodes a logical qubit in coherent superpositions of periodically displaced squeezed vacuum states, which form a periodic lattice in the phase space of a harmonic oscillator. The logical codewords are the $+1$ eigenstates of the stabilizer operators
\begin{equation}
    \hat{S}_\mr{x} = e^{2i\sqrt{\pi} \hat{x}}, \quad \hat{S}_\mr{p} = e^{-2i\sqrt{\pi} \hat{p}},
    \label{eq:gkp_stabilizers}
\end{equation}
where $\hat{x} = \frac{1}{\sqrt{2}}(\bop + \bod)$ and $\hat{p} = \frac{i}{\sqrt{2}}(\bod - \bop)$ are the dimensionless quadrature operators. Errors appear as small displacements in position or momentum, and we detect them by measuring the stabilizers. Ideal GKP states extend infinitely in phase space and carry infinite energy, which makes them unphysical, so practical implementations use finite-energy approximations~\cite{Gottesman2001,CampagneIbarcq2020,Sivak2023}. Following Ref.~\cite{Puri2019}, we construct approximate codewords as
\begin{align}
    \zerol_\mr{b} &= N_0 \sum_{n=-1}^{1} \binom{2}{n+1} \hat{D}(\sqrt{2\pi}\,n)\, \hat{S}_r \ket{0}_\mr{b},  \label{eq:gkp_zerol}\\
    \onel_\mr{b} &= N_1 \sum_{n=-1}^{1} \binom{2}{n+1} \hat{D}(i\sqrt{2\pi}\,n)\, \hat{S}_r \ket{0}_\mr{b},
    \label{eq:gkp_onel}
\end{align}
where $N_{0,1}$ are normalization constants, $\hat{S}_r = e^{\frac{1}{2}(r^*\bop^2 - r\bop^{2\dagger})}$ is the squeezing operator with squeezing parameter $r$, and $\hat{D}(\epsilon) = e^{\epsilon \bod - \epsilon^* \bop}$ is the displacement operator with $\epsilon \in \mathbb{C}$.

To perform error detection, we engineer a conditional-displacement interaction between the ancilla transmon and the storage mode~\cite{Puri2019,Touzard2019,CampagneIbarcq2020,Sivak2023},
\begin{equation}
    \Vint/\hbar = \aod \aop (\beta \bod + \beta^* \bop),
    \label{eq:ancilla_interaction_gkp}
\end{equation}
where $\beta$ is the conditional-displacement interaction strength, so that $\etaint = |\beta|$ in the definition of Subsec.~\ref{subsec:heatmon_syndrome_extraction}. Evolving under this interaction for a time $T = \sqrt{\frac{\pi}{2}}\frac{1}{|\beta|}$ generates the conditional unitary~\cite{Puri2019}
\begin{align}
    U(T) &= \hat{D}\!\left(-i\sqrt{\frac{\pi}{2}}\right)\left[\frac{\hat{\sigma}_\mr{z,a}+1}{2}\hat{D}(i\sqrt{2\pi})+\frac{1-\hat{\sigma}_\mr{z,a}}{2}\right] \quad \text{for } \beta \in \mathbb{R}, \\
    U(T) &= \hat{D}\!\left(-\sqrt{\frac{\pi}{2}}\right)\left[\frac{\hat{\sigma}_\mr{z,a}+1}{2}\hat{D}(\sqrt{2\pi})+\frac{1-\hat{\sigma}_\mr{z,a}}{2}\right] \quad \text{for } \beta \in i\mathbb{R}.
    \label{eq:gkp_conditional_unitary}
\end{align}
These conditional displacements are the elementary operations of a GKP error-correction cycle.

We simulate a single error-correction cycle targeting the $\hat{p}$-quadrature stabilizer $\hat{S}_\mr{p}$. In contrast to the previous two encodings, we do not characterize the performance through the backaction and the error-detection infidelity. Instead, we use the Holevo phase variance of the storage-mode state $\rhob$~\cite{Holevo84},
\begin{equation}
    V_\mr{x,p}(\rhob) = \left|\mathrm{tr}\left(\rhob\,\hat{S}_{\mr{x,p}}\right)\right|^{-2}-1,
    \label{eq:holevo_variance}
\end{equation}
which quantifies how well $\rhob$ satisfies the stabilizer condition and vanishes for an ideal GKP codeword. We initialize the joint system in
\begin{equation}
    \ket{\psi_0} = \ket{\psi_+}_\mr{a} \otimes \zerol_\mr{b},
\end{equation}
from which we obtain the initial Holevo variance $V_\mr{x,p}^0 \equiv V_\mr{x,p}(\rho_\mr{b,0})$, with $\rho_\mr{b,0} = \zerol_\mr{b}\bra{0_\mr{L}}_\mr{b}$ the initial density matrix of the storage mode. Evolving the joint state for a time $T$ under the master equation of Eq.~\eqref{eq:ancilla_me}, with $\beta = i|\beta|$ and the parameters of Table~\ref{tab:ancilla_sim_gkp}, yields the post-measurement Holevo variances conditioned on the ancilla outcome~\cite{Puri2019},
\begin{equation}
    V_{\mr{x}}' = p_+ V_\mr{x}(\rho_{\mr{b,+}}) + p_- V_{\mr{x}}(\rho_{\mr{b,-}}), \qquad V_{\mr{p}}' = p_+ V_{\mr{p}}(\rho_{\mr{b,+}}) + p_- V_{\mr{p}}(\rho_{\mr{b,-}}),
\end{equation}
where $\rho_{\mr{b,\pm}}$ is the reduced density matrix of the storage mode conditioned on the ancilla being measured in $\ket{\psi_\pm}_\mr{a}$, and $p_\pm$ are the corresponding outcome probabilities. We quantify the outcome of an error-correction round through the relative change in the Holevo variance,
\begin{equation}
    \frac{\Delta V_\mr{x,p}}{V_\mr{x,p}^0} \equiv \frac{V_\mr{x,p}' - V_\mr{x,p}^0}{V_\mr{x,p}^0}.
    \label{eq:holevo_relative_change}
\end{equation}

\begin{figure}
    \centering
    \includegraphics[width=\linewidth]{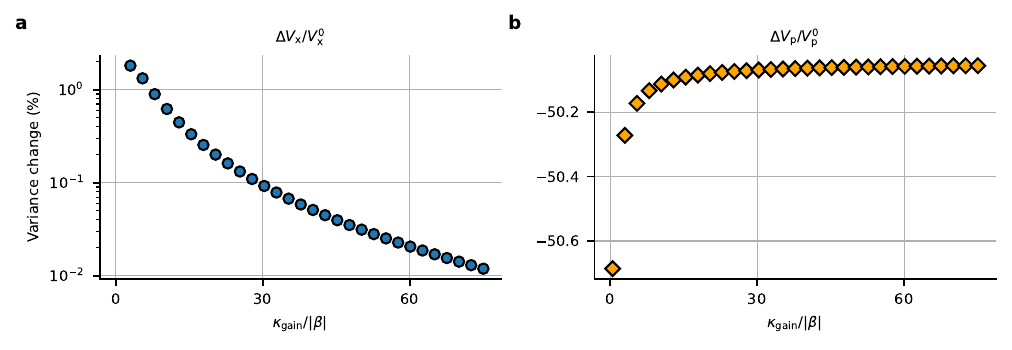}
    \caption{\textbf{Gottesman-Kitaev-Preskill code.} \textbf{a,} Relative change in the Holevo phase variance of the $\hat{x}$ quadrature, $\Delta V_\mr{x} / V_\mr{x}^0$, as a function of $\kappagain/|\beta|$ for an $\hat{S}_\mr{p}$ stabilizer measurement. \textbf{b,} Relative change in the Holevo phase variance of the $\hat{p}$ quadrature, $\Delta V_\mr{p} / V_\mr{p}^0$, as a function of $\kappagain/|\beta|$. Simulation parameters are listed in Table~\ref{tab:ancilla_sim_gkp}.}
    \label{fig:ancilla_gkp}
\end{figure}

Figure~\ref{fig:ancilla_gkp} shows $\Delta V_\mr{x,p}/V_\mr{x,p}^0$ as a function of the engineered gain rate $\kappagain$. In an ideal error-correction round, the Holevo phase variance of the targeted quadrature $\hat{p}$ decreases while that of the conjugate quadrature $\hat{x}$ remains unchanged, that is $\Delta V_\mr{p}/V_\mr{p}^0 < 0$ and $\Delta V_\mr{x}/V_\mr{x}^0 = 0$. The simulation shows that for $\kappagain \ll |\beta|$ ancilla loss errors propagate to the storage mode and increase $\Delta V_\mr{x}/V_\mr{x}^0$. In the regime $\kappagain \gg |\beta|$, by contrast, loss errors no longer propagate to the logical qubit, and the error-correction round substantially reduces $\Delta V_\mr{p}/V_\mr{p}^0$ while leaving $\Delta V_\mr{x}/V_\mr{x}^0$ unaffected.

\begin{table}
    \caption{\textbf{Simulation parameters for the GKP code.} $N_\mr{a}$ and $N_\mr{b}$ are the Fock-space truncation dimensions of the ancilla transmon and of the storage mode, $\anharm$ is the transmon anharmonicity of Eq.~\eqref{eq:ancilla_hamiltonian}, $|\beta|$ the conditional-displacement rate of Eq.~\eqref{eq:ancilla_interaction_gkp}, $r$ the squeezing parameter of the finite-energy codewords of Eqs.~\eqref{eq:gkp_zerol} and~\eqref{eq:gkp_onel}, $\kappaqubit$ the single-photon loss rate of the ancilla, and $T$ the interaction time of one stabilizer measurement.}
    \label{tab:ancilla_sim_gkp}
    \begin{ruledtabular}
    \renewcommand{\arraystretch}{1.2}
    \begin{tabular}{lr}
        \textbf{Parameter} & \textbf{Value} \\
        \hline
        Ancilla truncation $N_\mr{a}$ & \Nasim \\
        Storage truncation $N_\mr{b}$ & \Nbsimgkp \\
        Anharmonicity $\anharm/2\pi$ & \anharmsimval \\
        Conditional-displacement rate $|\beta|/2\pi$ & \betagkpval \\
        Squeezing parameter $r$ & \rgkpval \\
        Ancilla loss rate $\kappaqubit$ & \kappaonesimval \\
        Interaction time $T$ & $\sqrt{\pi/2}\,/|\beta|$ \\
    \end{tabular}
    \end{ruledtabular}
\end{table}

\subsubsection{Discussion and limitations}
\label{subsubsec:ancilla_discussion}

These results show that the gain-engineered transmon can serve as an ancilla qubit for fault-tolerant error detection. For the binomial code, a single stabilizer measurement lowers the measurement backaction $\backact$ on the logical qubit below the value achieved with an undriven $\transgeh$ transmon ancilla, while the error-detection infidelity $\detinf$ stays above it because of the reduced Ramsey coherence time $\Ttwo{\mr{gf}}$ of the gain-engineered transmon. For both the binomial and the four-legged cat encodings, three repeated measurements with a majority vote bring both metrics below the values of a single measurement with an undriven ancilla. The infidelity $\detinf_3$ lies below that value over the whole simulated range of $\kappagain$, whereas the backaction $\backact_3$ crosses below it only for $\kappagain \gg \chidisp$. For the GKP encoding, taking an $\hat{S}_\mr{p}$ stabilizer measurement as a representative example, the gain-engineered ancilla reduces the Holevo phase variance $V_\mr{p}$ of the targeted quadrature while suppressing the backaction-induced increase in $V_\mr{x}$. The common mechanism behind all three improvements is the suppression of ancilla loss errors by the engineered gain, which prevents these errors from propagating to the logical qubit.

The model of Eq.~\eqref{eq:ancilla_diss} omits two ancilla noise channels, dephasing noise and thermal excitations, in order to isolate the effect of relaxation. As shown in Subsec.~\ref{subsec:heatmon_syndrome_extraction}, the dephasing operator commutes with the interaction Hamiltonian $\Vint$ of Eq.~\eqref{eq:Vint_ancilla} and therefore leaves the measurement backaction $\backact$ on the logical qubit unaffected. The repeated-measurement protocol introduced above mitigates the associated increase in $\detinf$, which makes dephasing a manageable source of errors. Thermal excitations, by contrast, generate errors that do not commute with $\Vint$ and could increase both $\detinf$ and $\backact$. Quantifying their impact is an important direction for future work.

A further simplification in the simulations is that we model the engineered gain as a Lindblad dissipator $\kappagain\diss{\fstateh\estatehb_\mr{a}}$ acting at a single rate (Eq.~\eqref{eq:ancilla_diss}), independent of the ancilla $\transefh$ transition frequency. In practice, the gain rate follows the cavity susceptibility, with linewidth $\kappacav$ centered on the resonance set by $\omegagain$ (Subsec.~\ref{subsec:frq_sel_gain}). However, the interaction Hamiltonian $\Vint$ shifts the ancilla $\transefh$ transition frequency conditional on the state of the storage mode. Therefore, the engineered gain must address several resonance frequencies, each offset by a multiple of the interaction strength $\etaint$. In principle, the gain bandwidth can be broadened so that $\etaint \ll \kappacav \ll \anharm$ would cover all shifted $\transefh$ frequencies while preserving the frequency selectivity that protects the $\transgeh$ and $\transfhh$ transitions. Alternatively, a comb of pump tones can be used to address each shifted frequency individually. Both routes benefit from replacing the readout cavity with a dedicated multi-stage lossy filter~\cite{Putterman2022,Thorbeck24,Direkci26}, which would also provide the larger $\kappagain$ that the regime $\kappagain \gg \etaint$ requires.

\end{document}